\documentclass[prd,reprint,superscriptaddress,altaffillsymbol,amsmath,nofootinbib]{revtex4-1}
\pdfoutput=1

\usepackage{ifthen}
\usepackage{epsfig}
\usepackage{booktabs} 
\usepackage{natbib}
\usepackage{wasysym}
\usepackage{feynmp}
\usepackage{slashed} 
\usepackage{bbm}
\usepackage{mathrsfs}
\usepackage{amssymb}
\usepackage{dsfont}
\usepackage[export]{adjustbox}
\newcommand{\boldmidrule}{\specialrule{1.5pt}{0em}{0em}}

\usepackage{color}

\newcommand{\beqra}{\begin{eqnarray}}
\newcommand{\eeqra}{\end{eqnarray}}
\newcommand{\beq}{\begin{equation}}
\newcommand{\eeq}{\end{equation}}
\newcommand{\dd}{\mathrm{d}}

\newcommand{\mDM}{m_\text{DM}}

\newcommand{\unit}[1]{\,\mathrm{#1}}
\newcommand{\gev}{\unit{GeV}}

\newcommand{\ii}{\mathrm{i}}
\newcommand{\ee}{\mathrm{e}}
\newcommand{\hc}{\mathrm{h.c.}}
\renewcommand{\epsilon}{\varepsilon}

\newcommand{\reffig}[1]{Fig.~\ref{#1}}
\newcommand{\refeq}[1]{Eq.~(\ref{#1})}
\newcommand{\reftab}[1]{Tab.~\ref{#1}}

\newcommand{\Ref}{Ref.}
\newcommand{\Refs}{Refs.}
\newcommand{\app}{Appendix}
\renewcommand{\vec}[1]{\mathbf{#1}}
\renewcommand{\bar}{\overline}

\begin{document}

\title{Compatibility of a dark matter discovery at XENONnT/LZ with the WIMP thermal production mechanism}

\author{Riccardo Catena}
\email{catena@chalmers.se}
\affiliation{Chalmers University of Technology, Department of Physics, SE-412 96 G\"oteborg, Sweden}

\author{Jan Conrad}
\email{conrad@fysik.su.se}
\affiliation{Oskar Klein Centre, Department of Physics, Stockholm University, AlbaNova, Stockholm SE-10691, Sweden}

\author{Martin B.~Krauss}
\email{martin.krauss@chalmers.se}
\affiliation{Chalmers University of Technology, Department of Physics, SE-412 96 G\"oteborg, Sweden}

\begin{abstract}
The discovery of dark matter (DM) at XENONnT or LZ would place constraints on DM particle mass and coupling constants.~It is interesting to ask when these constraints can be compatible with the DM thermal production mechanism.~We address this question within the most general set of renormalisable models that preserve Lorentz and gauge symmetry, and that extend the Standard Model by one DM candidate of mass $m_{\rm DM}$ and one particle of mass $M_{\rm med}$ mediating DM-quark interactions.~Our analysis divides into two parts.~First, we postulate that XENONnT/LZ has detected $\mu_S\sim\mathcal{O}(100)$ signal events, and use this input to calculate the DM relic density, $\Omega_{\rm DM} h^2$.~Then, we identify the regions in the  $M_{\rm med} - \Omega_{\rm DM} h^2$ plane which are compatible with the observed signal and with current CMB data.~We find that for most of the models considered here, $\mathcal{O}(100)$ signal events at XENONnT/LZ and the DM thermal production are only compatible for resonant DM annihilations, i.e.~for $M_{\rm med}\simeq2 m_{\rm DM}$.~In this case, XENONnT/LZ would be able to simultaneously measure $m_{\rm DM}$ and $M_{\rm med}$.~We also discuss the dependence of our results on $m_{\rm DM}$, $\mu_S$ and the DM spin, and provide analytic expressions for annihilation cross-sections and mediator decay widths for all models considered in this study.
\end{abstract}

\maketitle

\section{Introduction}
\label{sec:introduction}
The evidence for the presence of dark matter (DM) in the Universe is based upon the observation of anomalous gravitational effects in astronomical and cosmological systems~\cite{Bertone:2016nfn}.~These systems range from stellar populations in the solar neighbourhood~\cite{Iocco:2015xga} to the large scale structure of the Universe~\cite{Springel:2006vs}.~The Cosmic Microwave Background (CMB) radiation is one of the key physical observables in this context~\cite{Mukhanov:2003xr}.~In particular, measurements of the CMB angular power spectrum performed by the Planck satellite set the DM relic density to critical density ratio to $\Omega_{\rm DM}h^2=0.1199 \pm 0.0022$, with a relative error of less than 2\%~\cite{Ade:2015xua}.~A particle in the few GeV up to about 300~TeV mass range with coupling constants at the weak scale is expected to be in thermal equilibrium in the early Universe, and to (chemically) decouple from the thermal bath with a relic density which is typically within a factor of a few from the observed value of $\Omega_{\rm DM}h^2$~\cite{Lee:1977ua,Griest:1989wd}.~This mechanism to generate $\Omega_{\rm DM}h^2$ via chemical decoupling is called DM thermal production.~Weakly Interacting Massive Particles (WIMPs) are a typical example of thermally produced DM particles, and by far the most extensively studied candidates for particle DM.~For a recent review on WIMPs as a DM candidate, see~\cite{Roszkowski:2017nbc,Arcadi:2017kky}.~Alternative production mechanisms, e.g., freeze-in and misalignment mechanism, are reviewed in~\cite{Hall:2009bx,Baer:2014eja}.~Here, we will focus on thermally produced WIMPs.

The WIMP relic density is obtained through the numerical solution of a collisional Boltzmann equation~\cite{Gondolo:1990dk}.~The collisional term in this equation takes into account particle physics processes affecting the rate of DM annihilation.~These include resonances and co-annihilations~\cite{Edsjo:1997bg}.~Motivated by increasingly accurate CMB data on $\Omega_{\rm DM}h^2$~\cite{Ade:2015xua}, predictions for the DM relic density have improved dramatically in recent years and, besides the already mentioned phenomena, now also include, e.g., next-to-leading order QCD corrections~\cite{Harz:2016dql} and the Sommerfeld enhancement of the DM annihilation cross-section due to possible DM self-interactions~\cite{Feng:2010zp}.~The impact of a non-standard cosmological expansion on the DM relic density, e.g.~induced by modifications of General Relativity on cosmological scales, has also been studied in detail~\cite{Catena:2004ba,Gelmini:2006pw}.~Computer programmes implementing accurate calculations of the DM relic density include {\sffamily darksusy}~\cite{Gondolo:2004sc}, {\sffamily MicrOMEGAs}~\cite{Belanger:2006is} and {\sffamily MadDM}~\cite{Backovic:2013dpa}.~See, e.g.,~\cite{Bergeron:2017rdm} for a recent comparison of relic density calculations performed using different computer programs.

Among the strategies designed for DM particle identification, the DM direct detection technique will be crucial in the next decade~\cite{Baudis:2012ig}.~It searches for nuclear recoil events induced by the scattering of Milky Way DM particles in low-background detectors.~See, e.g.,~\cite{Aprile:2017iyp,Cui:2017nnn,Undagoitia:2015gya} for a review on the current status of direct detection experiments.~The discovery of DM at direct detection experiments would constrain the DM particle mass and coupling constants.~Importantly, the same parameters are also constrained by CMB data on the DM relic density.~Therefore, it is interesting to ask whether the discovery of DM at XENONnT and the DM thermal production mechanism can be compatible.~Here we systematically address this question focusing on DM candidates primarily interacting with quarks.~Prospects for DM direct detection via DM-lepton interactions have been discussed in, e.g.~\cite{Essig:2011nj,DEramo:2016gos,Emken:2017erx}.

The aim of this work is to determine under what circumstances the DM thermal production mechanism is compatible with the discovery of DM at XENONnT/LZ.~In the analysis we assume that XENONnT/LZ has detected $\mu_{\rm S}\sim\mathcal{O}(100)$ signal events.~A comparable number of events is expected for DM-nucleus scattering cross-sections just below current exclusion limits.~We then use this information on $\mu_S$ as an input to compute the DM relic density.~The latter is computed within the most general set of renormalisable models compatible with Lorentz and gauge symmetry extending the Standard Model by one DM candidate and one particle mediating DM-quark interactions.~For most of the models considered here, we find that $\mu_{\rm S}\sim\mathcal{O}(100)$ signal events at XENONnT/LZ are compatible with DM thermal production only for a narrow range of mediator masses, i.e.~for $M_{\rm med}\simeq2m_{\rm DM}$, where $M_{\rm med}$ and $m_{\rm DM}$ are the mediator and DM particle mass, respectively.~In this case, a signal at XENONnT/LZ would determine the value of $M_{\rm med}$ and $m_{\rm DM}$ simultaneously, as long as DM is a WIMP.~In the analysis, we also relate the compatibility of DM thermal production and DM discovery at XENONnT/LZ to the DM spin and interaction properties.~Finally, we discuss the dependence of our results on $m_{\rm DM}$ and $\mu_S$.

The paper is organised as follows.~In Sec.~\ref{sec:theory} we review the set of single-mediator DM models considered in this study.~For each model the relic density is computed as described in Sec.~\ref{sec:relic} under the assumption that XENONnT or LZ has detected DM, constraining the model parameters as illustrated in Sec.~\ref{sec:DD}.~We present our results in Sec.~\ref{sec:results}, discuss their dependence on $m_{\rm DM}$ and $\mu_S$ in Sec.~\ref{sec:discussion}, and conclude in Sec.~\ref{sec:conclusion}.~For all models considered in this study, analytic expressions for annihilation cross-sections and mediator decay widths are listed in the Appendices. 

\section{Theoretical Framework}
\label{sec:theory}

We calculate DM relic density and DM-nucleus scattering cross-sections within the theoretical framework introduced in~\cite{Dent:2015zpa}.~This framework consists of ``simplified models'' extending the Standard Model by one DM candidate and one particle that mediates the interactions of DM with quarks.~This approach is usually sufficient to obtain a good understanding of the DM particle phenomenology, especially in the analysis of DM direct detection experiments and in the calculation of the DM relic density.~Simplified models for DM have also extensively been used in the experimental analysis and theoretical interpretation of LHC data, e.g.,~\cite{Abercrombie:2015wmb}, and to design strategies to extract DM particle spin~\cite{Catena:2017wzu,Baum:2017kfa,Capdevilla:2017doz,Kamon:2017yfx} and particle/antiparticle nature~\cite{Queiroz:2016sxf,Kavanagh:2017hcl} from future data.~Generically, simplified models are characterised by the nature of the DM particle, which can be a scalar $S$, a fermion $\chi$ or a vector $X^\mu$.~If DM has spin 1/2, we assume that the field $\chi$ is a Dirac rather than Majorana spinor.~Independently of its spin, DM is taken to be neutral with respect to the Standard Model gauge group, but can carry charge under some additional gauge or discrete symmetry group.~In this case, the scalar $S$ and the vector $X^\mu$ must be complex fields.~For example, assuming that DM is odd and all other particles are even under a new $\mathcal{Z}_2$ symmetry would guarantee that DM is stable on cosmological time-scales.~With the above notation, the Lagrangian of a simplified model for fermionic DM interacting with quarks $q$ through a vector mediator $G_\mu$, e.g.~a $Z'$-boson, can be written as follows
\begin{align}
 \mathcal{L} \ &= \mathcal{L}_\text{SM} + \ii \bar\chi \slashed{D}\chi - m_\chi \bar \chi \chi \nonumber \\ &
 -\frac{1}{4} \mathcal{G}_{\mu\nu}\mathcal{G}^{\mu\nu} + \frac{1}{2} M_{G}^2 G_\mu {G}^\mu \nonumber\\ &
 + \ii \bar q \slashed{D} q - m_q \bar q q \nonumber\\ &
 - \lambda_3 \bar \chi \gamma^\mu \chi G_\mu - \lambda_4 \bar \chi \gamma^\mu \gamma^5 \chi G_\mu \nonumber\\ &
 - h_3 \bar q \gamma^\mu q G_\mu - h_4 \bar q \gamma^\mu \gamma^5 q G_\mu \,,
 \label{eq:LagrGchi}
\end{align}
where a sum over all quark flavors $(q=u,d,c,s,b,t)$ is understood in the terms involving quark bilinears, and $\lambda_3$,  $\lambda_4$, $h_3$, $h_4$ are dimensionless constants.~Here we assume universal quark-mediator couplings, i.e.~the same $h_3$ and $h_4$ for all quark flavors.~Lagrangians for all models introduced in~\cite{Dent:2015zpa} and considered here are listed in \app~\ref{app:Lagrangians}.~For a given model, we will only consider cases where two couplings  are different from zero at the same time.

In the non-relativistic limit, the models in \app~\ref{app:Lagrangians} match onto (linear combinations of) non-relativistic operators for DM-nucleon interactions.~These operators are invariant under Hermitian conjugation and Galilean transformations, i.e.~constant shifts of particle velocities, but can in certain cases violate $T$ symmetry, as shown in~\cite{Fitzpatrick:2012ix}.~They have matrix elements between incoming and outgoing DM-nucleon states which can be expressed as combinations of the basic invariants under theses symmetries,
\begin{equation}
 \ii \vec q\,,\quad
 \vec v^\perp \equiv \vec v + \frac{\vec q}{2 \mu_N}\,,\quad
 \vec S_\chi\,,\quad\text{and} \quad
 \vec S_N\,,
\end{equation}
where $\vec v$ is the DM-nucleon relative velocity, $\vec q$ the three-dimensional momentum transfer in the DM-nucleus scattering, and $\vec S_{\rm N}$ and $\vec S_{\chi}$ the spin of the DM and of the nucleon, respectively.~DM-nucleon interaction operators have systematically been classified in~\cite{Fitzpatrick:2012ix}, focusing on spin $\le 1/2$ DM.~This classification has subsequently been extended to spin 1 DM in~\cite{Dent:2015zpa}.~At linear order in the DM-nucleon relative velocity and at second order in the momentum transfer, 16 independent operators can be constructed, although not all of them appear as leading operators in the non-relativistic limit of our simplified models for DM-quark interactions.~The 16 operators are listed in \reftab{tab:operators}.

For each simplified model in \app~\ref{app:Lagrangians}, the Hamiltonian for DM-nucleon interactions can be expressed as follows
\begin{equation}
 \mathcal{H}_\text{int} = \sum_{N=n,p} \sum_i c_i^{(N)} \hat{\mathcal{O}}_i^{(N)} \,,
\label{eq:H}
\end{equation}
where $N=p$ and $N=n$ denotes coupling to protons and neutrons, respectively, the index $i=1,\dots,18$ ($i\neq2$ and $i\neq16$; see caption in Tab.~\ref{tab:operators}) labels the interaction type, and the coupling constants $c_i^{(N)}$ have dimension mass to the power of $-2$.~From Eq.~(\ref{eq:H}), one can calculate cross-sections for DM-nucleus scattering as described in \Refs~\cite{Fitzpatrick:2012ix,Anand:2013yka,Catena:2015uha} in detail.~Prospects for DM particle detection in this framework have been studied in~\cite{Catena:2014hla,Catena:2014epa,Catena:2015vpa,Kavanagh:2016pyr,Catena:2016sfr,Catena:2016tlv,Catena:2017wzu}, and limits on the coupling constants  $c_i^{(N)}$ have been derived in, e.g.,~\cite{DelNobile:2013sia,Catena:2014uqa,Gresham:2014vja,Catena:2015uua,Catena:2015iea,Catena:2016kro,Aprile:2017aas},

\begin{table}[t]
    \centering
    \begin{ruledtabular}	
    \begin{tabular}{l}
    \toprule
    \boldmidrule
	 \toprule
	  \toprule
        $\hat{\mathcal{O}}_1 = \mathds{1}_{\chi}\mathds{1}_N$  \\
        $\hat{\mathcal{O}}_3 = i{\bf{\hat{S}}}_N\cdot\left(\frac{{\bf{\hat{q}}}}{m_N}\times{\bf{\hat{v}}}^{\perp}\right)\mathds{1}_\chi$ \\
        $\hat{\mathcal{O}}_4 = {\bf{\hat{S}}}_{\chi}\cdot {\bf{\hat{S}}}_{N}$ \\
        $\hat{\mathcal{O}}_5 = i{\bf{\hat{S}}}_\chi\cdot\left(\frac{{\bf{\hat{q}}}}{m_N}\times{\bf{\hat{v}}}^{\perp}\right)\mathds{1}_N$ \\
        $\hat{\mathcal{O}}_6 = \left({\bf{\hat{S}}}_\chi\cdot\frac{{\bf{\hat{q}}}}{m_N}\right) \left({\bf{\hat{S}}}_N\cdot\frac{\hat{{\bf{q}}}}{m_N}\right)$\\
        $\hat{\mathcal{O}}_7 = {\bf{\hat{S}}}_{N}\cdot {\bf{\hat{v}}}^{\perp}\mathds{1}_\chi$ \\
        $\hat{\mathcal{O}}_8 = {\bf{\hat{S}}}_{\chi}\cdot {\bf{\hat{v}}}^{\perp}\mathds{1}_N$  \\
        $\hat{\mathcal{O}}_9 = i{\bf{\hat{S}}}_\chi\cdot\left({\bf{\hat{S}}}_N\times\frac{{\bf{\hat{q}}}}{m_N}\right)$ \\
        $\hat{\mathcal{O}}_{10} = i{\bf{\hat{S}}}_N\cdot\frac{{\bf{\hat{q}}}}{m_N}\mathds{1}_\chi$   \\
        $\hat{\mathcal{O}}_{11} = i{\bf{\hat{S}}}_\chi\cdot\frac{{\bf{\hat{q}}}}{m_N}\mathds{1}_N$   \\
        $\hat{\mathcal{O}}_{12} = {\bf{\hat{S}}}_{\chi}\cdot \left({\bf{\hat{S}}}_{N} \times{\bf{\hat{v}}}^{\perp} \right)$  \\
        $\hat{\mathcal{O}}_{13} =i \left({\bf{\hat{S}}}_{\chi}\cdot {\bf{\hat{v}}}^{\perp}\right)\left({\bf{\hat{S}}}_{N}\cdot \frac{{\bf{\hat{q}}}}{m_N}\right)$ \\
        $\hat{\mathcal{O}}_{14} = i\left({\bf{\hat{S}}}_{\chi}\cdot \frac{{\bf{\hat{q}}}}{m_N}\right)\left({\bf{\hat{S}}}_{N}\cdot {\bf{\hat{v}}}^{\perp}\right)$ \\
        $\hat{\mathcal{O}}_{15} = -\left({\bf{\hat{S}}}_{\chi}\cdot \frac{{\bf{\hat{q}}}}{m_N}\right)\left[ \left({\bf{\hat{S}}}_{N}\times {\bf{\hat{v}}}^{\perp} \right) \cdot \frac{{\bf{\hat{q}}}}{m_N}\right] $  \\
        $\hat{\mathcal{O}}_{17}=i \frac{{\bf{\hat{q}}}}{m_N} \cdot \mathbf{\mathcal{S}} \cdot {\bf{\hat{v}}}^{\perp} \mathds{1}_N$ \\
$\hat{\mathcal{O}}_{18}=i \frac{{\bf{\hat{q}}}}{m_N} \cdot \mathbf{\mathcal{S}}  \cdot {\bf{\hat{S}}}_{N}$ \\
    \bottomrule
      \bottomrule
      \boldmidrule
    \end{tabular}
    \end{ruledtabular}	
    \caption{Quantum mechanical operators defining the non-relativistic effective theory of DM-nucleon interactions~\cite{Fan:2010gt,Fitzpatrick:2012ix}.~Here we adopt the notation introduced in Sec.~\ref{sec:theory}.~Canonical spin-independent and spin-dependent interactions correspond to the operators $\hat{\mathcal{O}}_{1}$ and $\hat{\mathcal{O}}_{4}$, respectively.~Operator $\hat{\mathcal{O}}_{2}$ is quadratic in ${\bf{\hat{v}}}^{\perp}$ and $\hat{\mathcal{O}}_{16}$ is a linear combination of $\hat{\mathcal{O}}_{12}$ and $\hat{\mathcal{O}}_{15}$~\cite{Anand:2013yka}.~These are therefore not reported in the table.~The operators $\hat{\mathcal{O}}_{17}$ and $\hat{\mathcal{O}}_{18}$ only arise for spin 1 WIMPs, and $\mathbf{\mathcal{S}}$ is a symmetric combination of spin 1 WIMP polarisation vectors~\cite{Dent:2015zpa}.~For simplicity, we omit the nucleon index in the operator definitions.}
    \label{tab:operators}
\end{table}
\begin{table}[t]
    \centering
    \begin{ruledtabular}	
\begin{tabular}{lccc}
    \toprule
  \boldmidrule
  Spin 0 DM & Coeff. & Scalar med. & Vector med. \\ \addlinespace[.4em]
   & $c_1$ & $\frac{h_1^N g_1}{M^2_\Phi}$ & $-2 \frac{h_3^N g_4}  { M_G^2}$ \\ \addlinespace[.4em]
   & $c_7$ & & $4 \frac{h_4^N g_4}  { M_G^2}$   \\ \addlinespace[.4em]
   & $c_{10}$ & $\frac{h_2^N g_1}{M^2_\Phi}$ & \\
   \bottomrule
    \toprule
    \boldmidrule
    spin 1/2 DM & Coeff. & Scalar med. & Vector med. \\ \addlinespace[.4em]
    &$c_1$ & $\frac{h_1^N \lambda_1}{M^2_\Phi}$ & $-\frac{h_3^N \lambda_3}{M_G^2}$\\ \addlinespace[.4em]
    &$c_4$ & & $4\frac{h_4^N \lambda_4}{M_G^2}$\\ \addlinespace[.4em]
    &$c_6$ & $\frac{h_2^N \lambda_2}{M_\Phi^2}\frac{m_N}{m_\chi}$ & \\ \addlinespace[.4em]
    &$c_7$ & & $2\frac{h_4^N \lambda_3}{M_G^2}$\\ \addlinespace[.4em]
    &$c_8$ & & $-2\frac{h_3^N \lambda_4}{M_G^2}$\\ \addlinespace[.4em]
    &$c_9$ & & $-2\frac{h_4^N \lambda_3}{M_G^2}\frac{m_N}{m_\chi} - 2\frac{h_3^N \lambda_4}{M_G^2} $  \\ \addlinespace[.4em]   
    &$c_{10}$ & $\frac{h_2^N \lambda_1}{M^2_\Phi}$ & \\
    &$c_{11}$ & $-\frac{h_1^N \lambda_2}{M_\Phi^2}\frac{m_N}{m_\chi}$ & \\
   \bottomrule
    \toprule
 \boldmidrule   
 Spin 1 DM & Coeff. & Scalar med. & Vector med. \\ \addlinespace[.4em]
&   $c_1$ & $\frac{b_1 h_1^N}{M^2_\Phi}$ & $-2 \frac{h_3^N b_5}  { M_G^2}$ \\ \addlinespace[.4em]
 &  $c_4$ & & $-4\frac{h_4^N \Re(b_7)}{M_G^2} + \frac{q^2}{m_X m_N} \frac{h_3^N \Im(b_6)}{M_G^2}$ \\ \addlinespace[.4em]
 &  $c_5$ &  & $-\frac{m_N}{m_X}\frac{h_3^N \Im(b_6)}{M_G^2}$ \\ \addlinespace[.4em]
 &  $c_6$ &  & \hspace{-0.08 cm}$-\frac{m_N}{m_X}\frac{h_3^N \Im(b_6)}{M_G^2} $ \\ \addlinespace[.4em]
 & $c_7$ & & $4\frac{h_4^N b_5}  { M_G^2}$   \\ \addlinespace[.4em]
 & $c_8$ &  & $2\frac{h_3^N \Re(b_7)}{M_G^2}$ \\ \addlinespace[.4em]
 & $c_9$ &  & $-2\frac{m_N}{m_X}\frac{h_4^N \Im(b_6)}{M_G^2} + 2\frac{h_3^N \Re(b_7)}{M_G^2}$ \\ \addlinespace[.4em]
 & $c_{10}$ & $\frac{b_1 h_2^N}{M_\Phi^2}$ & \\ \addlinespace[.4em]
 & $c_{11}$ & & $-\frac{m_N}{m_X}\frac{h_3^N \Im(b_7)}{M_G^2}$ \\ \addlinespace[.4em]
 & $c_{14}$ & & $2 \frac{m_N}{m_X}\frac{h_4^N \Im(b_7)}{M_G^2}$ \\
    \bottomrule
     \bottomrule
    \boldmidrule
    \end{tabular}    
    \end{ruledtabular}	
    \caption{Relation between coupling constants of the simplified models in \app~\ref{app:Lagrangians} (with only two coupling constants different from zero at the time; see text below Eq.~(\ref{eq:Meff})) and coefficients of the non-relativistic operators in \reftab{tab:operators} in the proton/neutron basis.~In the second column we omit the index $N$ for simplicity.} 
    \label{tab:coeffs}
\end{table}

The coupling constants $c_i^{(N)}$ in Eq.~(\ref{eq:H}) are directly related to the coupling constants of the simplified models in \app~\ref{app:Lagrangians} as illustrated in Tab.~\ref{tab:coeffs} and described in \Ref~\cite{Dent:2015zpa}.~The couplings to nucleons in Tab.~\ref{tab:coeffs}, $h_i^N$, are related to the quark level couplings by nucleon form factors~\cite{Dent:2015zpa,Agrawal:2010fh,Dienes:2013xya},
\begin{subequations}
 \label{eq:formfac}
 \begin{align}
 &h_1^n = 11.93 \, h_1	& &h_1^p = 12.31 \, h_1 \\
 &h_2^n = -0.07 \, h_2	& &h_2^p = -0.28 \, h_2  \\
 &h_3^{n,p} = 3 \, h_3 &
 &h_4^{n,p} = 0.33 \, h_4\,.
\end{align}
\end{subequations}
These values can have large uncertainties. We will briefly discuss how these uncertainties effect our results in Sec.~\ref{sec:discussion}.

Only leading contributions to the $c_i^{(N)}$ coefficients are considered in Tab.~\ref{tab:coeffs}\footnote{Some of the coefficients in Tab.~\ref{tab:coeffs} differ from those in the published version of Ref.~\cite{Dent:2015zpa}.~Currently, a revised version of Ref.~\cite{Dent:2015zpa} is in preparation.~In the revised version, the new coefficients will agree with Tab.~\ref{tab:coeffs} and~\cite{DelNobile:2013sia}.}.~In particular, we neglect momentum-dependent chiral effective field theory corrections~\cite{Bishara:2016hek,Hoferichter:2015ipa} and renormalisation group effects leading to operator evolution~\cite{Crivellin:2014qxa}.~Furthermore, we do not consider charged mediators and models generating the two additional operators introduced in~\cite{Dent:2015zpa}, $\hat{\mathcal{O}}_{17}$ and $\hat{\mathcal{O}}_{18}$.~The latter arise in the non-relativistic limit of simplified models with $b_6=\Re(b_6)$.

\section{Dark matter thermal production}
\label{sec:relic}

In the early Universe, the time evolution of the DM space density, $n$, is described by the Boltzmann equation~\cite{Gondolo:1990dk},
\begin{equation}
\dot{n}+3 Hn = - \langle \sigma v_{\rm M\o l} \rangle \left( n^2 - n_{\rm eq}^2 \right) \,,
\label{eq:B1}
\end{equation}
where $H$ is the expansion rate of the Universe, $\sigma$ is the invariant cross-section for DM annihilation into Standard Model particles, a dot denotes derivative with respect to the coordinate time, $n_{\rm eq}$ is the DM density at thermal equilibrium, and 
\begin{equation}
v_{\rm M\o l} = \frac{\sqrt{(p_1\cdot p_2)^2 - m_{\rm DM}^4}}{E_1 E_2}\,,
\label{eq:Moller}
\end{equation}
is the M\o ller velocity.~In Eq.~(\ref{eq:Moller}), $E_1$ and $E_2$ ($p_1$ and $p_2$) are the energies (four-momenta) of the annihilating DM particles.~Angle brackets in Eq.~(\ref{eq:B1}) represent an average over the phase-space density $F$ of the DM particles in the initial state,
\begin{equation}
 \langle \sigma v_{\rm M\o l} \rangle = \frac{\int {\rm d}^3 \mathbf{p}_1\int {\rm d}^3 \mathbf{p}_2 \, \left(\sigma v_{\rm M\o l}\right) \, e^{-E_1/T} \,e^{-E_2/T} }{\int {\rm d}^3 \mathbf{p}_1\int {\rm d}^3 \mathbf{p}_2  \, e^{-E_1/T} \,e^{-E_2/T} } \,
\label{eq:sigma}
\end{equation}
where $T$ is the time dependent CMB temperature, $\mathbf{p}_1$ and $\mathbf{p}_2$ are three-dimensional momenta and for the phase-space density $F$ we assumed a Boltzmann distribution.~Eq.~(\ref{eq:sigma}) is equivalent to the single-integral formula~\cite{Gondolo:1990dk} 
\begin{equation}
 \langle \sigma v_{\rm M\o l} \rangle = \int_0^\infty {\rm d}\epsilon \, \mathscr{K}(x,\epsilon)\, \sigma v_{\rm lab} \,,
 \label{eq:sigmavmol}
\end{equation}
where $x=m_{\rm DM}/T$, 
\begin{equation}
\mathscr{K}(x,\epsilon)= \frac{2x}{K_2^2(x)}\,\epsilon^{1/2} (1+2\epsilon) K_1(2x\sqrt{1+\epsilon})\,,
\end{equation}
$K_1$ and $K_2$ are the first two modified Bessel functions of the second kind, $\epsilon=(s-4m_{\rm DM}^2)/4m_{\rm DM}^2$ is the total kinetic energy per unit mass in the lab frame, and the DM-DM relative speed in the lab frame, $v_{\rm lab}$, is given by $v_{\rm lab}=2\epsilon^{1/2} (1+\epsilon)^{1/2}/(1+2\epsilon)$.~In general, for a two-particle final state, 
\begin{equation}
\sigma v_{\rm lab}= \frac{1}{64 \pi^2 (s-2m_{\rm DM}^2)} \beta_f \int {\rm d} \Omega \bar{|\mathscr{M}|^2}
\label{eq:sigmav}
\end{equation}
where
\begin{equation}
\beta_f= \left[ 1 - \frac{(m_3+m_4)^2}{s} \right]^{1/2} \left[ 1 - \frac{(m_3-m_4)^2}{s} \right]^{1/2}\,,
\end{equation}
$m_3$ and $m_4$ are the masses of the two particles in the final state, $s=(p_1+p_2)^2$, $\mathscr{M}$ is the invariant amplitude, ${\rm d}\Omega = {\rm d}\phi \,{\rm d}\hspace{-0.05cm}\cos\hspace{-0.05cm}\theta$, and $\theta$ and $\phi$ are centre-of-mass scattering angle and associated azimuthal angle, respectively.~Here we denote the square modulus of $\mathcal{M}$ averaged (summed) over initial (final) spin states by $\bar{|\mathscr{M}|^2}$.~Explicit expressions for $\bar{|\mathscr{M}|^2}$ for all simplified models in Appendix~\ref{app:Lagrangians} are listed in Appendix~\ref{app:M}.

In terms of DM abundance, $Y=n/\mathscr{S}$, where $\mathscr{S}$ is the total entropy density of the Universe, Eq.~(\ref{eq:B1}) reads as follows,
\begin{equation}
\frac{{\rm d}Y}{{\rm d}x} = -\left(\frac{45}{\pi} G\right)^{-1/2} \frac{g_*^{1/2} m_{\rm DM}}{x^2} \langle \sigma v_{\rm M\o l} \rangle \left( Y^2 - Y_{\rm eq}^2 \right), 
\label{eq:B2}
\end{equation}
where $G$ is Newton's gravitational constant, and $Y_{\rm eq}$ is given by
\begin{equation}
Y_{\rm eq}=\frac{45 g}{4\pi^4} \frac{x^2 K_2(x)}{h_{\rm eff}(m_{\rm DM}/x)} \,.
\end{equation}
In the above expression, $g$ is the number of DM spin states and
\begin{equation}
g_*^{1/2} = \frac{h_{\rm eff}}{g_{\rm eff}^{1/2}} \left( 1 + \frac{1}{3} \frac{T}{h_{\rm eff}} \frac{{\rm d}h_{\rm eff}}{{\rm d}T} \right)\,,
\end{equation}
where $g_{\rm eff}$ and $h_{\rm eff}$ are the effective number of degrees of freedom for the total energy and entropy densities of the Universe, respectively.~In the following, we will assume $g_*^{1/2} \simeq h_{\rm eff}/g_{\rm eff}^{1/2}~\simeq 9.5$, which is a fairly good approximation for temperatures above the QCD phase transition~\cite{Gondolo:1990dk}.

Numerical integration of Eq.~(\ref{eq:B1}) shows that there is a critical temperature, $T_f$, such that for $x< x_f\equiv m_{\rm DM}/T_f$, $Y(x)=Y_{\rm eq}(x)$, whereas for $x\ge x_f$, $Y(x)\neq Y_{\rm eq}(x)$.~The critical temperature $T_f$ is called freeze-out temperature and can be estimated by solving for $x=x_f$ the equation
\begin{equation}
 \frac{45 g}{4\pi^4} \frac{K_2(x)m_{\rm DM} \sqrt{g_*} }{h_{\rm eff}(m_{\rm DM}/x)} \langle \sigma v_{M\o l}\rangle \delta (\delta+2) = \sqrt{\left(\frac{45}{\pi}G\right)}\frac{K_1(x)}{K_2(x)} \,,
\label{eq:xf}
\end{equation}
which follows from Eq.~(\ref{eq:B1}) with $\Delta\equiv Y-Y_{\rm eq}=\delta Y_{\rm eq}$, $\delta\in\mathbb{R}$, and ${\rm d}\Delta/{\rm dx}=0$.~Setting ${\rm d}\Delta/{\rm dx}$ to zero at $x~=~x_f$ is a good approximation since $Y=Y_{\rm eq}$ for $x< x_f$, and ${\rm d}\Delta/{\rm dx}$ is a smooth function of $x$.~Here we set the parameter $\delta$ to 1.5, a value found comparing results for $x_f$ obtained by solving Eq.~(\ref{eq:B2}) with estimates based upon Eq.~(\ref{eq:xf})~\cite{Gondolo:1990dk}.

For $x\ge x_f$ an approximate solution to Eq.~(\ref{eq:B2}) can be found by setting to zero the $Y_{\rm eq}^2$ term in the right hand side.~Indeed, after freeze-out, $Y_{\rm eq}^2$ is small compared to $Y^2$.~In terms of $T$, the approximate solution reads as follows
\begin{equation}
\frac{1}{Y_0} = \frac{1}{Y_f} +\left(\frac{45}{\pi} G \right)^{-1/2} \int_{T_0}^{T_f} {\rm d} T\, g_*^{1/2} \langle \sigma v_{\rm M\o l} \rangle \,,
\end{equation}
where $Y_0=Y(m_{\rm DM}/T_0)$, $Y_f=Y(m_{\rm DM}/T_f)$ and $T_0$ is the present CMB temperature.~From $Y_0$, one can estimate the DM relic density in units of the critical density $\rho_c$ as follows $\Omega_{\rm DM}=m_{\rm DM} \mathscr{S}_0 Y_0/\rho_c$, where $\mathscr{S}_0$ is the present entropy density.~By evaluating $\Omega_{\rm DM}$ explicitly, one finds~\cite{Gondolo:1990dk}
\begin{equation}
\label{eq:Omega_h2}
\Omega_{\rm DM} h^2= 2.8282 \times 10^8 \left( \frac{m_{\rm DM}}{\rm GeV} \right) \mathscr{T}^3 Y_0 \,,
\end{equation}
where $\mathscr{T} = T_0/(2.75~{\rm K})$, $h=H_0/100$, and $H_0$ is the Hubble constant.

\begin{table}[t]
\centering
    \begin{ruledtabular}
    \begin{tabular}{lcccc}
    \toprule
\boldmidrule	
 Spin 0 DM & Op. & $g_q$ & $g_\text{DM}$ & $M_\text{eff}$ [GeV]\\
&1	&$h_1$	&$g_1$	&14564.484	\\
&1	&$h_3$	&$g_4$	&10260.217	\\
&7	&$h_4$	&$g_4$	&4.509	\\
&10	&$h_2$	&$g_1$	&10.706	\\
 \toprule
\boldmidrule
 Spin 1/2 DM & Op. & $g_q$ & $g_\text{DM}$ & $M_\text{eff}$ [GeV]\\
&1	&$h_1$	&$\lambda_1$	&14564.484	\\
&1	&$h_3$	&$\lambda_3$	&7255.068	\\
&4	&$h_4$	&$\lambda_4$	&147.354	\\
&6	&$h_2$	&$\lambda_2$	&0.286	\\
&7	&$h_4$	&$\lambda_3$	&3.188	\\
&8	&$h_3$	&$\lambda_4$	&225.159	\\
&10	&$h_2$	&$\lambda_1$	&10.706	\\
&11	&$h_1$	&$\lambda_2$	&351.589	\\
\toprule
\boldmidrule
Spin 1/2 DM & Op. & $g_q$ & $g_\text{DM}$ & $M_\text{eff}$ [GeV]\\
&1	&$h_1$	&$b_1$	&14564.484	\\
&1	&$h_3$	&$b_5$	&10260.216	\\ 
&4	&$h_4$	&$\Re$($b_7$)	&188.302	\\
&5	&$h_3$	&$\Im$($b_6$)	&6.946	\\
&7	&$h_4$	&$b_5$	&4.509	\\ 
&8	&$h_3$	&$\Re$($b_7$)	&287.728	\\
&9	&$h_4$	&$\Im$($b_6$)	&3.674	\\
&10	&$h_2$	&$b_1$	&10.706	\\
&11	&$h_3$	&$\Im$($b_7$)	&223.794	\\
&14 	&$h_4$	&$\Im$($b_7$) 	&0.201	\\
  \boldmidrule
\bottomrule
 \end{tabular}    	
 \end{ruledtabular}	
\caption{Benchmark points ideally producing 150 nuclear recoil events at XENONnT/LZ for $m_\chi = 50\gev$ and $g_q=g_\text{DM}=0.1$~\cite{Baum:2017kfa}.~We find these values by integrating Eq.~(\ref{eq:rateE}) from 5~keV to 45~keV in order to allow for a direct comparison of our results with~\cite{Dent:2015zpa}.~The second column shows the leading non-relativistic operator for the benchmark model.~Third and and fourth columns report $g_q$ and $g_{\rm DM}$, i.e.~the coupling constants for the DM-DM-mediator and $\bar{q}$-$q$-mediator vertices, respectively.}
\label{tab:benchmarks}
\end{table}

\begin{figure}[t]
 \begin{center}
  \includegraphics[width=\linewidth]{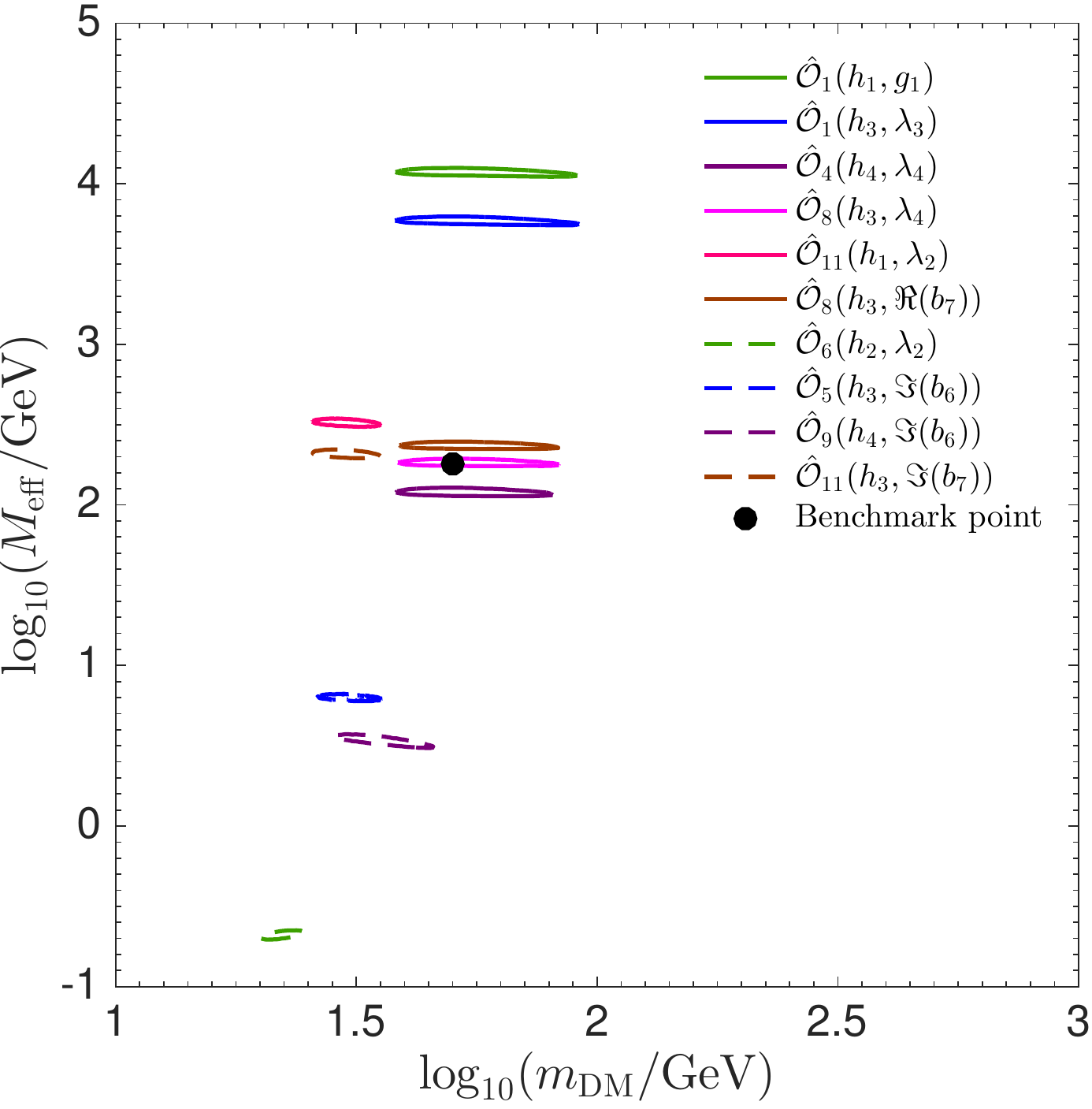}
 \end{center}
\caption{2D 95\% confidence intervals in the $m_{\rm DM} - M_{\rm eff}$ plane ($g_q=g_{\rm DM}=0.1$) obtained by fitting selected models from Tab.~\ref{tab:benchmarks} to data simulated from model $\hat{\mathcal{O}}_8(h_3,\lambda_4)$.~In the simulation we assume $m_{\rm DM}=50$~GeV and $\mu_{\rm S}=100$.~When model $\hat{\mathcal{O}}_8(h_3,\lambda_4)$ is fitted to the data, the error on $M_{\rm eff}$ is small compared to the uncertainties on $\Omega_{\rm DM} h^2$ arising from the fact that $M_{\rm med}$, $g_q$ and $g_{\rm DM}$ cannot be constrained independently.~The figure also shows the bias on the best fit value for $M_{\rm eff}$ arising when a model different from $\hat{\mathcal{O}}_8(h_3,\lambda_4)$ is fitted to the data.\label{fig:mm}}
\end{figure}

\begin{figure}[t]
\begin{center}
\includegraphics[width=0.49\textwidth]{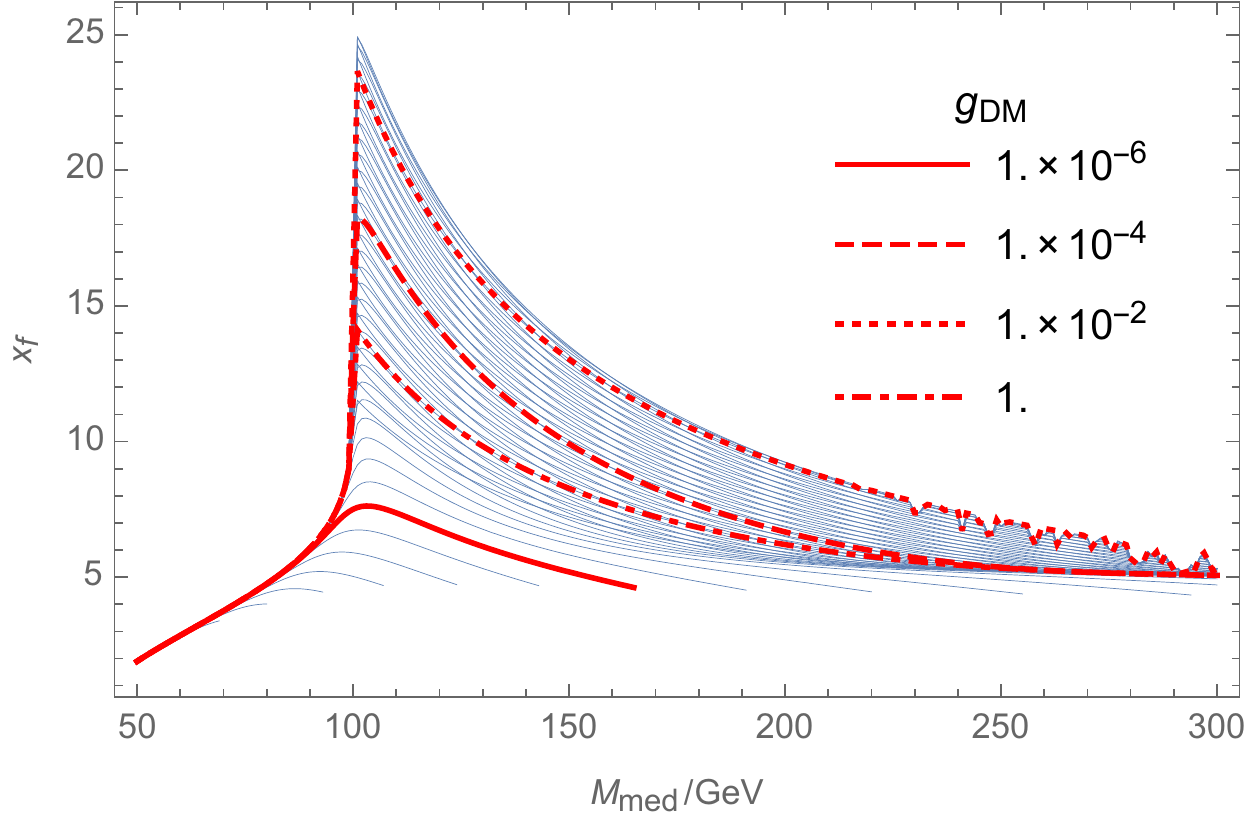}
\end{center}
\caption{Inverse freeze-out temperature, $x_f=m_{\rm DM}/T_f$, as a function of the mediator mass $M_{\rm med}$ for different values of $g_{\rm DM}$ and $m_{\rm DM}=50$~GeV.~The underlying DM model is $\hat{\mathcal{O}}_1(h_1,g_1)$.~The coupling constant $g_{q}$ has been set to the value ideally producing 150 signal events at XENONnT/LZ.~The envelope of the family of curves in the figure identifies a region in the $M_{\rm med}-x_f$ plane which is compatible with the observed signal at XENONnT/LZ.~All curves peak at mediator masses around 100~GeV, as expected for kinematical reasons:~for $M_{\rm med}=100$~GeV, the DM annihilation into a $\bar{q}q$ pair is resonant.\label{fig:xf}}  
\end{figure}

\begin{figure*}[t]
\begin{center}
\begin{minipage}[t]{0.49\linewidth}
\centering
\hspace{0.25 cm}\includegraphics[width=0.943\textwidth]{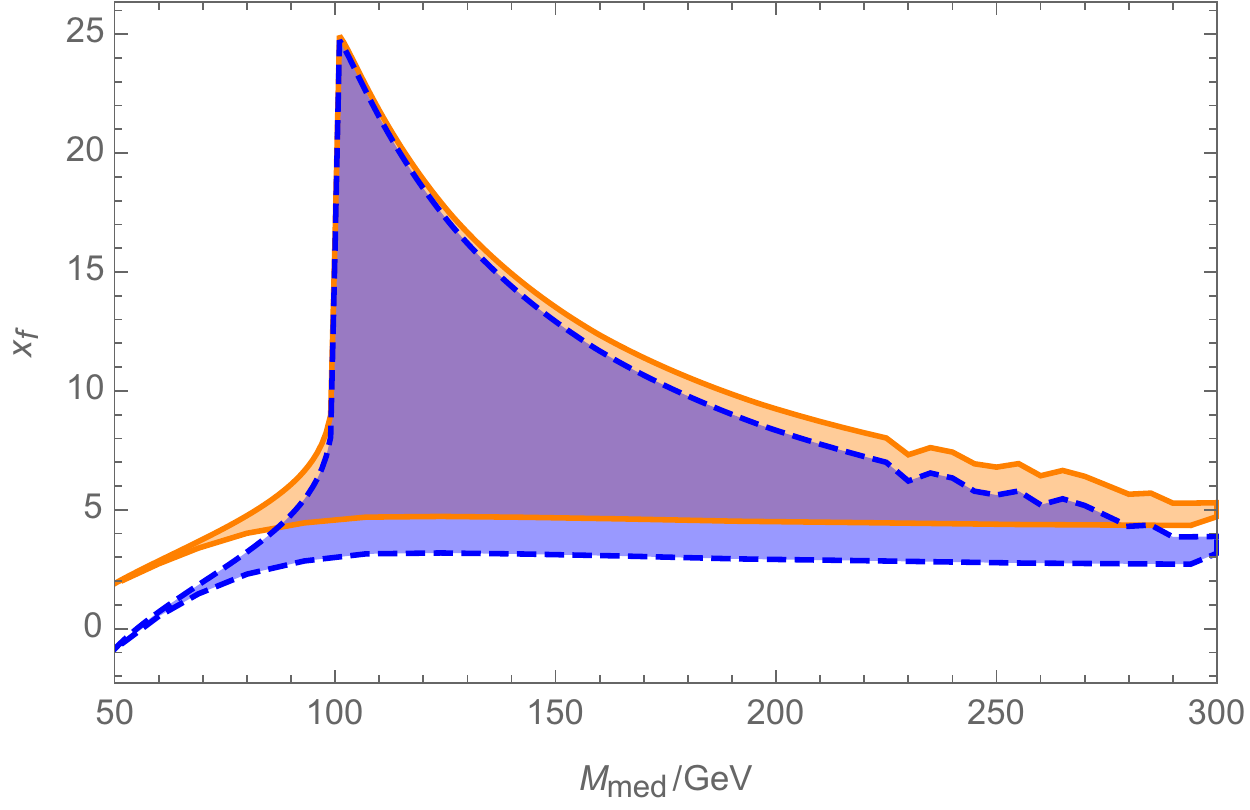}
\end{minipage}
\begin{minipage}[t]{0.49\linewidth}
\centering
\hspace{0.40 cm}\includegraphics[width=0.943\textwidth]{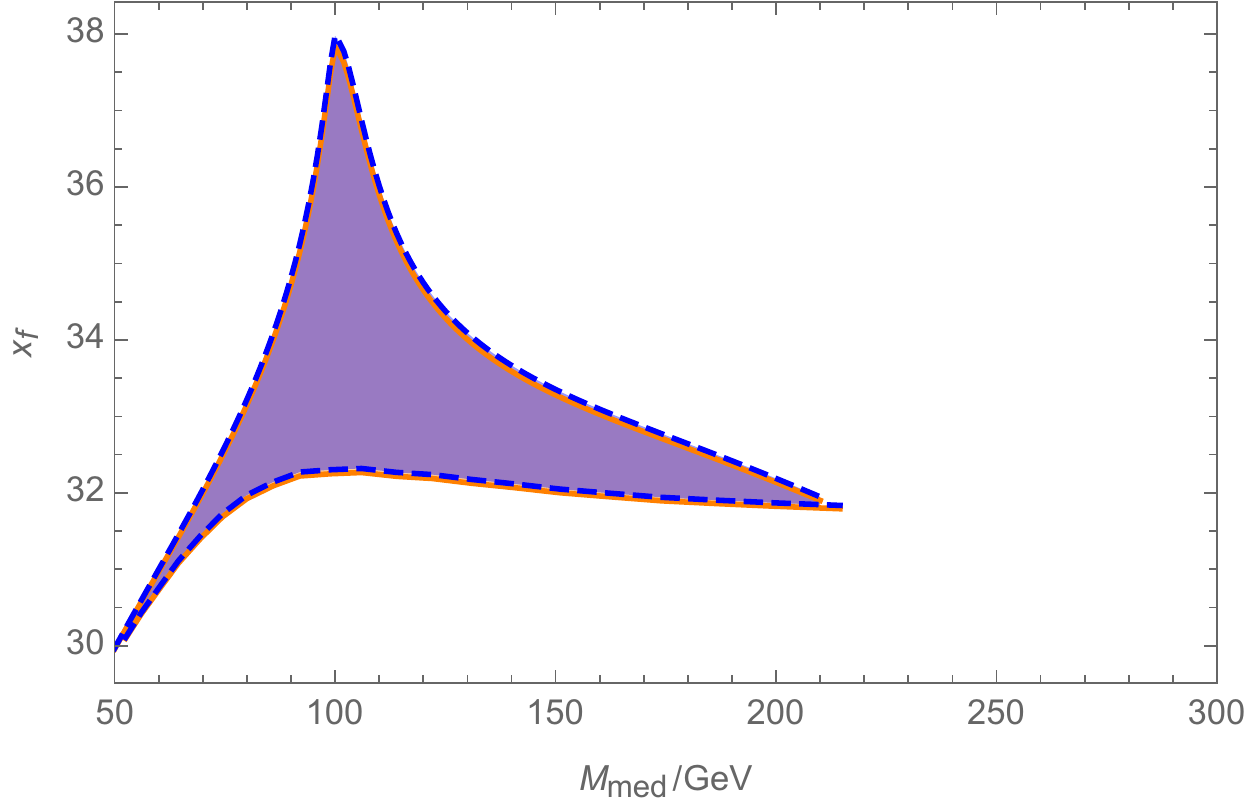}
\end{minipage}
\begin{minipage}[t]{0.49\linewidth}
\centering
\hspace{-0.25 cm}\includegraphics[width=0.995\textwidth]{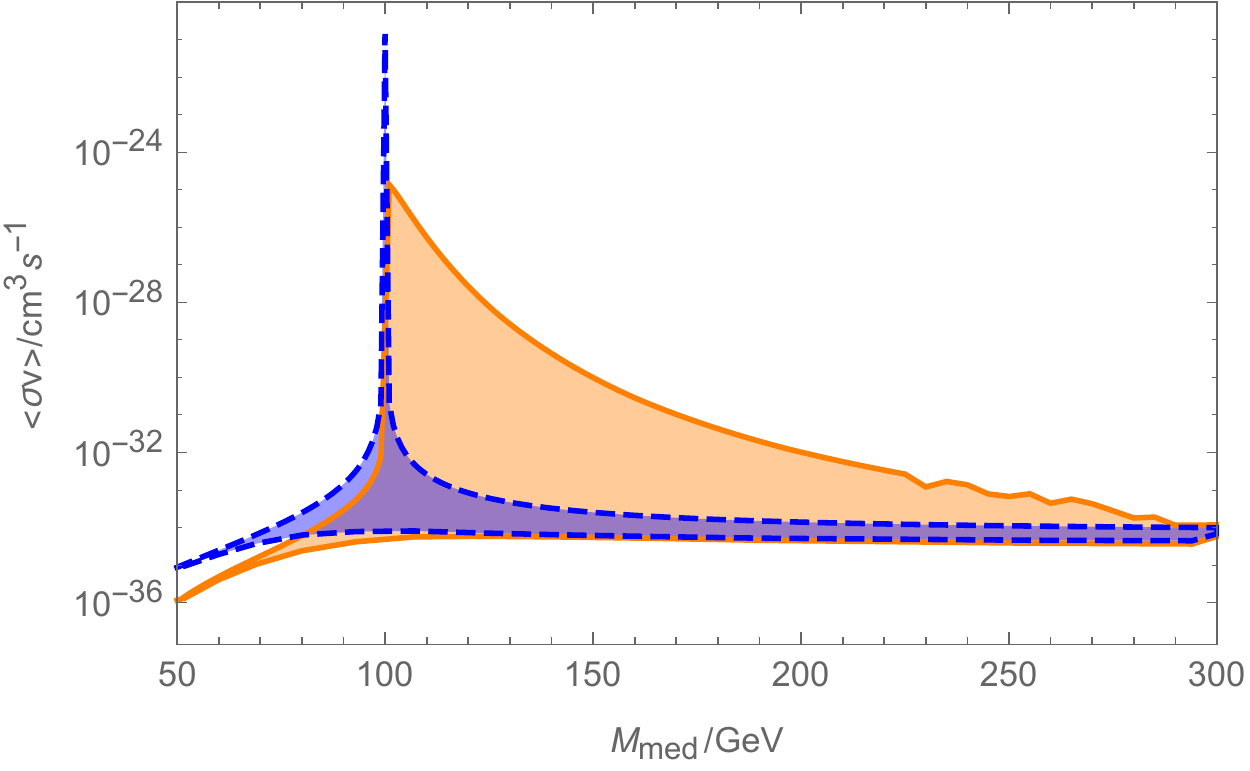}
\end{minipage}
\begin{minipage}[t]{0.49\linewidth}
\centering
\includegraphics[width=\textwidth]{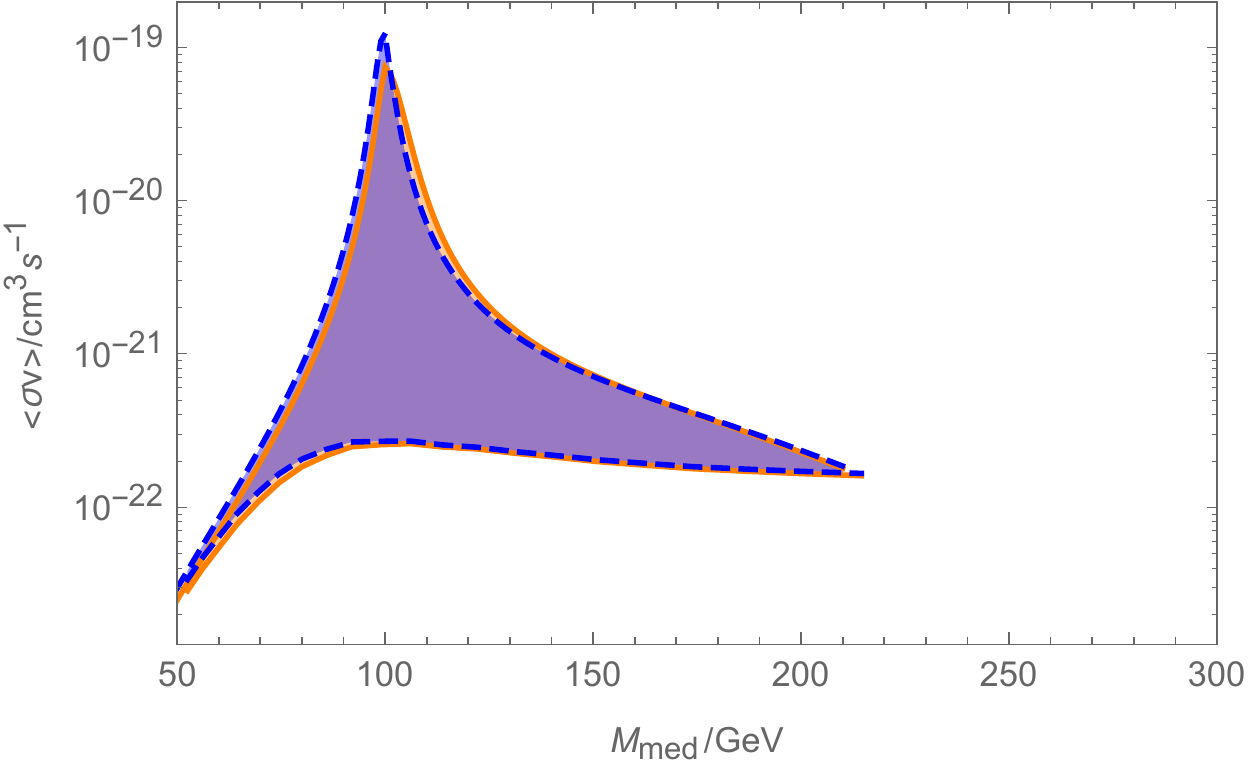}
\end{minipage}
\begin{minipage}[t]{0.49\linewidth}
\centering
\includegraphics[width=0.98\textwidth]{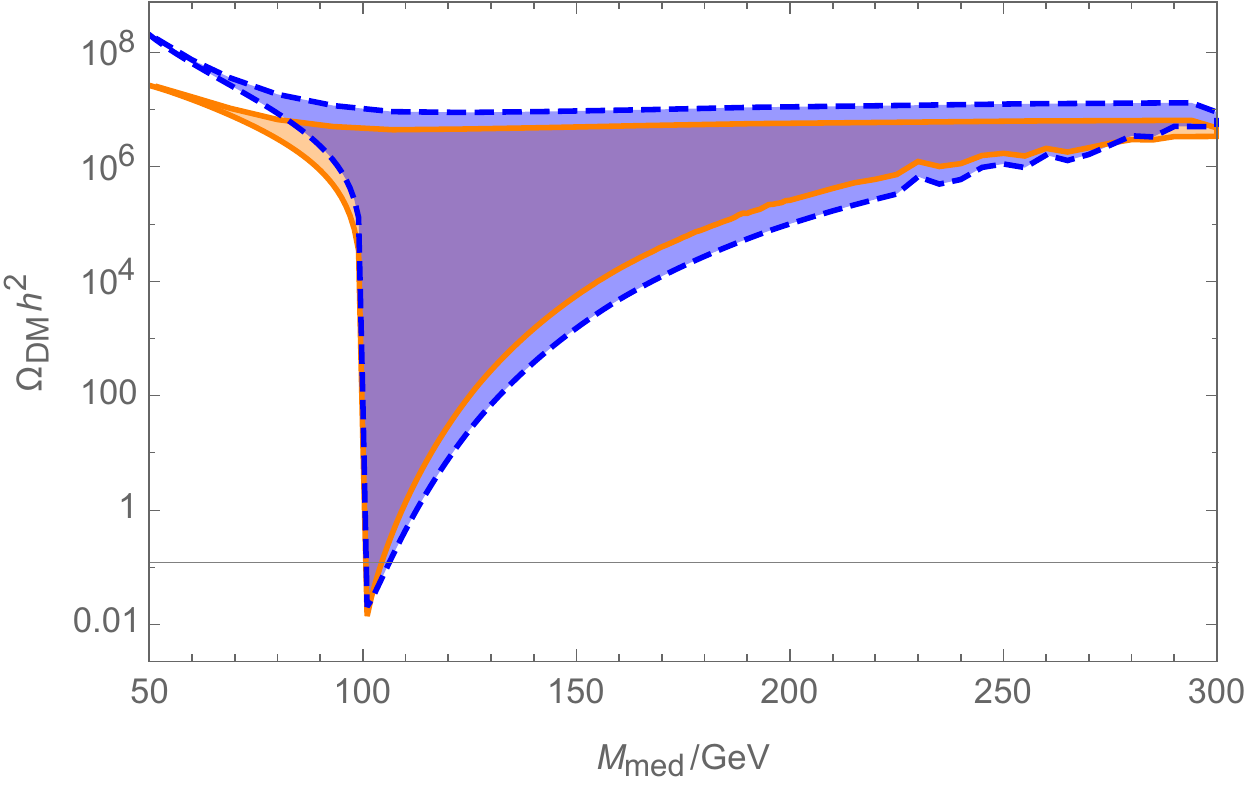}
\end{minipage}
\begin{minipage}[t]{0.49\linewidth}
\centering
\includegraphics[width=\textwidth]{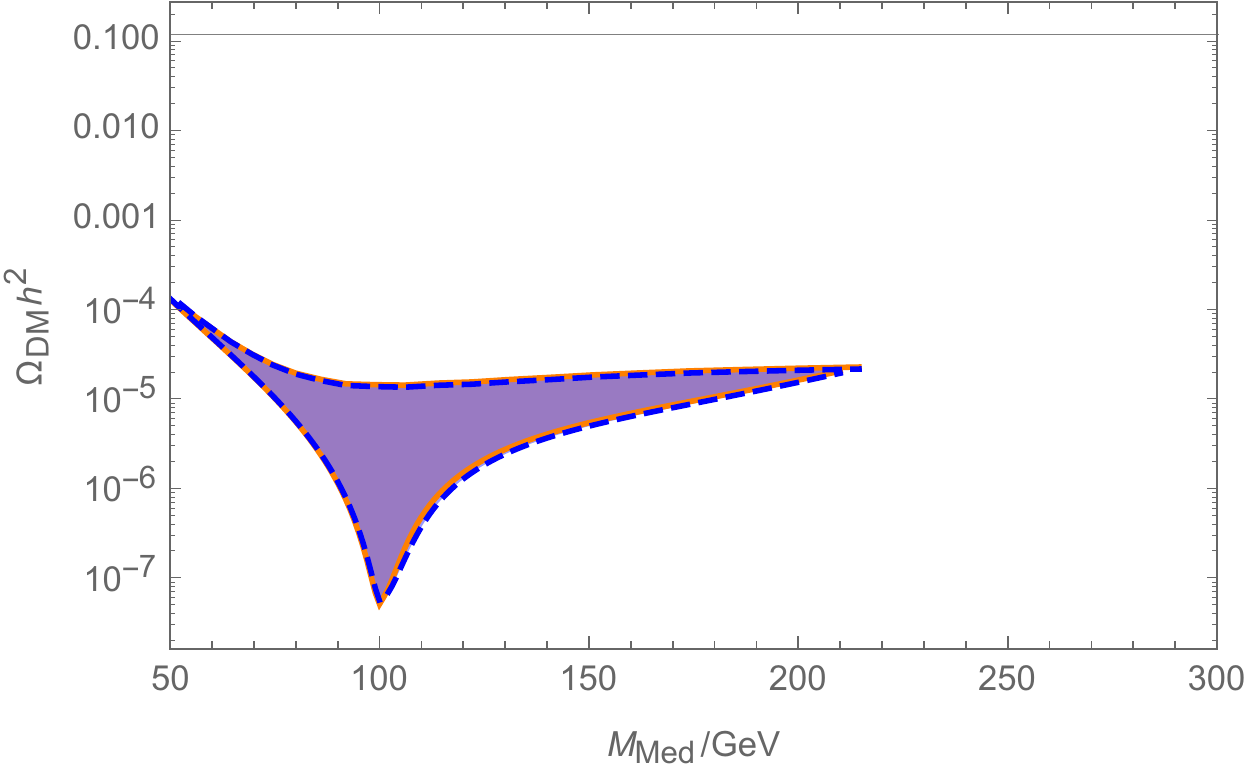}
\end{minipage}
\end{center}
\caption{Comparison of models $\hat{\mathcal{O}}_1(h_1,g_1)$ (left) and $\hat{\mathcal{O}}_{10}(h_2,g_1)$ (right) in the $M_{\rm med}-x_f$,  $M_{\rm med}-\langle \sigma v_{\rm M\o l} \rangle$ and $M_{\rm med}-\Omega_{\rm DM} h^2$ planes.~In all panels we only report the envelopes of the family of curves generated by the $g_{\rm DM}$ parameter.~The six panels show both exact results (orange areas) and estimates based on the approximations that are introduced in Appendix~\ref{app:app} (blue areas).~The solid horizontal line in the $M_{\rm med}-\Omega_{\rm DM} h^2$ plane corresponds to the Planck best fit value for $\Omega_{\rm DM}h^2$.\label{fig:O1O10}} \end{figure*}

\begin{figure*}[t]
\begin{center}
\begin{minipage}[t]{0.469\linewidth}
\centering
\includegraphics[width=\textwidth]{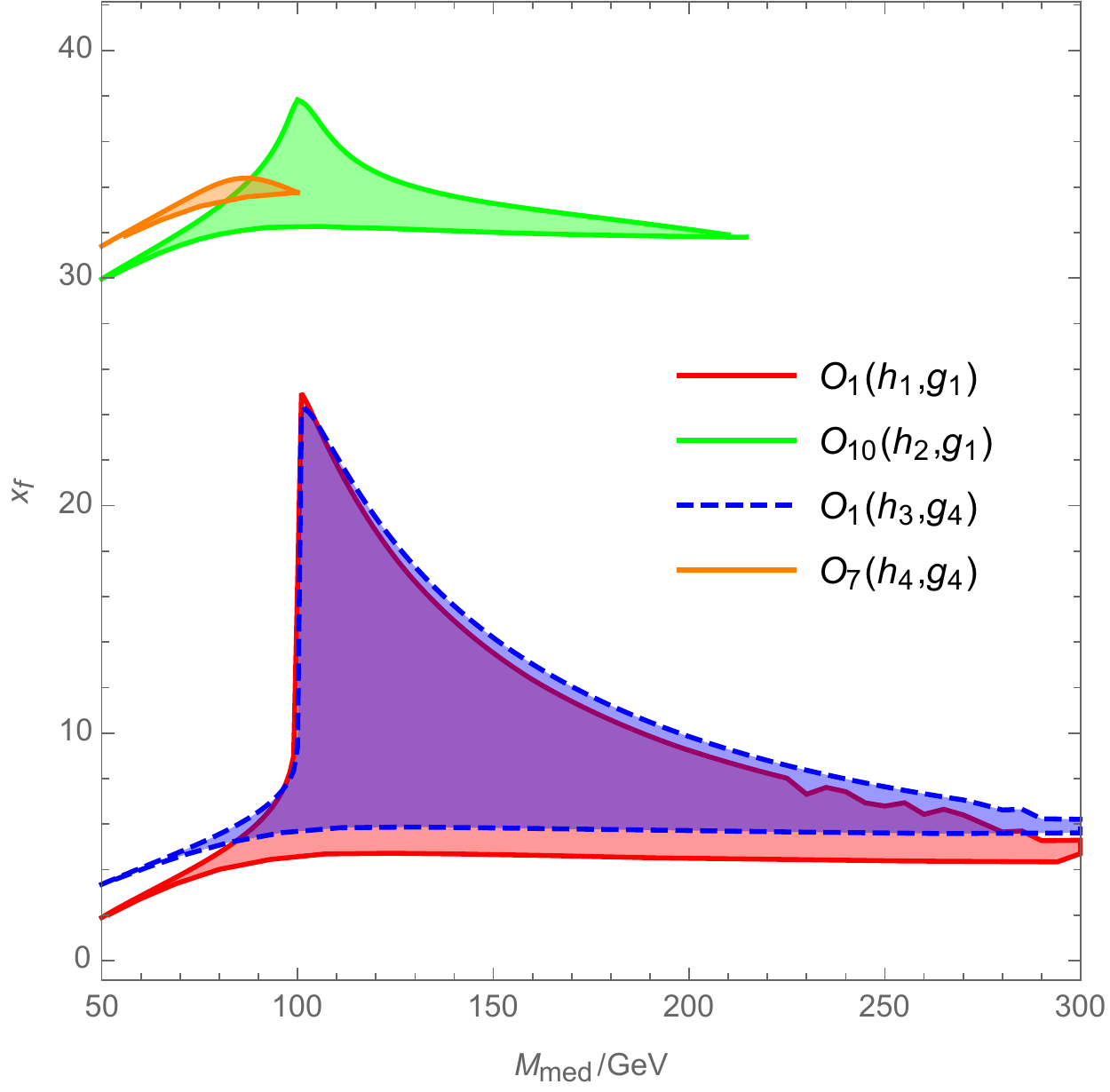}
\end{minipage}
\begin{minipage}[t]{0.494\linewidth}
\centering
\includegraphics[width=\textwidth]{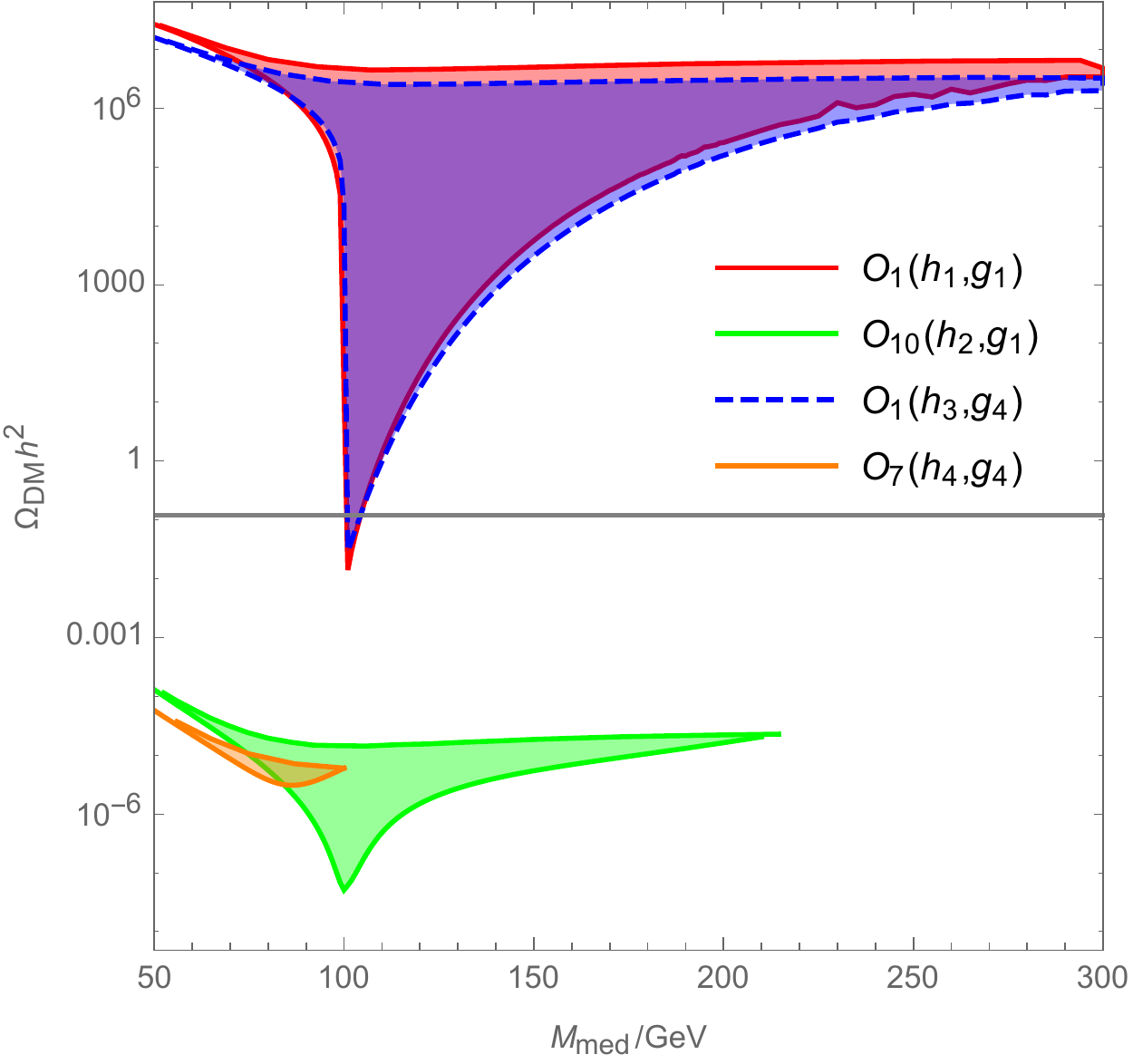}
\end{minipage}
\end{center}
\caption{The left (right) panel in the figure shows the $x_f$ ($\Omega_{\rm DM} h^2$) envelope as a function of $M_{\rm med}$ for the simplified models corresponding to spin 0 DM.~Envelopes are constructed by requiring that an idealized version of XENON/LZ (see Sec.~{\ref{sec:results}}) has detected 150 signal events.~In the right panel, the solid horizontal line in the $M_{\rm med}-\Omega_{\rm DM} h^2$ plane corresponds to the Planck best fit value for $\Omega_{\rm DM}h^2$.\label{fig:scalar}}  
\end{figure*}

\section{Dark matter detection at XENONNT/LZ}
\label{sec:DD}

In this section we review the equations that characterise DM direct detection at XENONnT/LZ.~The aim is to illustrate which DM parameters can be constrained in case of signal detection.~These constraints will then determine whether or not DM detection and thermal production are compatible for the simplified models in Appendix~\ref{app:Lagrangians}.

The number of observable photoelectrons per DM interaction in the XENONnT/LZ detector is denoted by S1.~The expected rate of DM interactions per unit detector mass is~\cite{Aprile:2011hx}
\begin{eqnarray}
\frac{{\rm d}R}{{\rm d}{\rm S1}} &=& \zeta(S1) \sum_{n=1}^{\infty}\mathscr{G}({\rm S1}| n,\sqrt{n}\hat{\sigma}) 
 \int_0^\infty {\rm d}E\, \frac{{\rm d}R}{{\rm d}E} 
 \mathscr{P}(n|\nu(E)) \,, \nonumber\\
\label{eq:rateS1}
\end{eqnarray}
where $\mathscr{P}$ is a Poisson distribution of mean $\nu(E)$, and $\nu(E)$ is the number of expected photoelectrons when a nuclear recoil energy $E$ is deposited in the DM-nucleus interaction.~The integer $n$ is the number of actually produced photoelectrons in the scattering process.~Finally, $\mathscr{G}$ is a Gaussian distribution of mean $n$ and variance $\sqrt{n}\hat{\sigma}$.~Here we extract the function $\nu(E)$ from Fig.~(13) in~\cite{Aprile:2015uzo}, set the single-photoelectron resolution of the XENONnT/LZ photomultipliers, $\hat{\sigma}$, to $\hat{\sigma}=0.4$, and assume a constant acceptance, $\zeta(S1)\simeq 0.4$~\cite{Aprile:2015uzo}.~The rate of nuclear recoil events per unit detector mass, ${\rm d}R/{\rm d}E$, reads as follows
\begin{equation}
\frac{{\rm d}R}{{\rm d}E} = \sum_T  \frac{\xi_T\rho_{\rm DM}}{m_T m_{\rm DM}} \int_{|\mathbf{v}|>v_{\rm min}} \hspace{-0.24cm} {\rm d}^3\mathbf{v}\, |\mathbf{v}|f(\mathbf{v}+\mathbf{v}_{\oplus}) \frac{{\rm d}\sigma_T(E,|\mathbf{v}|)}{{\rm d}E} \,.
\label{eq:rateE}
\end{equation}
where $v_{\rm min}$ is the minimum DM velocity required to deposit an energy $E$ in the detector.~In the experimental analysis, a secondary scintillation signal produced by electrons generated in the DM scattering and drifted to the top of the XENONnT/LZ detector by an electric field is used for background discrimination (i.e.~S2 signal).~For simplicity, here we neglect the S2 signal -- a simplification motivated by the fact that S1 and S2 are anti-correlated~\cite{Aprile:2015uzo}.

In Eq.~(\ref{eq:rateE}), $\mathbf{v}_\oplus$ is the velocity of the Earth in the galactic rest frame, and $\rho_{\rm DM}$ is the local DM density.~Here we set $\rho_{\rm DM}=0.3$~GeV~cm$^{-3}$ and assume a Gaussian distribution truncated at the galactic escape velocity $v_{\rm esc}=533$~km~s$^{-1}$ for $f$, the DM velocity distribution in the galactic rest frame boosted to the detector rest frame.~These are standard assumptions within the so-called Standard Halo Model, although larger values for the local DM density are favoured by astronomical data, e.g.~\cite{Catena:2009mf,Salucci:2010qr,Catena:2011kv,Pato:2015dua,Benito:2016kyp,Green:2017odb}.~For each simplified model in Appendix~\ref{app:Lagrangians}, we calculate the differential cross-section ${\rm d}\sigma_T(E,|\mathbf{v}|)/{\rm d}E$ using nuclear response functions implemented in {\sffamily DMFormFactor}~\cite{Anand:2013yka}.~We extend the sum in Eq.~(\ref{eq:rateE}) to the seven most abundant Xenon isotopes, with masses and mass fractions denoted here by $m_T$ and $\xi_T$, respectively.~Finally, we compute the number of signal events, $\mu_S$, integrating Eq.~(\ref{eq:rateS1}) from S1=3 to S1=70, and multiplying the result by the exposure, $\epsilon$.~For the latter, we assume $\varepsilon=20$~ton$\times$year\footnote{An analysis extending to S1=180 has recently been published by XENON100~\cite{Aprile:2017aas}.}.

The detection of DM particles at XENONnT/LZ would place constraints on the DM particle mass, $m_{\rm DM}$, and on the combinations of parameters listed in Tab.~\ref{tab:coeffs} for the models described in Appendix~\ref{app:Lagrangians}.~Equivalently, it would place constraints on $m_{\rm DM}$ and on the effective mass $M_{\rm eff}$,
\begin{equation}
M_{\rm eff} \equiv 0.1 \frac{M_{\rm med}}{\sqrt{g_q g_{\rm DM}}}\,,
\label{eq:Meff}
\end{equation}
where $M_{\rm med}$ denotes one of the mediator masses, $m_\phi$ or $m_{G}$, and $g_q$ and $g_{\rm DM}$ are the coupling constants for the DM-DM-mediator and $\bar{q}$-$q$-mediator vertices, respectively.~When a simplified model is charactered by the coupling constants $g_q$ and $g_{\rm DM}$ and by the leading non-relativistic operator $\hat{\mathcal{O}}_i$, it will here be denoted by $\hat{\mathcal{O}}_i(g_q,g_{\rm DM})$.~For simplicity, from here onwards we will omit the index $N$ in the definition of the non-relativistic operators.

Assuming $m_{\rm DM}=50$~GeV and $\mu_S=100$ signal events at XENONnT/LZ, $m_{\rm DM}$ and $M_{\rm eff}$ can be extracted from the data with uncertainties of the order of 20\%, e.g.~\cite{Baum:2017kfa}.~This is illustrated in Fig.~\ref{fig:mm} for the case of data simulated from model $\hat{\mathcal{O}}_8(h_3,\lambda_4)$ (in this calculation we assume the background model in~\cite{Baum:2017kfa}, as described in detail in~\cite{Aprile:2015uzo}).~The relic density $\Omega_{\rm DM} h^2$ depends on the reconstructed values for $m_{\rm DM}$ and $M_{\rm eff}$.~However, $\Omega_{\rm DM} h^2$ in general depends on $M_{\rm med}$, $g_q$ and $g_{\rm DM}$ separately, and not on their combination $M_{\rm eff}$.~Therefore, an error of about 20\% on $M_{\rm eff}$ is negligible compared to the uncertainties on $\Omega_{\rm DM} h^2$ arising from the fact that $M_{\rm med}$, $g_q$ and $g_{\rm DM}$ cannot be constrained independently.~Consequently, from here onwards we will assume that a signal at XENONnT/LZ would set $m_{\rm DM}$ and $M_{\rm eff}$ to their true values.

\section{Compatibility of direct detection and thermal production}
\label{sec:results}

In this section we calculate the DM relic density for the models in Appendix~\ref{app:Lagrangians}, showing when it is compatible with the detection of $\mathcal{O}(100)$ signal events at XENONnT/LZ.~The models in Appendix~\ref{app:Lagrangians} are characterised by four parameters:~$g_q$, $g_{\rm DM}$, $M_{\rm med}$ and $m_{\rm DM}$ (as already anticipated we focus on scenarios where only two coupling constants are different from zero at the same time, see Tab.~\ref{tab:coeffs}).~Here we set $m_{\rm DM}$ and $M_{\rm eff}=0.1 M_{\rm med}/{\sqrt{g_{q} g_{\rm DM}}}$ to reference values that we assume to be reconstructed from a hypothetical signal at XENONnT/LZ.~Due to parameter degeneracies, it is not possible to associate a unique value of $\Omega_{\rm DM} h^2$ to a $(m_{\rm DM},M_{\rm med})$ pair.~Therefore, we compute $\Omega_{\rm DM} h^2$ as follows.~We set $m_{\rm DM}$ to its reference value and vary $M_{\rm med}$ in an interval of interest around $m_{\rm DM}$.~We then set $g_q$ to $\min(0.01 M_{\rm med}^2/(g_{\rm DM}M_{\rm eff}^2),\sqrt{4\pi})$, as required by the XENONnT/LZ input, and vary $g_{\rm DM}$ in the $(10^{-7},\sqrt{4\pi})$ range, which implies perturbative values for the coupling constants.~Furthermore, we impose that the mediator decay width is less than the mediator mass.~Through this procedure, we are able to map a signal at XENONnT/LZ into a region in the $M_{\rm med}- x_f$,  $M_{\rm med} - \langle \sigma v_{\rm M\o l} \rangle$ and $M_{\rm med}- \Omega_{\rm DM} h^2$ planes.~We perform this calculation for all models in Tab.~\ref{tab:coeffs}, assuming $m_{\rm DM}=50$~GeV as a reference value if not otherwise specified.~For $M_{\rm eff}$ we assume the value that would {\it ideally} produce 150 nuclear recoils when $g_q=g_{\rm DM}=0.1$.~We estimate this value by integrating Eq.~(\ref{eq:rateE}) from 5~keV to 45~keV in order to allow for a direct comparison of our results with those in~\cite{Dent:2015zpa}.~The values of $M_{\rm eff}$ found in this way are listed in Tab.~\ref{tab:benchmarks} for all models in Tab.~\ref{tab:coeffs}.~These values correspond to $\mu_S\sim \mathcal{O}(100)$ events at XENONnT/LZ, as one can verify using Eq.~(\ref{eq:rateS1}).~In Sec.~\ref{sec:discussion}, we will comment on the dependence of our results on the assumed number of signal events.

\begin{figure*}[t]
\begin{center}
\begin{minipage}[t]{0.49\linewidth}
\centering
\includegraphics[width=\textwidth]{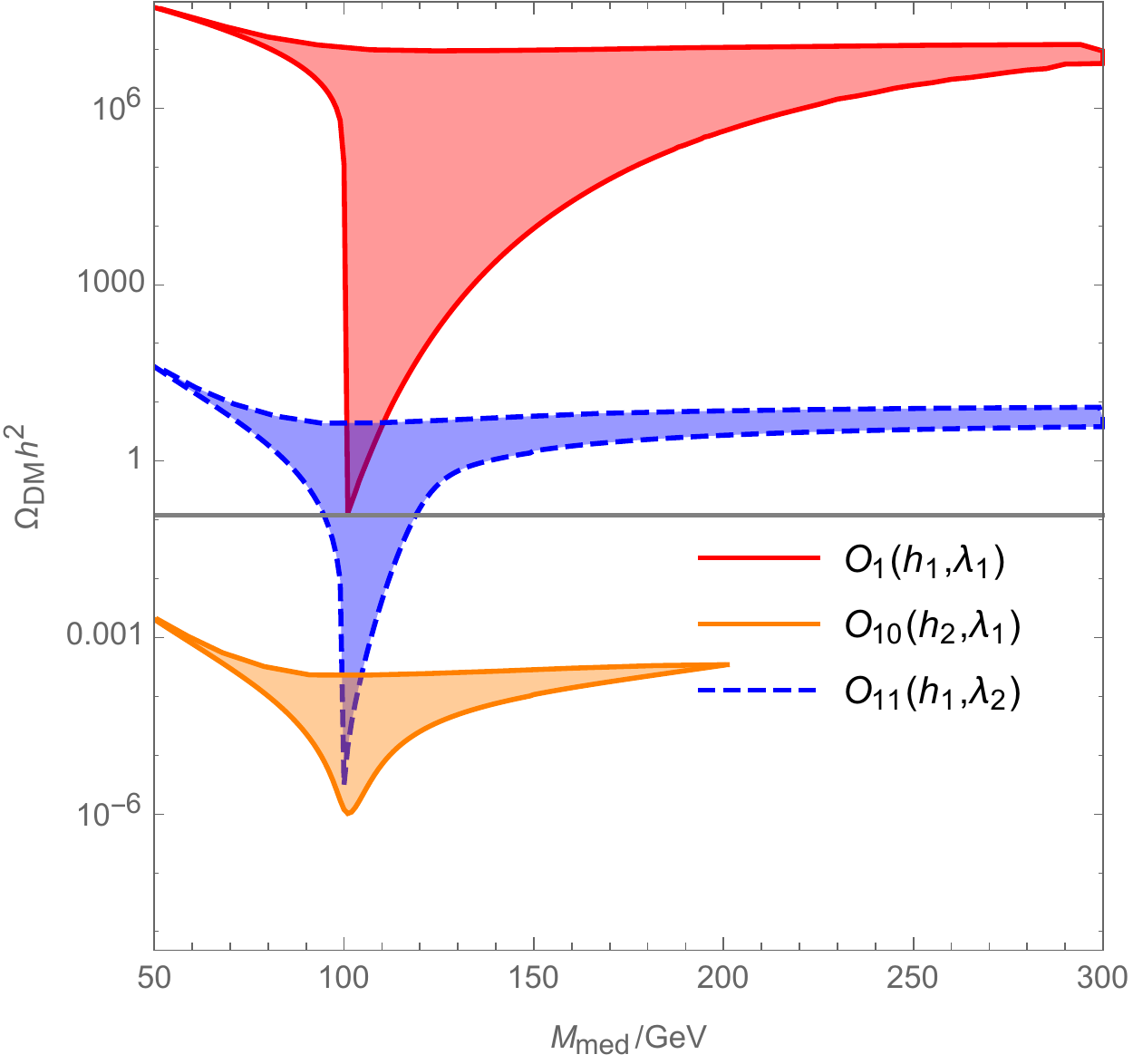}
\end{minipage}
\begin{minipage}[t]{0.49\linewidth}
\centering
\includegraphics[width=\textwidth]{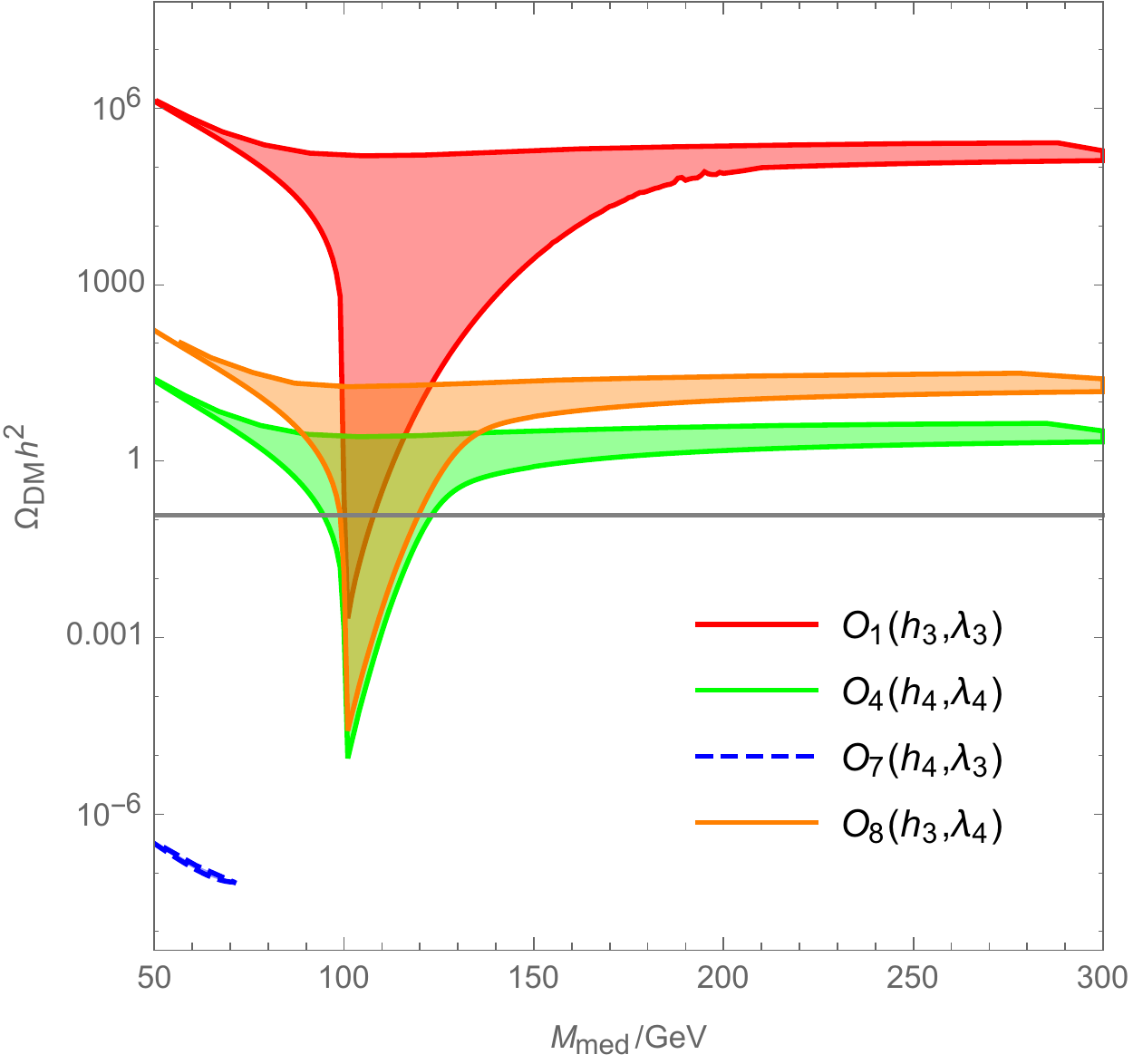}
\end{minipage}
\end{center}
\caption{Same as for Fig.~\ref{fig:scalar}, but now for spin 1/2 DM.~The left panel refers to models with scalar mediators, while the right panel to models with vector mediators.~Model $\hat{\mathcal{O}}_{6}(h_2,\lambda_2)$ does not appear in the left panel of the figure since in this case DM discovery at XENONnT/LZ and thermal production are never compatible for $M_{\rm med}>50$~GeV and $\mu_S\sim \mathcal{O}(100)$.\label{fig:fermion}}  
\end{figure*}

Fig.~\ref{fig:xf} shows $x_f$ as a function of $M_{\rm med}$ for different values of $g_{\rm DM}$ and for $m_{\rm DM}=50$~GeV.~The underlying model is $\hat{\mathcal{O}}_1(h_1,g_1)$, and $g_{q}$ has been set to the XENONnT/LZ input.~The envelope of the family of curves in the figure identifies a region in the $M_{\rm med}-x_f$ plane which is compatible with the observed signal at XENONnT/LZ.~All curves peak at mediator masses around 100~GeV, as expected for kinematical reasons:~for $M_{\rm med}=100$~GeV, the DM annihilation into a $\bar{q}q$ pair is resonant.~We obtain analogous curves in the $M_{\rm med}-\langle \sigma v_{\rm M\o l} \rangle$ and $M_{\rm med}-\Omega_{\rm DM} h^2$ planes.~For large $M_{\rm med}$, the $x_f$ envelope tends to a line, since in this limit scattering and annihilation cross-section depend on the same combination of model parameters (Effective Field Theory limit).

Fig.~\ref{fig:O1O10} compares models $\hat{\mathcal{O}}_1(h_1,g_1)$ and $\hat{\mathcal{O}}_{10}(h_2,g_1)$ in the $M_{\rm med}-x_f$,  $M_{\rm med}-\langle \sigma v_{\rm M\o l} \rangle$ and $M_{\rm med}-\Omega_{\rm DM} h^2$ planes.~In all panels we only report the envelopes of the family of curves generated by the $g_{\rm DM}$ parameter.~For both models, and in all panels, we observe the expected resonance around $M_{\rm med}=100$~GeV, and a plateau corresponding to the Effective Field Theory limit for larger mediator masses.~In all panels we report both exact results and estimates based upon a non-relativistic approximation for $\langle \sigma v_{\rm M\o l} \rangle$ introduced in Appendix~\ref{app:app}.~For model $\hat{\mathcal{O}}_{10}(h_2,g_1)$, the non-relativistic approximation is very good, whereas it breaks down in the case of model $\hat{\mathcal{O}}_1(h_1,g_1)$ due to the presence of a sharp resonance in annihilation cross-section.

The left (right) panel in Fig.~\ref{fig:scalar} shows the $x_f$ ($\Omega_{\rm DM} h^2$) envelope as a function of $M_{\rm med}$ for the simplified models corresponding to spin 0 DM.~In the figure, models are labelled according to the leading non-relativistic operator for DM-nucleon interactions.~Models $\hat{\mathcal{O}}_7(h_4,g_4)$ and $\hat{\mathcal{O}}_{10}(h_2,g_1)$ are not compatible with the thermal production mechanism for any value of $M_{\rm med}$, yielding a value for $\Omega_{\rm DM} h^2$ ($x_f$) much smaller (larger) than the observed one.~On the other hand, models $\hat{\mathcal{O}}_1(h_1,g_1)$ and $\hat{\mathcal{O}}_{1}(h_3,g_4)$ generate values for $\Omega_{\rm DM} h^2$ which are in general too large.~However, for $M_{\rm med}\sim 100$~GeV, i.e. at resonance, direct detection and thermal production can be compatible for models $\hat{\mathcal{O}}_1(h_1,g_1)$ and $\hat{\mathcal{O}}_{1}(h_3,g_4)$.~This result can be interpreted as follows:~if DM has spin 0 and is produced thermally, the detection of $\mathcal{O}(100)$ signal events at XENONnT would simultaneously determine the DM particle mass, and the mediator mass and spin (at least within the simplified model framework considered in this analysis).

Similarly, Fig.~\ref{fig:fermion} shows regions in the $M_{\rm med}-\Omega_{\rm DM} h^2$ plane corresponding to the detection of $\mathcal{O}(100)$ signal events at XENONnT/LZ for the case of spin 1/2 DM.~In this case, there are 5 models for which thermal production and direct detection can be compatible, namely:~$\hat{\mathcal{O}}_1(h_1,\lambda_1)$, $\hat{\mathcal{O}}_1(h_3,\lambda_3)$, $\hat{\mathcal{O}}_4(h_4,\lambda_4)$, $\hat{\mathcal{O}}_8(h_3,\lambda_4)$ and $\hat{\mathcal{O}}_{11}(h_1,\lambda_2)$.~For these 5 models, direct detection and thermal production can only be compatible at resonance.~Notice that model $\hat{\mathcal{O}}_{4}(h_4,\lambda_4)$ (green area in the right panel) corresponds to the canonical spin-dependent interaction.

\begin{figure*}[t]
\begin{center}
\begin{minipage}[t]{0.49\linewidth}
\centering
\includegraphics[width=\textwidth]{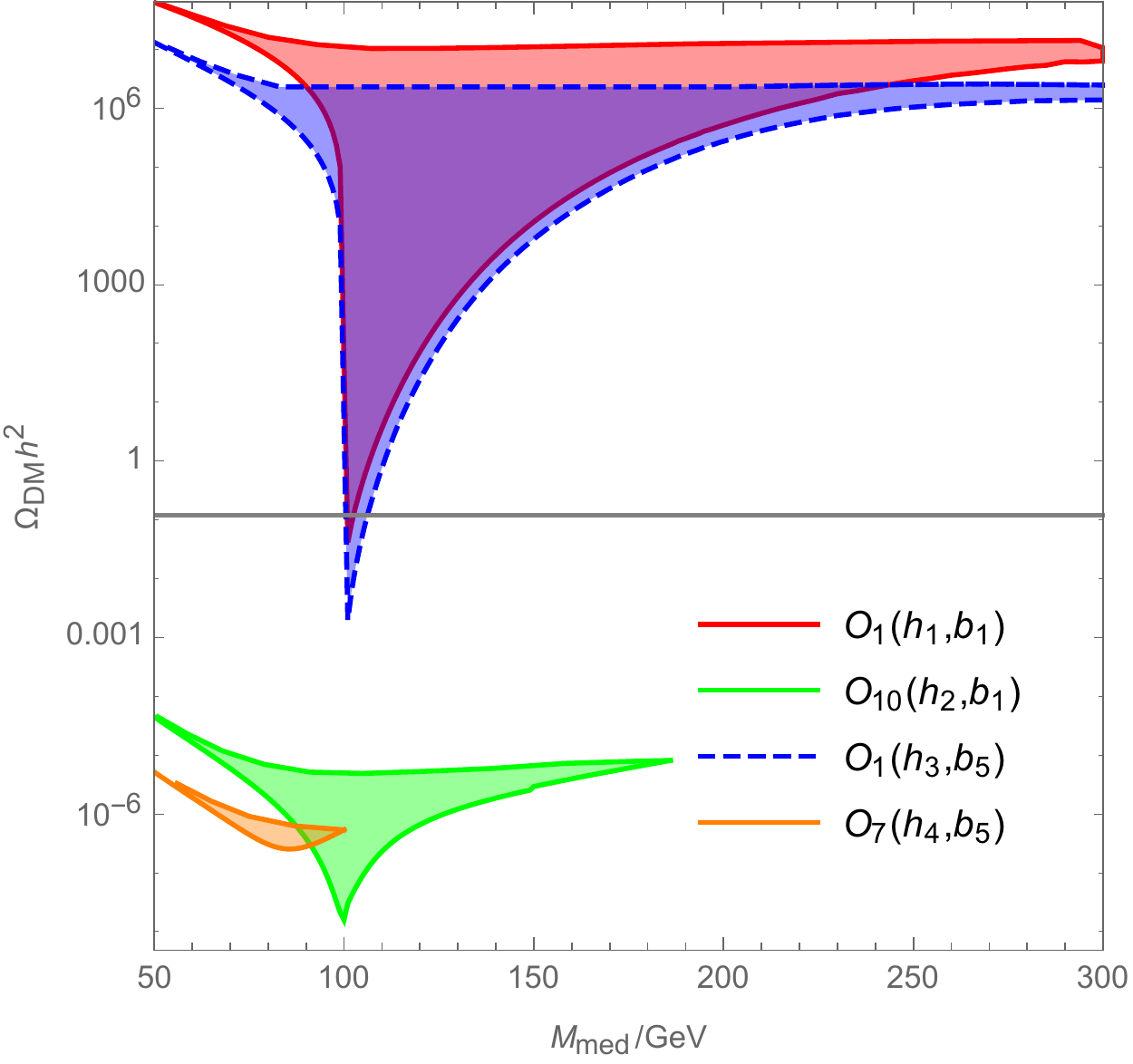}
\end{minipage}
\begin{minipage}[t]{0.49\linewidth}
\centering
\includegraphics[width=\textwidth]{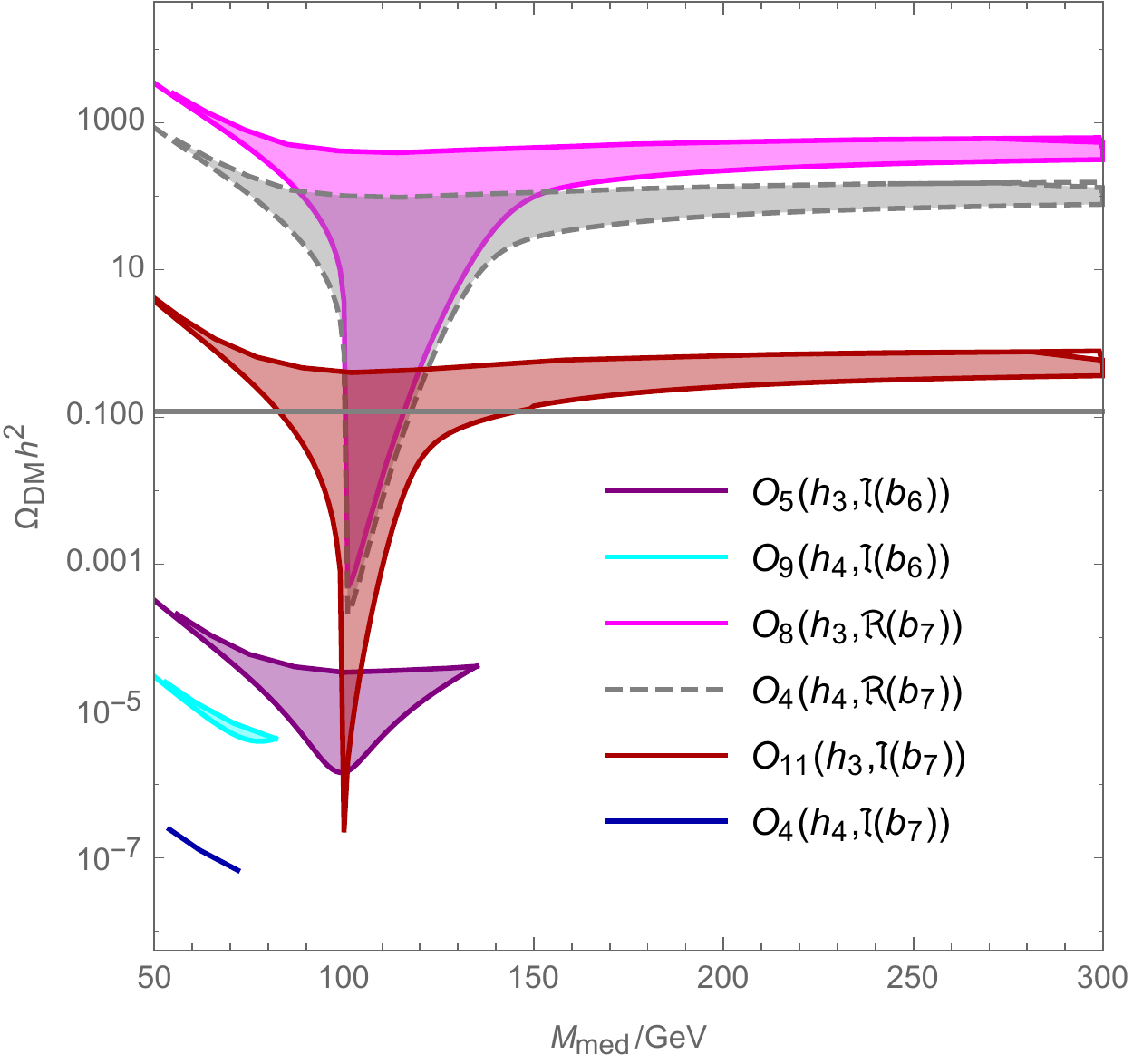}
\end{minipage}
\end{center}
\caption{Same as for Fig.~\ref{fig:scalar}, but now for spin 1 DM.~The left panel refers to CP preserving models, while the right panel corresponds to CP violating models.\label{fig:vector}}  
\end{figure*}

Finally, Fig.~\ref{fig:vector} shows the regions in the $M_{\rm med}-\Omega_{\rm DM} h^2$ plane corresponding to the detection of $\mathcal{O}(100)$ signal events at XENONnT/LZ for simplified models where DM has spin 1.~Also in this case, five models can yield the correct relic density and simultaneously explain the detection of $\mathcal{O}(100)$ signal events at XENONnT/LZ.~The five models are $\hat{\mathcal{O}}_1(h_1,b_1)$, $\hat{\mathcal{O}}_1(h_3,b_5)$, $\hat{\mathcal{O}}_4(h_4,\Re(b_7))$, $\hat{\mathcal{O}}_8(h_3,\Re(b_7))$ and $\hat{\mathcal{O}}_{11}(h_3,\Im(b_7))$.~For the $\hat{\mathcal{O}}_1(h_1,b_1)$ and $\hat{\mathcal{O}}_1(h_3,b_5)$ models, the predicted compatibility regions are identical.

\section{Discussion}
\label{sec:discussion}
In this section we briefly comment on the dependence of our results on DM particle mass and number of signal events.

\begin{figure*}[t]
\begin{center}
\begin{minipage}[t]{0.49\linewidth}
\centering
\includegraphics[width=\textwidth]{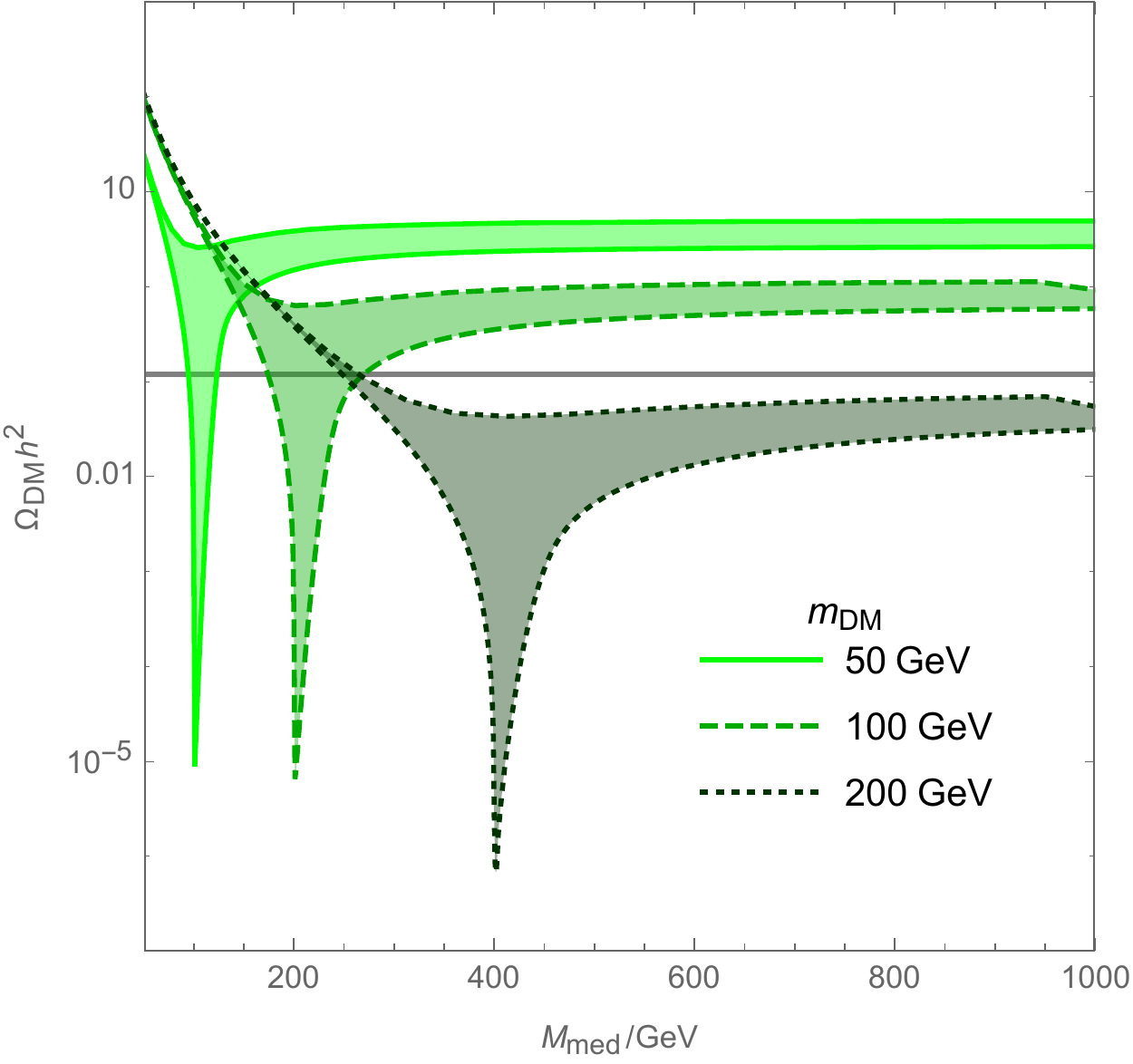}
\end{minipage}
\begin{minipage}[t]{0.49\linewidth}
\centering
\includegraphics[width=\textwidth]{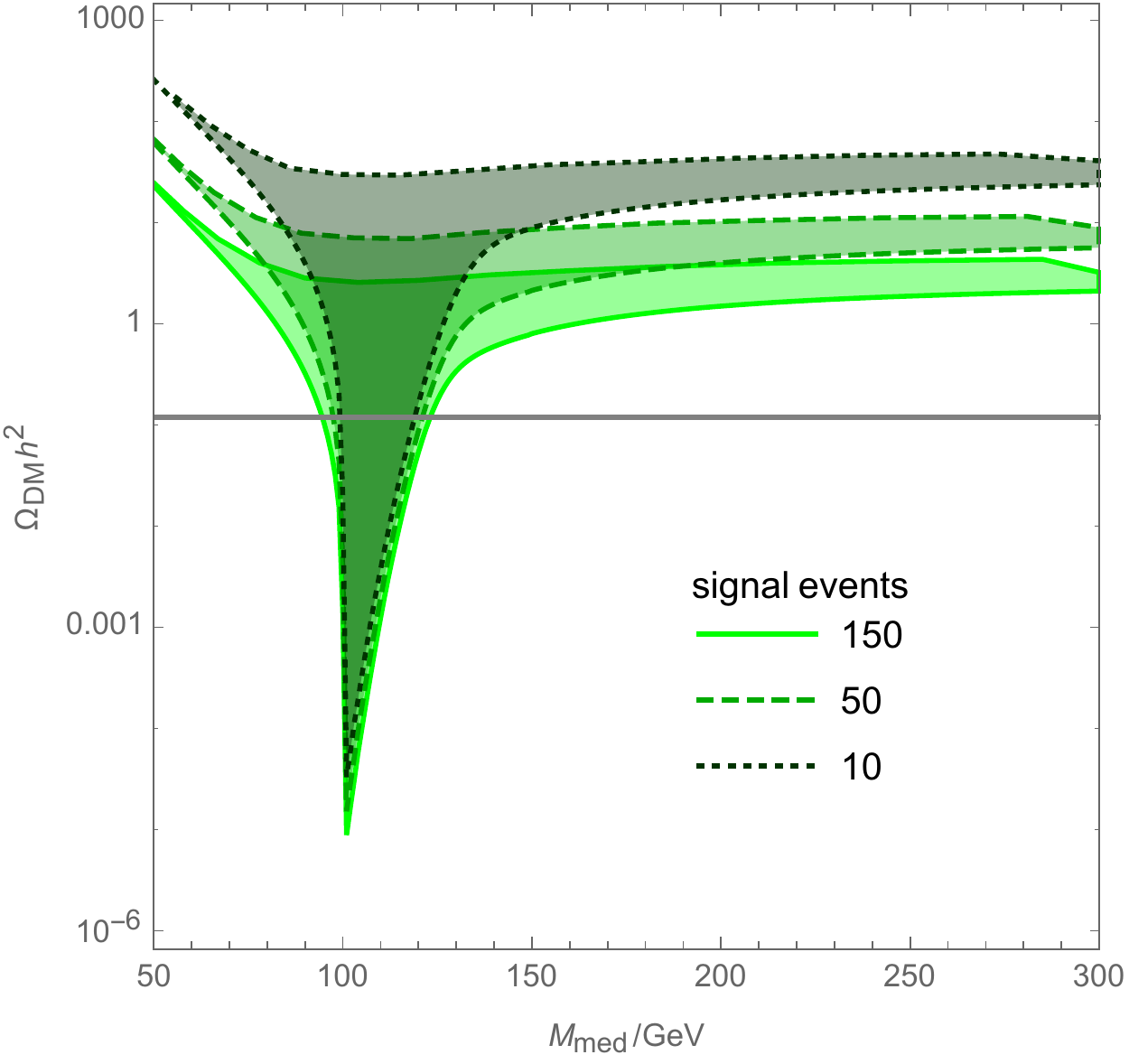}
\end{minipage}
\end{center}
\caption{The left panel shows $\Omega_{\rm DM} h^2$ as a function of $M_{\rm med}$ for model $\hat{\mathcal{O}}_{4}(h_4,\lambda_4)$, and for three masses:~$m_{\rm DM}=50, 100$ and 200~GeV.~For this model we assume 150 nuclear recoil events (in an ideal experiment with $\zeta=1$, see Sec.~\ref{sec:DD}).~By increasing $m_{\rm DM}$ envelopes move towards the bottom right corner in the $M_{\rm med} - \Omega_{\rm DM} h^2$ plane.~The right panel, shows the impact of decreasing $\mu_S$ on our results.~Here we focus on model $\hat{\mathcal{O}}_{4}(h_4,\lambda_4)$, but the results illustrated in the figure are general:~decreasing the number of signal events at XENONnT/LZ moves the $\Omega_{\rm DM} h^2$ envelopes towards larger values of $\Omega_{\rm DM} h^2$, since the larger is $\mu_S$ the smaller must be $M_{\rm eff}$.\label{fig:newassumptions}}  
\end{figure*}

Fig.~\ref{fig:newassumptions}, left panel, shows $\Omega_{\rm DM} h^2$ as a function of $M_{\rm med}$ for model $\hat{\mathcal{O}}_{4}(h_4,\lambda_4)$, and for different values of the dark matter particle mass, $m_{\rm DM}$.~For a given mediator mass, we construct an envelope for $\Omega_{\rm DM} h^2$ by varying $g_{\rm DM}$ as explained in the previous section.~For both models we assume 150 nuclear recoil events (in an ideal experiment with $\zeta=1$, see Sec.~\ref{sec:DD}) and consider three masses:~$m_{\rm DM}=50, 100$ and 200~GeV.~Above 200~GeV, we assume that errors on the reconstructed DM particle mass would be too large to be neglected in deriving the $\Omega_{\rm DM} h^2$ envelope.~We focus on model $\hat{\mathcal{O}}_{4}(h_4,\lambda_4)$ to illustrate a general result:~by increasing the DM particle mass envelopes move towards the bottom right corner in the $M_{\rm med} - \Omega_{\rm DM} h^2$ plane.~This occurs partly because the DM annihilation is resonant at $M_{\rm med}=2 m_{\rm DM}$, and partly because larger values of $m_{\rm DM}$ must be compensated by increasing $M_{\rm eff}$ in order to keep the number of signal events constant.

Fig.~\ref{fig:newassumptions}, right panel, shows the impact of decreasing $\mu_S$ on our results.~Here we focus on model $\hat{\mathcal{O}}_{4}(h_4,\lambda_4)$, but the results illustrated in Fig.~\ref{fig:mm} are general:~decreasing the number of signal events at XENONnT/LZ moves the $\Omega_{\rm DM} h^2$ envelopes towards larger values of $\Omega_{\rm DM} h^2$, since the larger is $\mu_S$ the smaller must be $M_{\rm eff}$.~In the specific case of model $\hat{\mathcal{O}}_{4}(h_4,\lambda_4)$, this can broaden the range of $M_{\rm med}$ values for which thermal production and direct detection are compatible.

The nucleon form factors given in \refeq{eq:formfac} can have large uncertainties. While the vector couplings are determined by gauge invariance, the other couplings can carry uncertainties of up to $30\%$~\cite{Agrawal:2010fh,Dienes:2013xya,Bishara:2016hek}. This corresponds to uncertainties in the benchmarks listed in \reftab{tab:benchmarks} of up to $15\%$. Varying the input of our numerical simulations for several of the previously discussed models accordingly, we find that these uncertainties will only lead to minor shifts of the predicted regions for DM relic density (less than one order of magnitude).

\section{Conclusion}
\label{sec:conclusion}
We determined under what circumstances  the detection of $\mathcal{O}(100)$ signal events at XENONnT/LZ can be compatible with the DM thermal production mechanism.~The relic density calculation was performed within the most general set of renormalisable models that preserve Lorentz and gauge symmetry, and that extend the Standard Model by one DM candidate and one particle mediating DM-quark interactions.~In this calculation, the detection of $\mathcal{O}(100)$ signal events at XENONnT/LZ was used as an input constraining the underlying model parameters.

Agreement between DM thermal production and detection of $\mathcal{O}(100)$ signal events at XENONnT/LZ was translated into compatibility regions in the $M_{\rm med} - \Omega_{\rm DM} h^2$ plane.~Deriving these compatibility regions, we also required that the coupling constants $g_{\rm DM}$ and $g_q$ were perturbative, i.e.~$\le \sqrt{4\pi}$, and that the mediator decay width was smaller than the mediator mass.~For spin 0 DM, we found that DM thermal production and detection at XENONnT/LZ can be compatible only for models $\hat{\mathcal{O}}_1(h_1,g_1)$ and $\hat{\mathcal{O}}_{1}(h_3,g_4)$ in the case of $\mathcal{O}(100)$ signal events.~For spin 1/2 DM, there are 5 models that can reconcile detection and production, namely:~$\hat{\mathcal{O}}_1(h_1,\lambda_1)$, $\hat{\mathcal{O}}_1(h_3,\lambda_3)$, $\hat{\mathcal{O}}_4(h_4,\lambda_4)$, $\hat{\mathcal{O}}_8(h_3,\lambda_4)$ and $\hat{\mathcal{O}}_{11}(h_1,\lambda_2)$.~Finally, for spin 1 DM, there are 5 models that can make DM direct detection and thermal production compatible, namely:~$\hat{\mathcal{O}}_1(h_1,b_1)$, $\hat{\mathcal{O}}_1(h_3,b_5)$, $\hat{\mathcal{O}}_4(h_4,\Re(b_7))$, $\hat{\mathcal{O}}_8(h_3,\Re(b_7))$ and $\hat{\mathcal{O}}_{11}(h_3,\Im(b_7))$.

By increasing the DM particles mass, compatibility regions move towards the bottom right corner in the $M_{\rm med} - \Omega_{\rm DM} h^2$ plane.~In the same plane, they move upwards if the number of signal events decreases.~Interestingly, for thermal DM models yielding a correct DM relic density only at resonance, direct detection experiments will be able to simultaneously reconstruct DM and mediator mass, in case of signal detection.~For these models $M_{\rm med}\simeq2 m_{\rm DM}$.~Whether or not this value for the mediator mass is excluded by the LHC searches for new physics beyond the Standard Models is a model dependent question, which crucially depends on the UV completion of the simplified models discussed in this work.~To address this question goes beyond the scope of the present analysis.~However, in the large mediator mass limit, $M_{\rm med}\gg m_{\rm DM}$, the impact of a XENONnT/LZ signal on future LHC mono-jet searches has recently been studied in~\cite{Baum:2017kfa} for the same set of simplified models considered in this work.

We complemented this analysis by providing analytic expressions for annihilation cross-sections and mediator decay widths for all models considered in this study (see Appendix~\ref{app:M}).
 
\acknowledgments This work has been supported by the Knut and Alice Wallenberg Foundation and is performed in the context of the Swedish Consortium for Direct Detection of Dark Matter (SweDCube).~We would like to thank Katherine Freese and Sebastian Baum for interesting discussions on topics related to this work, and for their feedback on a first draft of this article.~This investigation has also been supported by the Munich Institute for Astro- and Particle Physics (MIAPP) within the Deutsche Forschungsgemeinschaft (DFG) cluster of excellence ``Origin and Structure of the Universe''.~Finally, we would like to thank the participants of the programme ``Astro-, Particle and Nuclear Physics of Dark Matter Direct Detection'', hosted by MIAPP in 2017, for many valuable discussions. 

\appendix
\section{Lagrangians of simplified Models}
\label{app:Lagrangians}
In this appendix we list the Lagrangians considered in our analysis~\cite{Dent:2015zpa}. 

\subsection{Scalar dark matter}
\noindent Scalar and pseudoscalar mediator:
\begin{eqnarray}
\mathcal{L}_{S\phi q} &=& \partial_\mu S^\dagger\partial^\mu S - m_S^2S^\dagger S - \frac{\lambda_S}{2}(S^\dagger S)^2 \nonumber\\
&+&\frac{1}{2}\partial_\mu\phi\partial^\mu\phi - \frac{1}{2}m_\phi^2\phi^2 -\frac{m_\phi\mu_1}{3}\phi^3-\frac{\mu_2}{4}\phi^4 \nonumber\\
&+& \ii\bar{q}\slashed{D} q - m_q \bar q q \nonumber\\
&-&g_1m_SS^\dagger S\phi -\frac{g_2}{2}S^\dagger S\phi^2-h_1\bar q q\phi-ih_2\bar{q}\gamma^5q\phi\,. \nonumber\\ 
\end{eqnarray}
Vector and axial-vector mediator:
\begin{eqnarray}
\mathcal{L}_{SGq} &=& \partial_{\mu}S^{\dagger}\partial^{\mu}S -m_S^2 S^{\dagger}S -\frac{\lambda_S}{2}(S^{\dagger}{S})^2  \nonumber \\
&-&\frac{1}{4}\mathcal{G}_{\mu\nu}\mathcal{G}^{\mu\nu} + \frac{1}{2}m_G^2G_{\mu}G^{\mu} -\frac{\lambda_G}{4}(G_{\mu}G^{\mu})^2 \nonumber \\
&+&i\bar{q}\slashed{D}q -m_q\bar{q}q  \nonumber\\
&-&\frac{g_3}{2}S^{\dagger}SG_{\mu}G^{\mu} -ig_4(S^{\dagger}\partial_{\mu}S-\partial_{\mu}S^{\dagger}S)G^{\mu} \nonumber\\
&-&h_3(\bar{q}\gamma_{\mu}q	)G^{\mu}-h_4(\bar{q}\gamma_{\mu}\gamma^5q)G^{\mu}\,.
\end{eqnarray}

\subsection{Fermionic dark matter}
\noindent Scalar and pseudoscalar mediator:
\begin{eqnarray}
\mathcal{L}_{\chi\phi q} &=& \ii\bar{\chi}\slashed{D}\chi - m_{\chi}\bar{\chi}\chi \nonumber\\
&+&\frac{1}{2}\partial_\mu\phi\partial^\mu\phi - \frac{1}{2}m_\phi^2\phi^2 -\frac{m_\phi\mu_1}{3}\phi^3-\frac{\mu_2}{4}\phi^4 \nonumber\\
&+& \ii\bar{q}\slashed{D} q - m_q \bar q q \nonumber\\	
&-&\lambda_1\phi\bar{\chi}\chi -i\lambda_2\phi\bar{\chi}\gamma^{5}\chi-h_1\phi\bar q q-ih_2\phi\bar{q}\gamma^5q\,. \nonumber\\
\end{eqnarray}
Vector and axial-vector mediator:
\begin{eqnarray}
\mathcal{L}_{\chi Gq} &=& \ii\bar{\chi}\slashed{D}\chi - m_\chi\bar{\chi}\chi \nonumber\\
&-&\frac{1}{4}\mathcal{G}_{\mu\nu}\mathcal{G}^{\mu\nu}+\frac{1}{2}m_{G}^2G_{\mu}G^{\mu}\nonumber\\
&+& \ii\bar{q}\slashed{D} q - m_q \bar q \nonumber\\
&-&\lambda_{3}\bar\chi\gamma^\mu\chi G_{\mu}-\lambda_{4}\bar\chi\gamma^\mu\gamma^5\chi G_{\mu}\nonumber\\
&-&h_3\bar{q}\gamma_{\mu}qG^{\mu}-h_4\bar{q}\gamma_{\mu}\gamma^{5}qG^{\mu}\,.
\end{eqnarray}

\subsection{Vector dark matter}
\noindent Scalar and pseudoscalar mediator:
\begin{eqnarray}
\mathcal{L}_{X\phi q}&=&-\frac{1}{2}{\mathcal{X}}_{\mu\nu}^{\dagger}\mathcal{X}^{\mu\nu}+m_{X}^2X_{\mu}^{\dagger}X^{\mu}-\frac{\lambda_{X}}{2}(X_{\mu}^{\dagger}X^{\mu})^2 \nonumber \\
&+&\frac{1}{2}(\partial_{\mu}\phi)^2-\frac{1}{2}m_{\phi}^2\phi^2-\frac{m_\phi \mu_1}{3}\phi^3-\frac{\mu_2}{4}\phi^4 \nonumber\\
&+&\ii\bar{q}\slashed{D}q-m_{q}\bar{q}q \nonumber\\
&-&b_1m_X\phi X_{\mu}^{\dagger}X^{\mu}-\frac{b_{2}}{2}\phi^2X_{\mu}^{\dagger}X^{\mu}  \nonumber\\ 
&-&h_1\phi\bar{q}q-ih_2\phi\bar{q}\gamma^{5}q\,.  \nonumber\\ 
\end{eqnarray}
Vector and axial-vector mediator:
\begin{eqnarray}
\mathcal{L}_{XGq}&=& -\frac{1}{2}\mathcal{X}^{\dagger}_{\mu\nu}\mathcal{X}^{\mu\nu}+m_{X}^2X^{\dagger}_{\mu}X^{\mu}-\frac{\lambda_{X}}{2}(X_{\mu}^{\dagger}X^{\mu})^2 \nonumber\\
&-&\frac{1}{4}\mathcal{G}_{\mu\nu}\mathcal{G}^{\mu\nu}+\frac{1}{2}m_{G}^2G_\mu^2-\frac{\lambda_G}{4}(G_\mu G^\mu)^2\nonumber \\
&+&i\bar{q}\slashed{D}q-m_{q}\bar{q}q\nonumber\\
&-&\frac{b_3}{2}G_\mu^2(X^{\dagger}_\nu X^{\nu}) -\frac{b_{4}}{2}(G^{\mu}G^{\nu})(X^{\dagger}_{\mu}X_{\nu}) \nonumber\\ 
&-&\left[ib_{5}X_{\nu}^{\dagger}\partial_{\mu}X^{\nu}G^\mu+b_{6}X_{\mu}^{\dagger}\partial^\mu X_{\nu}G^{\nu}\right.\nonumber\\ 
&+& \left.b_{7}\epsilon_{\mu\nu\rho\sigma}(X^{\dagger\mu}\partial^{\nu}X^{\rho})G^{\sigma} +\hc \right]\nonumber\\
&-&h_3G_\mu\bar{q}\gamma^\mu q - h_4 G_\mu\bar{q}\gamma^\mu\gamma^{5}q\,. 
\end{eqnarray}

\section{Amplitudes}
\label{app:M}
The differential decay width, $\dd \Gamma$, of a particle at rest with mass $M$ into two particles with identical mass $m$ is given by
\begin{equation}
 \dd \Gamma =	 \frac{1}{2M} \bar{|\mathcal{M}|^2} \frac{\dd^3 \mathbf{p}_1}{(2\pi)^3 2 E_1}\frac{\dd^3 \mathbf{p}_2}{(2\pi^3)2 E_2}(2\pi)^4 \delta(k-p_1-p_2) \,,
\label{eq:dGamma}
\end{equation}
where $\mathcal{M}$ is the corresponding amplitude, $k$ is the sum of the incoming four-momenta, and $p_i$ and $E_i$, $i=1,2$, are the final state four-momenta and energies, respectively.~If $\mathcal{M}$ is isotropic, one can integrate Eq.~(\ref{eq:dGamma}) over $\dd^3 \mathbf{p}_1\dd^3 \mathbf{p}_2$ and obtain
\begin{align}
 \Gamma  = \frac{ \bar{|\mathcal{M}|^2}}{8\pi M^2} \sqrt{\left(\frac{M}{2}\right)^2-m^2}\,.
\end{align}
The relation between DM annihilation cross-section in the centre-of-mass frame times lab frame relative velocity and the corresponding Feynman amplitude is given in Eq.~(\ref{eq:sigmav}).

For all simplified models in Appendix~\ref{app:Lagrangians}, we list the modulus squared of the amplitudes for mediator decay and DM annihilation.~Expressions can be found in different subsections depending on the DM spin.~Results presented in this appendix were cross-checked analytically using FeynCalc~\cite{Mertig:1990an,Shtabovenko:2016sxi} and through numerical calculations performed with our modified version of WHIZARD~\cite{Kilian:2007gr}.~In the amplitudes we included the averaging over initial state spin and polarisation, where applicable.~The amplitudes including quarks are for one flavor and without color factors.

\subsection{Scalar dark matter}
\label{app:spin0}

\subsubsection{Scalar and pseudoscalar mediator}
In the case of scalar DM and scalar or pseudo-scalar mediator, the relevant terms in the Lagrangian are $-g_1 m_S S^\dagger S\phi$, $-h_1\bar q q\phi$ and $-i h_2\bar q \gamma_5 q\phi$.~Assuming that only the coupling constants $(g_1,h_1)$ are different from zero, we obtain
\begin{align}
\label{eq:g1h1_1}
 |\mathcal{M}(\phi \rightarrow S^\dagger S)|^2 &= g_1^2 \mDM^2 \,, \\
\label{eq:g1h1_2}
 |\mathcal{M}(\phi \rightarrow \bar q q)|^2 
  &= h_1^2 (2 M^2 - 8 m_q^2) \,,\\
  \label{eq:g1h1_3}
 |\mathcal{M}(S^\dagger S \rightarrow \bar q q)|^2 &=h_1^2 g_1^2\frac{(2s - 8 m_q^2) \mDM^2 }{|s-M^2+\ii M \Gamma|^2} \,,
\end{align}
and when only the coupling consants $(g_1,h_2)$ are different from zero, we find
\begin{align}
 |\mathcal{M}(\phi \rightarrow S^\dagger S)|^2 &= g_1^2 \mDM^2 \,, \\
 \label{eq:g1h2_2}
 |\mathcal{M}(\phi \rightarrow \bar q q)|^2
 &= 2 h_2^2 M^2 \,,\\
 |\mathcal{M}(S^\dagger S \rightarrow \bar q q)|^2 &= h_2^2 g_1^2\frac{2s\,  \mDM^2 }{|s-M^2+\ii M \Gamma|^2}\,.
\end{align}

\subsubsection{Vector and axial-vector mediator}
In this case, the relevant terms in the Lagrangian are  $-ig_4(S^{\dagger}\partial_{\mu}S-\partial_{\mu}S^{\dagger}S)G^{\mu}$,
$-h_3(\bar{q}\gamma_{\mu}q)G^{\mu}$ and $-h_4(\bar{q}\gamma_{\mu}\gamma^5q)G^{\mu}$.
For the decay of vector mediators, we have to average over the initial polarisation states, which requires the identity 
\begin{align}
 \frac{1}{3} \sum_a (\epsilon^{\mu}_a)^* \epsilon^\nu_a = \frac{1}{3}\left(g^{\mu\nu} - \frac{k^\mu k^\nu}{M^2} \right)\,.
\end{align}
When only the coupling constants $(h_3,g_4)$ are different from zero, we obtain
\begin{align}
 |\mathcal{M}(G \rightarrow S^\dagger S)|^2 &= 
 g_4^4 (M^2 - 4\mDM^2)\,,\\
  |\mathcal{M}(G \rightarrow \bar q q)|^2 
  & = \frac{h_3^2 }{3}(4M^2+8m_q^2)\,,\\
 |\mathcal{M}(S^\dagger S \rightarrow \bar q q)|^2 &=2 h_3^2 g_4^2 \nonumber\\ &\times 
 \frac{(s-4\mDM^2)[s-(s-4m_q^2)\cos^2\theta]}{|s-M^2+\ii M \Gamma|^2} \,. 
\end{align}
When only the coupling constants $(h_4,g_4)$ are different from zero, we find
\begin{align}
 |\mathcal{M}(G \rightarrow S^\dagger S)|^2 &= 
 g_4^4 (M^2 - 4 \mDM^2)\\
 |\mathcal{M}(G \rightarrow \bar q q)|^2
 & = \frac{h_4^2 }{3}(4M^2-16m_q^2) \,,\\
|\mathcal{M}(S^\dagger S \rightarrow \bar q q)|^2 &=2 h_3^2 g_4^2 \nonumber\\ &\times 
 \frac{(s-4\mDM^2)(s-4m_q^2)(1-\cos^2\theta)}{|s-M^2+\ii M \Gamma|^2} \,.
\end{align}

\subsection{Fermionic dark matter}

\subsubsection{Scalar and pseudoscalar mediator}
In the case of fermionic DM  and scalar or pseudo-scalar mediator, the relevant terms in the Lagrangian are $-\lambda_1\phi\bar{\chi}\chi$, $-i\lambda_2\phi\bar{\chi}\gamma^{5}\chi$, $-h_1\phi\bar q q$ and $-ih_2\phi\bar{q}\gamma^5q$.~When only the coupling constants ($\lambda_1,h_1$) are different from zero, we find 
\begin{align}
\label{eq:l1h1_1}
|\mathcal{M}(\phi \rightarrow \bar \chi \chi)|^2 &= 2 \lambda_1^2 (M^2-4\mDM)\,, \\
\label{eq:l1h1_2}
 |\mathcal{M}(\bar \chi \chi \rightarrow \bar q q)|^2 &= h_1^2 \lambda_1^2 \frac{(s-4\mDM)(s-4m_q^2)}{|s-M^2+\ii M \Gamma|^2}\,.
\end{align}
If only the coupling constants ($\lambda_2,h_2$) are different from zero, we obtain
\begin{align}
\label{eq:l2h1_1}
|\mathcal{M}(\phi \rightarrow \bar \chi \chi)|^2 &= 2 \lambda_2^2 M^2  \,,\\
\label{eq:l2h1_2}
|\mathcal{M}(\bar \chi \chi \rightarrow \bar q q)|^2 &= h_2^2 \lambda_2^2 \frac{s^2}{|s-M^2+\ii M \Gamma|^2}\,.
\end{align}
When only the coupling constants ($\lambda_1,h_2$) are different from zero, we find
\begin{equation}
 |\mathcal{M}(\bar \chi \chi \rightarrow \bar q q)|^2 = h_2^2 \lambda_1^2 \frac{s(s-4\mDM)}{|s-M^2+\ii M \Gamma|^2}\,,
\end{equation}
and for $\lambda_2\neq0$ and $h_1\neq0$ we obtain
\begin{equation}
 |\mathcal{M}(\bar \chi \chi \rightarrow \bar q q)|^2 = h_1^2 \lambda_2^2 \frac{s(s-4m_q^2)}{|s-M^2+\ii M \Gamma|^2}\,.
\end{equation}
Amplitudes for the process $\phi \rightarrow \bar q q$ with $h_1\neq 0$ and $h_2\neq 0$, and for the process $\phi\rightarrow \bar{\chi}\chi$ with $\lambda_1\neq 0$ and $\lambda_2\neq 0$ can be found in Eqs.~(\ref{eq:g1h1_2}), (\ref{eq:g1h2_2}), (\ref{eq:l1h1_1}) and (\ref{eq:l2h1_1}), respectively.

\subsubsection{Vector and axial-vector mediator}
In the case of vector or axial-vector mediator, the relevant terms in the Lagrangian are $-\lambda_{3}\bar\chi\gamma^\mu\chi G_{\mu}$, $-\lambda_{4}\bar\chi\gamma^\mu\gamma^5\chi G_{\mu}$, $-h_3\bar{q}\gamma_{\mu}qG^{\mu}$, and $-h_4\bar{q}\gamma_{\mu}\gamma^{5}qG^{\mu}$.~When only the coupling constants ($\lambda_3,h_3$) are different from zero, we find
\begin{align}
|\mathcal{M}(G \rightarrow  \bar \chi \chi)|^2 &= \frac{\lambda_3^2 }{3}(4M^2+8\mDM^2) \,,\\
|\mathcal{M}(\bar \chi \chi \rightarrow \bar q q)|^2 &= 
 \lambda_3^2 h_3^2  \big\{s^2 + 4 s(\mDM^2+m_q^2) \nonumber \\ 
  &+ (s - 4 \mDM^2)(s-4 m_q^2) \cos^2\theta \big\} \nonumber \\ 
&\times 1 /|s-M^2+\ii M \Gamma|^2\,.
\end{align}
If only the coupling constants ($\lambda_4,h_4$) are different from zero, we obtain
\begin{align}
|\mathcal{M}(G \rightarrow  \bar \chi \chi)|^2 &= \frac{\lambda_4^2 }{3}(4M^2-16\mDM^2) \,,\\
|\mathcal{M}(\bar \chi \chi \rightarrow \bar q q)|^2 &= 
 \lambda_4^2 h_4^2  \big\{s^2 - 4 s(\mDM^2+m_q^2) \nonumber \\ 
  &+ (s - 4 \mDM^2)(s-4 m_q^2) \cos^2\theta \nonumber \\ 
& + 64 \mDM^2 m_q^2  \big\} \times 1 /|s-M^2+\ii M \Gamma|^2\,.
\end{align}
Finally, for $\lambda_3\neq0$ and $h_4\neq0$ we find
\begin{align}
|\mathcal{M}(\bar \chi \chi \rightarrow \bar q q)|^2 &= 
 \lambda_3^2 h_4^2  \big\{s^2 + 4 s(\mDM^2-m_q^2) \nonumber \\ 
  &+ (s - 4 \mDM^2)(s-4 m_q^2) \cos^2\theta \nonumber \\ 
& - 16 \mDM^2 m_q^2  \big\} \times 1 /|s-M^2+\ii M \Gamma|^2\,.
\end{align}
and for $\lambda_4\neq0$ and $h_3\neq0$ we obtain
\begin{align}
|\mathcal{M}(\bar \chi \chi \rightarrow \bar q q)|^2 &= 
 \lambda_4^2 h_3^2  \big\{s^2 - 4 s(\mDM^2-m_q^2) \nonumber \\ 
  &+ (s - 4 \mDM^2)(s-4 m_q^2) \cos^2\theta \nonumber \\ 
& - 16 \mDM^2 m_q^2  \big\} \times 1 /|s-M^2+\ii M \Gamma|^2\,.
\end{align}
Amplitudes for the decay of $G$ into $\bar q q$ can be found in Sec.~\ref{app:spin0}.

\subsection{Vector dark matter}

\subsubsection{Scalar and pseudoscalar mediator}
In the case of fermionic DM  and scalar or pseudo-scalar mediator, the relevant terms in the Lagrangian are $-b_1m_X\phi X_{\mu}^{\dagger}X^{\mu}$, $-h_1\phi\bar q q$ and $-ih_2\phi\bar{q}\gamma^5q$.~When only the coupling constants ($b_1,h_1$) are different from zero, we find 
\begin{align}
|\mathcal{M}(G \rightarrow  X^\dagger X)|^2 &= \frac{b_1^2}{3}\left(\frac{M^4}{4\mDM^2}-M^2+3\mDM^2\right) \,,\\
 |\mathcal{M}(X^\dagger X \rightarrow \bar q q)|^2 &= 
 \frac{b_1^2 h_1^2}{18 \mDM^2}(s^2-4s\mDM^2+12\mDM^4) \nonumber \\ 
&\times (s-4m_q^2) /|s-M^2+\ii M \Gamma|^2\,.
\end{align}
If only the coupling constants ($b_1,h_2$) are different from zero, we obtain
\begin{align}
 |\mathcal{M}(X^\dagger X \rightarrow \bar q q)|^2 &= 
 \frac{b_1^2 h_1^2}{18 \mDM^2}(s^2-4s\mDM^2+12\mDM^4) \nonumber \\ 
&\times s  /|s-M^2+\ii M \Gamma|^2\,.
\end{align}

\subsubsection{Vector and axial-vector mediator}

In the case of vector or axial-vector mediator, the relevant terms in the Lagrangian are $-(ib_{5}X_{\nu}^{\dagger}\partial_{\mu}X^{\nu}G^\mu + \hc)$, $-(b_{6}X_{\mu}^{\dagger}\partial^\mu X_{\nu}G^{\nu} + \hc)$, 
$-(b_{7}\epsilon_{\mu\nu\rho\sigma}(X^{\dagger\mu}\partial^{\nu}X^{\rho})G^{\sigma} + \hc)$, $-h_3\bar{q}\gamma_{\mu}qG^{\mu}$, and $-h_4\bar{q}\gamma_{\mu}\gamma^{5}qG^{\mu}$.~When only the coupling constants ($b_5,h_3$) are different from zero, we find
\begin{align}
|\mathcal{M}(G \rightarrow  X^\dagger X)|^2 &= \frac{b_5^2}{12\mDM^4}(M^6-8M^4\mDM^2 \nonumber\\
&+28M^2\mDM^4- 48\mDM^6) \,,\\
 |\mathcal{M}(X^\dagger X \rightarrow \bar q q)|^2 &= 
 \frac{b_5^2 h_3^2}{18\mDM^4} (s-4\mDM^2)\nonumber\\
 &\times(s^2-4s\mDM^2+12\mDM^4)  \nonumber \\ 
  &\times(s-(s-4m_q^2)\cos^2\theta)  \nonumber \\ 
&\times 1/|s-M^2+\ii M \Gamma|^2\,.
\end{align}
When only ($b_5,h_4$)  are different from zero, we find
\begin{align}
 |\mathcal{M}(X^\dagger X \rightarrow \bar q q)|^2 &= 
 \frac{b_5^2 h_3^2}{18\mDM^4} (s-4\mDM^2)\nonumber\\
 &\times(s^2-4s\mDM^2+12\mDM^4)  \nonumber \\ 
  &\times(s-4m_q^2-(s-4m_q^2)\cos^2\theta)  \nonumber \\ 
&\times 1/|s-M^2+\ii M \Gamma|^2\,.
\end{align}
For ($\Re(b_6),h_3$) different from zero, we find
\begin{align}
|\mathcal{M}(G \rightarrow  X^\dagger X)|^2 &= \frac{\Re(b_6)^2}{3\mDM^2}(M^4-4 M^2\mDM^2)  \,,\\
 |\mathcal{M}(X^\dagger X \rightarrow \bar q q)|^2 &= 
 \frac{\Re(b_6)^2 h_3^2}{9\mDM^2} (s-4\mDM^2) \nonumber \\ 
 &\times \left[(1+\cos^2\theta)s+4m_q^2(1-\cos^2\theta)\right] \nonumber \\ 
&\times s/|s-M^2+\ii M \Gamma|^2\,.
\end{align}
For ($\Re(b_6),h_4$)  different from zero, we obtain
\begin{align}
 |\mathcal{M}(X^\dagger X \rightarrow \bar q q)|^2 &= 
 \frac{\Re(b_6)^2 h_4^2}{9\mDM^2}  (s-4\mDM^2) \nonumber \\ 
 &\times \left[(1+\cos^2\theta)(s-4m_q^2)\mDM^2\right.\nonumber\\
 &\left.+2m_q^2(s-4\mDM^2)\right] \nonumber \\ 
&\times s/|s-M^2+\ii M \Gamma|^2\,.
\end{align}
When only ($\Im(b_6),h_3$)  are different from zero, we have
\begin{align}
|\mathcal{M}(G \rightarrow  X^\dagger X)|^2 &= \frac{\Im(b_6)^2}{12\mDM^4}(M^6-16 M^2\mDM^4)  \,,\\
 |\mathcal{M}(X^\dagger X \rightarrow \bar q q)|^2 &= 
 \frac{\Im(b_6)^2 h_3^2}{9\mDM^2} (s-4\mDM^2) \nonumber \\ 
 &\times\left[(s-2\mDM^2)(s-4m_q^2) (1-\cos^2\theta) \right.\nonumber\\
 &\left.+4s(\mDM^2+m_q^2)\right] \nonumber \\ 
&\times s/|s-M^2+\ii M \Gamma|^2\,.
\end{align}
If only ($\Im(b_6),h_4$)  are different from zero, we obtain
\begin{align}
 |\mathcal{M}(X^\dagger X \rightarrow \bar q q)|^2 &= 
 \frac{\Im(b_6)^2 h_4^2}{18\mDM^4} (s-4\mDM^2) \nonumber \\ 
 &\times(s-4m_q^2)\left[s + 2\mDM^2 \right. \nonumber \\ 
 &\left.-(s-2\mDM^2)\cos^2\theta\right]\nonumber \\ 
&\times s/|s-M^2+\ii M \Gamma|^2\,.
\end{align}
For ($\Re(b_7),h_3$)  different from zero, we find
\begin{align}
|\mathcal{M}(G \rightarrow  X^\dagger X)|^2 &= \frac{\Re(b_7)^2}{3\mDM^2}(M^4-8 M^2\mDM^2+16\mDM^4)  \,,\\
 |\mathcal{M}(X^\dagger X \rightarrow \bar q q)|^2 &= 
 \frac{\Re(b_7)^2 h_3^2}{9\mDM^2} (s-4\mDM^2)^2 \nonumber \\ 
 &\times \left[(1+\cos^2\theta)s+4m_q^2(1-\cos^2\theta)\right] \nonumber \\ 
&\times 1/|s-M^2+\ii M \Gamma|^2\,.
\end{align}
For ($\Re(b_7),h_4$)  different from zero, we have
\begin{align}
 |\mathcal{M}(X^\dagger X \rightarrow \bar q q)|^2 &= 
 \frac{\Re(b_7)^2 h_4^2}{9\mDM^2} (s-4\mDM^2) \nonumber \\ 
 &\times\left\{[s^2-4s(\mDM^2+m_q^2)] (1+\cos^2\theta)\right. \nonumber \\ 
&\left.+16\mDM^2 m_q^2(2+\cos^2\theta)\right\} \nonumber \\
&\times 1/|s-M^2+\ii M \Gamma|^2\,.
\end{align}
When only ($\Im(b_7),h_3$)  are different from zero, we find
\begin{align}
|\mathcal{M}(G \rightarrow  X^\dagger X)|^2 &= \frac{\Im(b_7)^2}{3\mDM^2}(M^4+2 M^2\mDM^2)  \,,\\
 |\mathcal{M}(X^\dagger X \rightarrow \bar q q)|^2 &= 
 \frac{\Im(b_7)^2 h_3^2}{9\mDM^2} \left[s^2 + 4s(\mDM^2+m_q^2)\right. \nonumber \\
 &\left.+(s-4\mDM^2)(s-4m_q^2)\cos^2\theta \right]\nonumber \\ 
&\times s/|s-M^2+\ii M \Gamma|^2\,.
\end{align}
Finally, for ($\Im(b_7),h_4$)  different from zero, we obtain
\begin{align}
 |\mathcal{M}(X^\dagger X \rightarrow \bar q q)|^2 &= 
 \frac{\Im(b_7)^2 h_4^2}{9\mDM^2} (s-4m_q^2) \nonumber \\ 
 &\times\left[s + 4\mDM^2 + (s-4\mDM^2)\cos^2\theta\right] \nonumber\\ 
&\times s /|s-M^2+\ii M \Gamma|^2\,.
\end{align}

\section{Non-relativistic approximation}
\label{app:app}
For the freeze out point and the relic abundance the approximations
\begin{align}
\label{eq:xf_app}
 x_f &= \ln \left(0.038 \frac{g}{\sqrt{h_\text{eff}}} M_\text{Pl} \mDM \sigma_0 \right) \,, 
\intertext{and} 
 \label{eq:Y0_app}
\qquad Y_0 &= 3.79 \frac{\sqrt{g_*}}{h_\text{eff}(T)} \frac{x_f}{ M_\text{Pl} \mDM \sigma_0 }
\end{align}
are commonly used, where $\sigma_0$ is the first term in the expansion of $\sigma v$ in powers of $v$. The approximated blue regions in \reffig{fig:O1O10} use $\sigma_0$ (middel panels) as well as \refeq{eq:xf_app} (upper panels) and \refeq{eq:Y0_app} together with 
\refeq{eq:Omega_h2} (lower panels), where in the latter two cases the correct value for $\langle \sigma v_{\rm M\o l}\rangle$ was used as input instead of $\sigma_0$. 

If DM annihilates into a $\bar q q$ pair through a resonance of narrow width, Eq.~(\ref{eq:sigmav}) can be approximated as follows~\cite{Gondolo:1990dk}
\begin{equation}
 \sigma v_\text{lab} = \frac{8 \pi^2}{\mDM} \frac{2J + 1}{(2S+1)^2} \gamma_R \delta(\epsilon - \epsilon_R) b_R(\epsilon_R)\,,
\label{eq:sigmavres}
\end{equation}
where $\gamma_R = m_R \Gamma_R / (4 m^2)$, $\epsilon_R = (m_R^2-4m^2)/(4m^2)$ and $b_R = B_R (1-B_R)\sqrt{1+\epsilon_R}/(\sqrt{\epsilon_R(1+2\epsilon_R)})$, with $m_R$, $\Gamma_R$ and $J$ mass, decay width and spin of the resonance, respectively.~In Eq.~(\ref{eq:sigmavres}), $S$ is the DM particle spin, and $B_R$ is the branching ratio for the resonance decay into a pair of DM particles.~Combining Eq.~(\ref{eq:sigmavres}) with Eq.~(\ref{eq:sigmavmol}), one finds
\begin{align}
 \langle \sigma v_{\rm M\o l}\rangle  &= \frac{16 \pi^2}{\mDM^2} \frac{2J + 1}{(2S+1)^2} \frac{x}{K_2^2(x)} K_1\left(2x\sqrt{1+\epsilon_R}\right)\sqrt{\epsilon_R} \nonumber\\ 
 &\times  \gamma_R b_R (\epsilon_R) \theta(\epsilon_R)\,.
   \label{eq:approx_delta_rel}
\end{align}
The non-relativistic limit of Eq.~(\ref{eq:approx_delta_rel}) yields 
\begin{align}
 \langle \sigma v_{\rm M\o l}\rangle &= \frac{16 \pi}{\mDM^2} \frac{2J + 1}{(2S+1)^2} \sqrt{\pi} x^\frac{3}{2} \ee^{-x \epsilon_R} \sqrt{\epsilon_R} \nonumber\\ 
 &\times \gamma_R b_R (\epsilon_R) \theta(\epsilon_R)\,.
 \label{eq:approx_delta_nonrel}
\end{align}


\begin{thebibliography}{69}%
\makeatletter
\providecommand \@ifxundefined [1]{%
 \@ifx{#1\undefined}
}%
\providecommand \@ifnum [1]{%
 \ifnum #1\expandafter \@firstoftwo
 \else \expandafter \@secondoftwo
 \fi
}%
\providecommand \@ifx [1]{%
 \ifx #1\expandafter \@firstoftwo
 \else \expandafter \@secondoftwo
 \fi
}%
\providecommand \natexlab [1]{#1}%
\providecommand \enquote  [1]{``#1''}%
\providecommand \bibnamefont  [1]{#1}%
\providecommand \bibfnamefont [1]{#1}%
\providecommand \citenamefont [1]{#1}%
\providecommand \href@noop [0]{\@secondoftwo}%
\providecommand \href [0]{\begingroup \@sanitize@url \@href}%
\providecommand \@href[1]{\@@startlink{#1}\@@href}%
\providecommand \@@href[1]{\endgroup#1\@@endlink}%
\providecommand \@sanitize@url [0]{\catcode `\\12\catcode `\$12\catcode
  `\&12\catcode `\#12\catcode `\^12\catcode `\_12\catcode `\%12\relax}%
\providecommand \@@startlink[1]{}%
\providecommand \@@endlink[0]{}%
\providecommand \url  [0]{\begingroup\@sanitize@url \@url }%
\providecommand \@url [1]{\endgroup\@href {#1}{\urlprefix }}%
\providecommand \urlprefix  [0]{URL }%
\providecommand \Eprint [0]{\href }%
\providecommand \doibase [0]{http://dx.doi.org/}%
\providecommand \selectlanguage [0]{\@gobble}%
\providecommand \bibinfo  [0]{\@secondoftwo}%
\providecommand \bibfield  [0]{\@secondoftwo}%
\providecommand \translation [1]{[#1]}%
\providecommand \BibitemOpen [0]{}%
\providecommand \bibitemStop [0]{}%
\providecommand \bibitemNoStop [0]{.\EOS\space}%
\providecommand \EOS [0]{\spacefactor3000\relax}%
\providecommand \BibitemShut  [1]{\csname bibitem#1\endcsname}%
\let\auto@bib@innerbib\@empty
\bibitem [{\citenamefont {Bertone}\ and\ \citenamefont
  {Hooper}(2016)}]{Bertone:2016nfn}%
  \BibitemOpen
  \bibfield  {author} {\bibinfo {author} {\bibfnamefont {G.}~\bibnamefont
  {Bertone}}\ and\ \bibinfo {author} {\bibfnamefont {D.}~\bibnamefont
  {Hooper}},\ }\href@noop {} {\bibfield  {journal} {\bibinfo  {journal}
  {Submitted to: Rev. Mod. Phys.}\ } (\bibinfo {year} {2016})},\ \Eprint
  {http://arxiv.org/abs/1605.04909} {arXiv:1605.04909 [astro-ph.CO]}
  \BibitemShut {NoStop}%
\bibitem [{\citenamefont {Iocco}\ \emph {et~al.}(2015)\citenamefont {Iocco},
  \citenamefont {Pato},\ and\ \citenamefont {Bertone}}]{Iocco:2015xga}%
  \BibitemOpen
  \bibfield  {author} {\bibinfo {author} {\bibfnamefont {F.}~\bibnamefont
  {Iocco}}, \bibinfo {author} {\bibfnamefont {M.}~\bibnamefont {Pato}}, \ and\
  \bibinfo {author} {\bibfnamefont {G.}~\bibnamefont {Bertone}},\ }\href
  {\doibase 10.1038/nphys3237} {\bibfield  {journal} {\bibinfo  {journal}
  {Nature Phys.}\ }\textbf {\bibinfo {volume} {11}},\ \bibinfo {pages} {245}
  (\bibinfo {year} {2015})},\ \Eprint {http://arxiv.org/abs/1502.03821}
  {arXiv:1502.03821 [astro-ph.GA]} \BibitemShut {NoStop}%
\bibitem [{\citenamefont {Springel}\ \emph {et~al.}(2006)\citenamefont
  {Springel}, \citenamefont {Frenk},\ and\ \citenamefont
  {White}}]{Springel:2006vs}%
  \BibitemOpen
  \bibfield  {author} {\bibinfo {author} {\bibfnamefont {V.}~\bibnamefont
  {Springel}}, \bibinfo {author} {\bibfnamefont {C.~S.}\ \bibnamefont {Frenk}},
  \ and\ \bibinfo {author} {\bibfnamefont {S.~D.~M.}\ \bibnamefont {White}},\
  }\href {\doibase 10.1038/nature04805} {\bibfield  {journal} {\bibinfo
  {journal} {Nature}\ }\textbf {\bibinfo {volume} {440}},\ \bibinfo {pages}
  {1137} (\bibinfo {year} {2006})},\ \Eprint
  {http://arxiv.org/abs/astro-ph/0604561} {arXiv:astro-ph/0604561 [astro-ph]}
  \BibitemShut {NoStop}%
\bibitem [{\citenamefont {Mukhanov}(2004)}]{Mukhanov:2003xr}%
  \BibitemOpen
  \bibfield  {author} {\bibinfo {author} {\bibfnamefont {V.~F.}\ \bibnamefont
  {Mukhanov}},\ }\bibfield  {booktitle} {\emph {\bibinfo {booktitle} {{The
  early universe: Confronting theory with observations. Proceedings, 8th
  Workshop, Peyresq Physics 8, Peyresq, France, June 21-27, 2003}}},\ }\href
  {\doibase 10.1023/B:IJTP.0000048168.90282.db} {\bibfield  {journal} {\bibinfo
   {journal} {Int. J. Theor. Phys.}\ }\textbf {\bibinfo {volume} {43}},\
  \bibinfo {pages} {623} (\bibinfo {year} {2004})},\ \Eprint
  {http://arxiv.org/abs/astro-ph/0303072} {arXiv:astro-ph/0303072 [astro-ph]}
  \BibitemShut {NoStop}%
\bibitem [{\citenamefont {Ade}\ \emph {et~al.}(2016)\citenamefont {Ade} \emph
  {et~al.}}]{Ade:2015xua}%
  \BibitemOpen
  \bibfield  {author} {\bibinfo {author} {\bibfnamefont {P.~A.~R.}\
  \bibnamefont {Ade}} \emph {et~al.} (\bibinfo {collaboration} {Planck}),\
  }\href {\doibase 10.1051/0004-6361/201525830} {\bibfield  {journal} {\bibinfo
   {journal} {Astron. Astrophys.}\ }\textbf {\bibinfo {volume} {594}},\
  \bibinfo {pages} {A13} (\bibinfo {year} {2016})},\ \Eprint
  {http://arxiv.org/abs/1502.01589} {arXiv:1502.01589 [astro-ph.CO]}
  \BibitemShut {NoStop}%
\bibitem [{\citenamefont {Lee}\ and\ \citenamefont
  {Weinberg}(1977)}]{Lee:1977ua}%
  \BibitemOpen
  \bibfield  {author} {\bibinfo {author} {\bibfnamefont {B.~W.}\ \bibnamefont
  {Lee}}\ and\ \bibinfo {author} {\bibfnamefont {S.}~\bibnamefont {Weinberg}},\
  }\href {\doibase 10.1103/PhysRevLett.39.165} {\bibfield  {journal} {\bibinfo
  {journal} {Phys. Rev. Lett.}\ }\textbf {\bibinfo {volume} {39}},\ \bibinfo
  {pages} {165} (\bibinfo {year} {1977})}\BibitemShut {NoStop}%
\bibitem [{\citenamefont {Griest}\ and\ \citenamefont
  {Kamionkowski}(1990)}]{Griest:1989wd}%
  \BibitemOpen
  \bibfield  {author} {\bibinfo {author} {\bibfnamefont {K.}~\bibnamefont
  {Griest}}\ and\ \bibinfo {author} {\bibfnamefont {M.}~\bibnamefont
  {Kamionkowski}},\ }\href {\doibase 10.1103/PhysRevLett.64.615} {\bibfield
  {journal} {\bibinfo  {journal} {Phys. Rev. Lett.}\ }\textbf {\bibinfo
  {volume} {64}},\ \bibinfo {pages} {615} (\bibinfo {year} {1990})}\BibitemShut
  {NoStop}%
\bibitem [{\citenamefont {Roszkowski}\ \emph {et~al.}(2017)\citenamefont
  {Roszkowski}, \citenamefont {Sessolo},\ and\ \citenamefont
  {Trojanowski}}]{Roszkowski:2017nbc}%
  \BibitemOpen
  \bibfield  {author} {\bibinfo {author} {\bibfnamefont {L.}~\bibnamefont
  {Roszkowski}}, \bibinfo {author} {\bibfnamefont {E.~M.}\ \bibnamefont
  {Sessolo}}, \ and\ \bibinfo {author} {\bibfnamefont {S.}~\bibnamefont
  {Trojanowski}},\ }\href@noop {} {\  (\bibinfo {year} {2017})},\ \Eprint
  {http://arxiv.org/abs/1707.06277} {arXiv:1707.06277 [hep-ph]} \BibitemShut
  {NoStop}%
\bibitem [{\citenamefont {Arcadi}\ \emph {et~al.}(2017)\citenamefont {Arcadi},
  \citenamefont {Dutra}, \citenamefont {Ghosh}, \citenamefont {Lindner},
  \citenamefont {Mambrini}, \citenamefont {Pierre}, \citenamefont {Profumo},\
  and\ \citenamefont {Queiroz}}]{Arcadi:2017kky}%
  \BibitemOpen
  \bibfield  {author} {\bibinfo {author} {\bibfnamefont {G.}~\bibnamefont
  {Arcadi}}, \bibinfo {author} {\bibfnamefont {M.}~\bibnamefont {Dutra}},
  \bibinfo {author} {\bibfnamefont {P.}~\bibnamefont {Ghosh}}, \bibinfo
  {author} {\bibfnamefont {M.}~\bibnamefont {Lindner}}, \bibinfo {author}
  {\bibfnamefont {Y.}~\bibnamefont {Mambrini}}, \bibinfo {author}
  {\bibfnamefont {M.}~\bibnamefont {Pierre}}, \bibinfo {author} {\bibfnamefont
  {S.}~\bibnamefont {Profumo}}, \ and\ \bibinfo {author} {\bibfnamefont
  {F.~S.}\ \bibnamefont {Queiroz}},\ }\href@noop {} {\  (\bibinfo {year}
  {2017})},\ \Eprint {http://arxiv.org/abs/1703.07364} {arXiv:1703.07364
  [hep-ph]} \BibitemShut {NoStop}%
\bibitem [{\citenamefont {Hall}\ \emph {et~al.}(2010)\citenamefont {Hall},
  \citenamefont {Jedamzik}, \citenamefont {March-Russell},\ and\ \citenamefont
  {West}}]{Hall:2009bx}%
  \BibitemOpen
  \bibfield  {author} {\bibinfo {author} {\bibfnamefont {L.~J.}\ \bibnamefont
  {Hall}}, \bibinfo {author} {\bibfnamefont {K.}~\bibnamefont {Jedamzik}},
  \bibinfo {author} {\bibfnamefont {J.}~\bibnamefont {March-Russell}}, \ and\
  \bibinfo {author} {\bibfnamefont {S.~M.}\ \bibnamefont {West}},\ }\href
  {\doibase 10.1007/JHEP03(2010)080} {\bibfield  {journal} {\bibinfo  {journal}
  {JHEP}\ }\textbf {\bibinfo {volume} {03}},\ \bibinfo {pages} {080} (\bibinfo
  {year} {2010})},\ \Eprint {http://arxiv.org/abs/0911.1120} {arXiv:0911.1120
  [hep-ph]} \BibitemShut {NoStop}%
\bibitem [{\citenamefont {Baer}\ \emph {et~al.}(2015)\citenamefont {Baer},
  \citenamefont {Choi}, \citenamefont {Kim},\ and\ \citenamefont
  {Roszkowski}}]{Baer:2014eja}%
  \BibitemOpen
  \bibfield  {author} {\bibinfo {author} {\bibfnamefont {H.}~\bibnamefont
  {Baer}}, \bibinfo {author} {\bibfnamefont {K.-Y.}\ \bibnamefont {Choi}},
  \bibinfo {author} {\bibfnamefont {J.~E.}\ \bibnamefont {Kim}}, \ and\
  \bibinfo {author} {\bibfnamefont {L.}~\bibnamefont {Roszkowski}},\ }\href
  {\doibase 10.1016/j.physrep.2014.10.002} {\bibfield  {journal} {\bibinfo
  {journal} {Phys. Rept.}\ }\textbf {\bibinfo {volume} {555}},\ \bibinfo
  {pages} {1} (\bibinfo {year} {2015})},\ \Eprint
  {http://arxiv.org/abs/1407.0017} {arXiv:1407.0017 [hep-ph]} \BibitemShut
  {NoStop}%
\bibitem [{\citenamefont {Gondolo}\ and\ \citenamefont
  {Gelmini}(1991)}]{Gondolo:1990dk}%
  \BibitemOpen
  \bibfield  {author} {\bibinfo {author} {\bibfnamefont {P.}~\bibnamefont
  {Gondolo}}\ and\ \bibinfo {author} {\bibfnamefont {G.}~\bibnamefont
  {Gelmini}},\ }\href {\doibase 10.1016/0550-3213(91)90438-4} {\bibfield
  {journal} {\bibinfo  {journal} {Nucl. Phys.}\ }\textbf {\bibinfo {volume}
  {B360}},\ \bibinfo {pages} {145} (\bibinfo {year} {1991})}\BibitemShut
  {NoStop}%
\bibitem [{\citenamefont {Edsjo}\ and\ \citenamefont
  {Gondolo}(1997)}]{Edsjo:1997bg}%
  \BibitemOpen
  \bibfield  {author} {\bibinfo {author} {\bibfnamefont {J.}~\bibnamefont
  {Edsjo}}\ and\ \bibinfo {author} {\bibfnamefont {P.}~\bibnamefont
  {Gondolo}},\ }\href {\doibase 10.1103/PhysRevD.56.1879} {\bibfield  {journal}
  {\bibinfo  {journal} {Phys. Rev.}\ }\textbf {\bibinfo {volume} {D56}},\
  \bibinfo {pages} {1879} (\bibinfo {year} {1997})},\ \Eprint
  {http://arxiv.org/abs/hep-ph/9704361} {arXiv:hep-ph/9704361 [hep-ph]}
  \BibitemShut {NoStop}%
\bibitem [{\citenamefont {Harz}\ \emph {et~al.}(2016)\citenamefont {Harz},
  \citenamefont {Herrmann}, \citenamefont {Klasen}, \citenamefont {Kovarik},\
  and\ \citenamefont {Steppeler}}]{Harz:2016dql}%
  \BibitemOpen
  \bibfield  {author} {\bibinfo {author} {\bibfnamefont {J.}~\bibnamefont
  {Harz}}, \bibinfo {author} {\bibfnamefont {B.}~\bibnamefont {Herrmann}},
  \bibinfo {author} {\bibfnamefont {M.}~\bibnamefont {Klasen}}, \bibinfo
  {author} {\bibfnamefont {K.}~\bibnamefont {Kovarik}}, \ and\ \bibinfo
  {author} {\bibfnamefont {P.}~\bibnamefont {Steppeler}},\ }\href {\doibase
  10.1103/PhysRevD.93.114023} {\bibfield  {journal} {\bibinfo  {journal} {Phys.
  Rev.}\ }\textbf {\bibinfo {volume} {D93}},\ \bibinfo {pages} {114023}
  (\bibinfo {year} {2016})},\ \Eprint {http://arxiv.org/abs/1602.08103}
  {arXiv:1602.08103 [hep-ph]} \BibitemShut {NoStop}%
\bibitem [{\citenamefont {Feng}\ \emph {et~al.}(2010)\citenamefont {Feng},
  \citenamefont {Kaplinghat},\ and\ \citenamefont {Yu}}]{Feng:2010zp}%
  \BibitemOpen
  \bibfield  {author} {\bibinfo {author} {\bibfnamefont {J.~L.}\ \bibnamefont
  {Feng}}, \bibinfo {author} {\bibfnamefont {M.}~\bibnamefont {Kaplinghat}}, \
  and\ \bibinfo {author} {\bibfnamefont {H.-B.}\ \bibnamefont {Yu}},\ }\href
  {\doibase 10.1103/PhysRevD.82.083525} {\bibfield  {journal} {\bibinfo
  {journal} {Phys. Rev.}\ }\textbf {\bibinfo {volume} {D82}},\ \bibinfo {pages}
  {083525} (\bibinfo {year} {2010})},\ \Eprint {http://arxiv.org/abs/1005.4678}
  {arXiv:1005.4678 [hep-ph]} \BibitemShut {NoStop}%
\bibitem [{\citenamefont {Catena}\ \emph {et~al.}(2004)\citenamefont {Catena},
  \citenamefont {Fornengo}, \citenamefont {Masiero}, \citenamefont {Pietroni},\
  and\ \citenamefont {Rosati}}]{Catena:2004ba}%
  \BibitemOpen
  \bibfield  {author} {\bibinfo {author} {\bibfnamefont {R.}~\bibnamefont
  {Catena}}, \bibinfo {author} {\bibfnamefont {N.}~\bibnamefont {Fornengo}},
  \bibinfo {author} {\bibfnamefont {A.}~\bibnamefont {Masiero}}, \bibinfo
  {author} {\bibfnamefont {M.}~\bibnamefont {Pietroni}}, \ and\ \bibinfo
  {author} {\bibfnamefont {F.}~\bibnamefont {Rosati}},\ }\href {\doibase
  10.1103/PhysRevD.70.063519} {\bibfield  {journal} {\bibinfo  {journal} {Phys.
  Rev.}\ }\textbf {\bibinfo {volume} {D70}},\ \bibinfo {pages} {063519}
  (\bibinfo {year} {2004})},\ \Eprint {http://arxiv.org/abs/astro-ph/0403614}
  {arXiv:astro-ph/0403614 [astro-ph]} \BibitemShut {NoStop}%
\bibitem [{\citenamefont {Gelmini}\ and\ \citenamefont
  {Gondolo}(2006)}]{Gelmini:2006pw}%
  \BibitemOpen
  \bibfield  {author} {\bibinfo {author} {\bibfnamefont {G.~B.}\ \bibnamefont
  {Gelmini}}\ and\ \bibinfo {author} {\bibfnamefont {P.}~\bibnamefont
  {Gondolo}},\ }\href {\doibase 10.1103/PhysRevD.74.023510} {\bibfield
  {journal} {\bibinfo  {journal} {Phys. Rev.}\ }\textbf {\bibinfo {volume}
  {D74}},\ \bibinfo {pages} {023510} (\bibinfo {year} {2006})},\ \Eprint
  {http://arxiv.org/abs/hep-ph/0602230} {arXiv:hep-ph/0602230 [hep-ph]}
  \BibitemShut {NoStop}%
\bibitem [{\citenamefont {Gondolo}\ \emph {et~al.}(2004)\citenamefont
  {Gondolo}, \citenamefont {Edsjo}, \citenamefont {Ullio}, \citenamefont
  {Bergstrom}, \citenamefont {Schelke},\ and\ \citenamefont
  {Baltz}}]{Gondolo:2004sc}%
  \BibitemOpen
  \bibfield  {author} {\bibinfo {author} {\bibfnamefont {P.}~\bibnamefont
  {Gondolo}}, \bibinfo {author} {\bibfnamefont {J.}~\bibnamefont {Edsjo}},
  \bibinfo {author} {\bibfnamefont {P.}~\bibnamefont {Ullio}}, \bibinfo
  {author} {\bibfnamefont {L.}~\bibnamefont {Bergstrom}}, \bibinfo {author}
  {\bibfnamefont {M.}~\bibnamefont {Schelke}}, \ and\ \bibinfo {author}
  {\bibfnamefont {E.~A.}\ \bibnamefont {Baltz}},\ }\href {\doibase
  10.1088/1475-7516/2004/07/008} {\bibfield  {journal} {\bibinfo  {journal}
  {JCAP}\ }\textbf {\bibinfo {volume} {0407}},\ \bibinfo {pages} {008}
  (\bibinfo {year} {2004})},\ \Eprint {http://arxiv.org/abs/astro-ph/0406204}
  {arXiv:astro-ph/0406204 [astro-ph]} \BibitemShut {NoStop}%
\bibitem [{\citenamefont {Belanger}\ \emph {et~al.}(2007)\citenamefont
  {Belanger}, \citenamefont {Boudjema}, \citenamefont {Pukhov},\ and\
  \citenamefont {Semenov}}]{Belanger:2006is}%
  \BibitemOpen
  \bibfield  {author} {\bibinfo {author} {\bibfnamefont {G.}~\bibnamefont
  {Belanger}}, \bibinfo {author} {\bibfnamefont {F.}~\bibnamefont {Boudjema}},
  \bibinfo {author} {\bibfnamefont {A.}~\bibnamefont {Pukhov}}, \ and\ \bibinfo
  {author} {\bibfnamefont {A.}~\bibnamefont {Semenov}},\ }\href {\doibase
  10.1016/j.cpc.2006.11.008} {\bibfield  {journal} {\bibinfo  {journal}
  {Comput. Phys. Commun.}\ }\textbf {\bibinfo {volume} {176}},\ \bibinfo
  {pages} {367} (\bibinfo {year} {2007})},\ \Eprint
  {http://arxiv.org/abs/hep-ph/0607059} {arXiv:hep-ph/0607059 [hep-ph]}
  \BibitemShut {NoStop}%
\bibitem [{\citenamefont {Backovic}\ \emph {et~al.}(2014)\citenamefont
  {Backovic}, \citenamefont {Kong},\ and\ \citenamefont
  {McCaskey}}]{Backovic:2013dpa}%
  \BibitemOpen
  \bibfield  {author} {\bibinfo {author} {\bibfnamefont {M.}~\bibnamefont
  {Backovic}}, \bibinfo {author} {\bibfnamefont {K.}~\bibnamefont {Kong}}, \
  and\ \bibinfo {author} {\bibfnamefont {M.}~\bibnamefont {McCaskey}},\ }\href
  {\doibase 10.1016/j.dark.2014.04.001} {\bibfield  {journal} {\bibinfo
  {journal} {Physics of the Dark Universe}\ }\textbf {\bibinfo {volume}
  {5-6}},\ \bibinfo {pages} {18} (\bibinfo {year} {2014})},\ \Eprint
  {http://arxiv.org/abs/1308.4955} {arXiv:1308.4955 [hep-ph]} \BibitemShut
  {NoStop}%
\bibitem [{\citenamefont {Bergeron}\ \emph {et~al.}(2017)\citenamefont
  {Bergeron}, \citenamefont {Sandick},\ and\ \citenamefont
  {Sinha}}]{Bergeron:2017rdm}%
  \BibitemOpen
  \bibfield  {author} {\bibinfo {author} {\bibfnamefont {P.}~\bibnamefont
  {Bergeron}}, \bibinfo {author} {\bibfnamefont {P.}~\bibnamefont {Sandick}}, \
  and\ \bibinfo {author} {\bibfnamefont {K.}~\bibnamefont {Sinha}},\
  }\href@noop {} {\  (\bibinfo {year} {2017})},\ \Eprint
  {http://arxiv.org/abs/1712.05491} {arXiv:1712.05491 [hep-ph]} \BibitemShut
  {NoStop}%
\bibitem [{\citenamefont {Baudis}(2012)}]{Baudis:2012ig}%
  \BibitemOpen
  \bibfield  {author} {\bibinfo {author} {\bibfnamefont {L.}~\bibnamefont
  {Baudis}},\ }\href {\doibase 10.1016/j.dark.2012.10.006} {\bibfield
  {journal} {\bibinfo  {journal} {Phys. Dark Univ.}\ }\textbf {\bibinfo
  {volume} {1}},\ \bibinfo {pages} {94} (\bibinfo {year} {2012})},\ \Eprint
  {http://arxiv.org/abs/1211.7222} {arXiv:1211.7222 [astro-ph.IM]} \BibitemShut
  {NoStop}%
\bibitem [{\citenamefont {Aprile}\ \emph
  {et~al.}(2017{\natexlab{a}})\citenamefont {Aprile} \emph
  {et~al.}}]{Aprile:2017iyp}%
  \BibitemOpen
  \bibfield  {author} {\bibinfo {author} {\bibfnamefont {E.}~\bibnamefont
  {Aprile}} \emph {et~al.} (\bibinfo {collaboration} {XENON}),\ }\href@noop {}
  {\  (\bibinfo {year} {2017}{\natexlab{a}})},\ \Eprint
  {http://arxiv.org/abs/1705.06655} {arXiv:1705.06655 [astro-ph.CO]}
  \BibitemShut {NoStop}%
\bibitem [{\citenamefont {Cui}\ \emph {et~al.}(2017)\citenamefont {Cui} \emph
  {et~al.}}]{Cui:2017nnn}%
  \BibitemOpen
  \bibfield  {author} {\bibinfo {author} {\bibfnamefont {X.}~\bibnamefont
  {Cui}} \emph {et~al.} (\bibinfo {collaboration} {PandaX-II}),\ }\href@noop {}
  {\  (\bibinfo {year} {2017})},\ \Eprint {http://arxiv.org/abs/1708.06917}
  {arXiv:1708.06917 [astro-ph.CO]} \BibitemShut {NoStop}%
\bibitem [{\citenamefont {Marrodán~Undagoitia}\ and\ \citenamefont
  {Rauch}(2016)}]{Undagoitia:2015gya}%
  \BibitemOpen
  \bibfield  {author} {\bibinfo {author} {\bibfnamefont {T.}~\bibnamefont
  {Marrodán~Undagoitia}}\ and\ \bibinfo {author} {\bibfnamefont
  {L.}~\bibnamefont {Rauch}},\ }\href {\doibase 10.1088/0954-3899/43/1/013001}
  {\bibfield  {journal} {\bibinfo  {journal} {J. Phys.}\ }\textbf {\bibinfo
  {volume} {G43}},\ \bibinfo {pages} {013001} (\bibinfo {year} {2016})},\
  \Eprint {http://arxiv.org/abs/1509.08767} {arXiv:1509.08767
  [physics.ins-det]} \BibitemShut {NoStop}%
\bibitem [{\citenamefont {Essig}\ \emph {et~al.}(2012)\citenamefont {Essig},
  \citenamefont {Mardon},\ and\ \citenamefont {Volansky}}]{Essig:2011nj}%
  \BibitemOpen
  \bibfield  {author} {\bibinfo {author} {\bibfnamefont {R.}~\bibnamefont
  {Essig}}, \bibinfo {author} {\bibfnamefont {J.}~\bibnamefont {Mardon}}, \
  and\ \bibinfo {author} {\bibfnamefont {T.}~\bibnamefont {Volansky}},\ }\href
  {\doibase 10.1103/PhysRevD.85.076007} {\bibfield  {journal} {\bibinfo
  {journal} {Phys. Rev.}\ }\textbf {\bibinfo {volume} {D85}},\ \bibinfo {pages}
  {076007} (\bibinfo {year} {2012})},\ \Eprint {http://arxiv.org/abs/1108.5383}
  {arXiv:1108.5383 [hep-ph]} \BibitemShut {NoStop}%
\bibitem [{\citenamefont {D'Eramo}\ \emph {et~al.}(2016)\citenamefont
  {D'Eramo}, \citenamefont {Kavanagh},\ and\ \citenamefont
  {Panci}}]{DEramo:2016gos}%
  \BibitemOpen
  \bibfield  {author} {\bibinfo {author} {\bibfnamefont {F.}~\bibnamefont
  {D'Eramo}}, \bibinfo {author} {\bibfnamefont {B.~J.}\ \bibnamefont
  {Kavanagh}}, \ and\ \bibinfo {author} {\bibfnamefont {P.}~\bibnamefont
  {Panci}},\ }\href {\doibase 10.1007/JHEP08(2016)111} {\bibfield  {journal}
  {\bibinfo  {journal} {JHEP}\ }\textbf {\bibinfo {volume} {08}},\ \bibinfo
  {pages} {111} (\bibinfo {year} {2016})},\ \Eprint
  {http://arxiv.org/abs/1605.04917} {arXiv:1605.04917 [hep-ph]} \BibitemShut
  {NoStop}%
\bibitem [{\citenamefont {Emken}\ \emph {et~al.}(2017)\citenamefont {Emken},
  \citenamefont {Kouvaris},\ and\ \citenamefont {Shoemaker}}]{Emken:2017erx}%
  \BibitemOpen
  \bibfield  {author} {\bibinfo {author} {\bibfnamefont {T.}~\bibnamefont
  {Emken}}, \bibinfo {author} {\bibfnamefont {C.}~\bibnamefont {Kouvaris}}, \
  and\ \bibinfo {author} {\bibfnamefont {I.~M.}\ \bibnamefont {Shoemaker}},\
  }\href {\doibase 10.1103/PhysRevD.96.015018} {\bibfield  {journal} {\bibinfo
  {journal} {Phys. Rev.}\ }\textbf {\bibinfo {volume} {D96}},\ \bibinfo {pages}
  {015018} (\bibinfo {year} {2017})},\ \Eprint
  {http://arxiv.org/abs/1702.07750} {arXiv:1702.07750 [hep-ph]} \BibitemShut
  {NoStop}%
\bibitem [{\citenamefont {Dent}\ \emph {et~al.}(2015)\citenamefont {Dent},
  \citenamefont {Krauss}, \citenamefont {Newstead},\ and\ \citenamefont
  {Sabharwal}}]{Dent:2015zpa}%
  \BibitemOpen
  \bibfield  {author} {\bibinfo {author} {\bibfnamefont {J.~B.}\ \bibnamefont
  {Dent}}, \bibinfo {author} {\bibfnamefont {L.~M.}\ \bibnamefont {Krauss}},
  \bibinfo {author} {\bibfnamefont {J.~L.}\ \bibnamefont {Newstead}}, \ and\
  \bibinfo {author} {\bibfnamefont {S.}~\bibnamefont {Sabharwal}},\ }\href
  {\doibase 10.1103/PhysRevD.92.063515} {\bibfield  {journal} {\bibinfo
  {journal} {Phys. Rev.}\ }\textbf {\bibinfo {volume} {D92}},\ \bibinfo {pages}
  {063515} (\bibinfo {year} {2015})},\ \Eprint
  {http://arxiv.org/abs/1505.03117} {arXiv:1505.03117 [hep-ph]} \BibitemShut
  {NoStop}%
\bibitem [{\citenamefont {Abercrombie}\ \emph {et~al.}(2015)\citenamefont
  {Abercrombie} \emph {et~al.}}]{Abercrombie:2015wmb}%
  \BibitemOpen
  \bibfield  {author} {\bibinfo {author} {\bibfnamefont {D.}~\bibnamefont
  {Abercrombie}} \emph {et~al.},\ }\href@noop {} {\  (\bibinfo {year}
  {2015})},\ \bibinfo {note} {fERMILAB-PUB-15-282-CD},\ \Eprint
  {http://arxiv.org/abs/1507.00966} {arXiv:1507.00966 [hep-ex]} \BibitemShut
  {NoStop}%
\bibitem [{\citenamefont {Catena}\ \emph {et~al.}(2017)\citenamefont {Catena},
  \citenamefont {Conrad}, \citenamefont {Döring}, \citenamefont {Ferella},\
  and\ \citenamefont {Krauss}}]{Catena:2017wzu}%
  \BibitemOpen
  \bibfield  {author} {\bibinfo {author} {\bibfnamefont {R.}~\bibnamefont
  {Catena}}, \bibinfo {author} {\bibfnamefont {J.}~\bibnamefont {Conrad}},
  \bibinfo {author} {\bibfnamefont {C.}~\bibnamefont {Döring}}, \bibinfo
  {author} {\bibfnamefont {A.~D.}\ \bibnamefont {Ferella}}, \ and\ \bibinfo
  {author} {\bibfnamefont {M.~B.}\ \bibnamefont {Krauss}},\ }\href@noop {} {\
  (\bibinfo {year} {2017})},\ \Eprint {http://arxiv.org/abs/1706.09471}
  {arXiv:1706.09471 [hep-ph]} \BibitemShut {NoStop}%
\bibitem [{\citenamefont {Baum}\ \emph {et~al.}(2017)\citenamefont {Baum},
  \citenamefont {Catena}, \citenamefont {Conrad}, \citenamefont {Freese},\ and\
  \citenamefont {Krauss}}]{Baum:2017kfa}%
  \BibitemOpen
  \bibfield  {author} {\bibinfo {author} {\bibfnamefont {S.}~\bibnamefont
  {Baum}}, \bibinfo {author} {\bibfnamefont {R.}~\bibnamefont {Catena}},
  \bibinfo {author} {\bibfnamefont {J.}~\bibnamefont {Conrad}}, \bibinfo
  {author} {\bibfnamefont {K.}~\bibnamefont {Freese}}, \ and\ \bibinfo {author}
  {\bibfnamefont {M.~B.}\ \bibnamefont {Krauss}},\ }\href@noop {} {\  (\bibinfo
  {year} {2017})},\ \Eprint {http://arxiv.org/abs/1709.06051} {arXiv:1709.06051
  [hep-ph]} \BibitemShut {NoStop}%
\bibitem [{\citenamefont {Capdevilla}\ \emph {et~al.}(2017)\citenamefont
  {Capdevilla}, \citenamefont {Delgado}, \citenamefont {Martin},\ and\
  \citenamefont {Raj}}]{Capdevilla:2017doz}%
  \BibitemOpen
  \bibfield  {author} {\bibinfo {author} {\bibfnamefont {R.~M.}\ \bibnamefont
  {Capdevilla}}, \bibinfo {author} {\bibfnamefont {A.}~\bibnamefont {Delgado}},
  \bibinfo {author} {\bibfnamefont {A.}~\bibnamefont {Martin}}, \ and\ \bibinfo
  {author} {\bibfnamefont {N.}~\bibnamefont {Raj}},\ }\href@noop {} {\
  (\bibinfo {year} {2017})},\ \Eprint {http://arxiv.org/abs/1709.00439}
  {arXiv:1709.00439 [hep-ph]} \BibitemShut {NoStop}%
\bibitem [{\citenamefont {Kamon}\ \emph {et~al.}(2017)\citenamefont {Kamon},
  \citenamefont {Ko},\ and\ \citenamefont {Li}}]{Kamon:2017yfx}%
  \BibitemOpen
  \bibfield  {author} {\bibinfo {author} {\bibfnamefont {T.}~\bibnamefont
  {Kamon}}, \bibinfo {author} {\bibfnamefont {P.}~\bibnamefont {Ko}}, \ and\
  \bibinfo {author} {\bibfnamefont {J.}~\bibnamefont {Li}},\ }\href {\doibase
  10.1140/epjc/s10052-017-5240-8} {\bibfield  {journal} {\bibinfo  {journal}
  {Eur. Phys. J.}\ }\textbf {\bibinfo {volume} {C77}},\ \bibinfo {pages} {652}
  (\bibinfo {year} {2017})},\ \Eprint {http://arxiv.org/abs/1705.02149}
  {arXiv:1705.02149 [hep-ph]} \BibitemShut {NoStop}%
\bibitem [{\citenamefont {Queiroz}\ \emph {et~al.}(2017)\citenamefont
  {Queiroz}, \citenamefont {Rodejohann},\ and\ \citenamefont
  {Yaguna}}]{Queiroz:2016sxf}%
  \BibitemOpen
  \bibfield  {author} {\bibinfo {author} {\bibfnamefont {F.~S.}\ \bibnamefont
  {Queiroz}}, \bibinfo {author} {\bibfnamefont {W.}~\bibnamefont {Rodejohann}},
  \ and\ \bibinfo {author} {\bibfnamefont {C.~E.}\ \bibnamefont {Yaguna}},\
  }\href {\doibase 10.1103/PhysRevD.95.095010} {\bibfield  {journal} {\bibinfo
  {journal} {Phys. Rev.}\ }\textbf {\bibinfo {volume} {D95}},\ \bibinfo {pages}
  {095010} (\bibinfo {year} {2017})},\ \Eprint
  {http://arxiv.org/abs/1610.06581} {arXiv:1610.06581 [hep-ph]} \BibitemShut
  {NoStop}%
\bibitem [{\citenamefont {Kavanagh}\ \emph
  {et~al.}(2017{\natexlab{a}})\citenamefont {Kavanagh}, \citenamefont
  {Queiroz}, \citenamefont {Rodejohann},\ and\ \citenamefont
  {Yaguna}}]{Kavanagh:2017hcl}%
  \BibitemOpen
  \bibfield  {author} {\bibinfo {author} {\bibfnamefont {B.~J.}\ \bibnamefont
  {Kavanagh}}, \bibinfo {author} {\bibfnamefont {F.~S.}\ \bibnamefont
  {Queiroz}}, \bibinfo {author} {\bibfnamefont {W.}~\bibnamefont {Rodejohann}},
  \ and\ \bibinfo {author} {\bibfnamefont {C.~E.}\ \bibnamefont {Yaguna}},\
  }\href {\doibase 10.1007/JHEP10(2017)059} {\bibfield  {journal} {\bibinfo
  {journal} {JHEP}\ }\textbf {\bibinfo {volume} {10}},\ \bibinfo {pages} {059}
  (\bibinfo {year} {2017}{\natexlab{a}})},\ \Eprint
  {http://arxiv.org/abs/1706.07819} {arXiv:1706.07819 [hep-ph]} \BibitemShut
  {NoStop}%
\bibitem [{\citenamefont {Fitzpatrick}\ \emph {et~al.}(2013)\citenamefont
  {Fitzpatrick}, \citenamefont {Haxton}, \citenamefont {Katz}, \citenamefont
  {Lubbers},\ and\ \citenamefont {Xu}}]{Fitzpatrick:2012ix}%
  \BibitemOpen
  \bibfield  {author} {\bibinfo {author} {\bibfnamefont {A.~L.}\ \bibnamefont
  {Fitzpatrick}}, \bibinfo {author} {\bibfnamefont {W.}~\bibnamefont {Haxton}},
  \bibinfo {author} {\bibfnamefont {E.}~\bibnamefont {Katz}}, \bibinfo {author}
  {\bibfnamefont {N.}~\bibnamefont {Lubbers}}, \ and\ \bibinfo {author}
  {\bibfnamefont {Y.}~\bibnamefont {Xu}},\ }\href {\doibase
  10.1088/1475-7516/2013/02/004} {\bibfield  {journal} {\bibinfo  {journal}
  {JCAP}\ }\textbf {\bibinfo {volume} {1302}},\ \bibinfo {pages} {004}
  (\bibinfo {year} {2013})},\ \Eprint {http://arxiv.org/abs/1203.3542}
  {arXiv:1203.3542 [hep-ph]} \BibitemShut {NoStop}%
\bibitem [{\citenamefont {Anand}\ \emph {et~al.}(2014)\citenamefont {Anand},
  \citenamefont {Fitzpatrick},\ and\ \citenamefont {Haxton}}]{Anand:2013yka}%
  \BibitemOpen
  \bibfield  {author} {\bibinfo {author} {\bibfnamefont {N.}~\bibnamefont
  {Anand}}, \bibinfo {author} {\bibfnamefont {A.~L.}\ \bibnamefont
  {Fitzpatrick}}, \ and\ \bibinfo {author} {\bibfnamefont {W.~C.}\ \bibnamefont
  {Haxton}},\ }\href {\doibase 10.1103/PhysRevC.89.065501} {\bibfield
  {journal} {\bibinfo  {journal} {Phys. Rev.}\ }\textbf {\bibinfo {volume}
  {C89}},\ \bibinfo {pages} {065501} (\bibinfo {year} {2014})},\ \Eprint
  {http://arxiv.org/abs/1308.6288} {arXiv:1308.6288 [hep-ph]} \BibitemShut
  {NoStop}%
\bibitem [{\citenamefont {Catena}\ and\ \citenamefont
  {Schwabe}(2015)}]{Catena:2015uha}%
  \BibitemOpen
  \bibfield  {author} {\bibinfo {author} {\bibfnamefont {R.}~\bibnamefont
  {Catena}}\ and\ \bibinfo {author} {\bibfnamefont {B.}~\bibnamefont
  {Schwabe}},\ }\href {\doibase 10.1088/1475-7516/2015/04/042} {\bibfield
  {journal} {\bibinfo  {journal} {JCAP}\ }\textbf {\bibinfo {volume} {1504}},\
  \bibinfo {pages} {042} (\bibinfo {year} {2015})},\ \Eprint
  {http://arxiv.org/abs/1501.03729} {arXiv:1501.03729 [hep-ph]} \BibitemShut
  {NoStop}%
\bibitem [{\citenamefont {Catena}(2014{\natexlab{a}})}]{Catena:2014hla}%
  \BibitemOpen
  \bibfield  {author} {\bibinfo {author} {\bibfnamefont {R.}~\bibnamefont
  {Catena}},\ }\href {\doibase 10.1088/1475-7516/2014/09/049} {\bibfield
  {journal} {\bibinfo  {journal} {JCAP}\ }\textbf {\bibinfo {volume} {1409}},\
  \bibinfo {pages} {049} (\bibinfo {year} {2014}{\natexlab{a}})},\ \Eprint
  {http://arxiv.org/abs/1407.0127} {arXiv:1407.0127 [hep-ph]} \BibitemShut
  {NoStop}%
\bibitem [{\citenamefont {Catena}(2014{\natexlab{b}})}]{Catena:2014epa}%
  \BibitemOpen
  \bibfield  {author} {\bibinfo {author} {\bibfnamefont {R.}~\bibnamefont
  {Catena}},\ }\href {\doibase 10.1088/1475-7516/2014/07/055} {\bibfield
  {journal} {\bibinfo  {journal} {JCAP}\ }\textbf {\bibinfo {volume} {1407}},\
  \bibinfo {pages} {055} (\bibinfo {year} {2014}{\natexlab{b}})},\ \Eprint
  {http://arxiv.org/abs/1406.0524} {arXiv:1406.0524 [hep-ph]} \BibitemShut
  {NoStop}%
\bibitem [{\citenamefont {Catena}(2015{\natexlab{a}})}]{Catena:2015vpa}%
  \BibitemOpen
  \bibfield  {author} {\bibinfo {author} {\bibfnamefont {R.}~\bibnamefont
  {Catena}},\ }\href {\doibase 10.1088/1475-7516/2015/07/026} {\bibfield
  {journal} {\bibinfo  {journal} {JCAP}\ }\textbf {\bibinfo {volume} {1507}},\
  \bibinfo {pages} {026} (\bibinfo {year} {2015}{\natexlab{a}})},\ \Eprint
  {http://arxiv.org/abs/1505.06441} {arXiv:1505.06441 [hep-ph]} \BibitemShut
  {NoStop}%
\bibitem [{\citenamefont {Kavanagh}\ \emph
  {et~al.}(2017{\natexlab{b}})\citenamefont {Kavanagh}, \citenamefont
  {Catena},\ and\ \citenamefont {Kouvaris}}]{Kavanagh:2016pyr}%
  \BibitemOpen
  \bibfield  {author} {\bibinfo {author} {\bibfnamefont {B.~J.}\ \bibnamefont
  {Kavanagh}}, \bibinfo {author} {\bibfnamefont {R.}~\bibnamefont {Catena}}, \
  and\ \bibinfo {author} {\bibfnamefont {C.}~\bibnamefont {Kouvaris}},\ }\href
  {\doibase 10.1088/1475-7516/2017/01/012} {\bibfield  {journal} {\bibinfo
  {journal} {JCAP}\ }\textbf {\bibinfo {volume} {1701}},\ \bibinfo {pages}
  {012} (\bibinfo {year} {2017}{\natexlab{b}})},\ \Eprint
  {http://arxiv.org/abs/1611.05453} {arXiv:1611.05453 [hep-ph]} \BibitemShut
  {NoStop}%
\bibitem [{\citenamefont {Catena}\ and\ \citenamefont
  {Kouvaris}(2016{\natexlab{a}})}]{Catena:2016sfr}%
  \BibitemOpen
  \bibfield  {author} {\bibinfo {author} {\bibfnamefont {R.}~\bibnamefont
  {Catena}}\ and\ \bibinfo {author} {\bibfnamefont {C.}~\bibnamefont
  {Kouvaris}},\ }\href {\doibase 10.1103/PhysRevD.94.023527} {\bibfield
  {journal} {\bibinfo  {journal} {Phys. Rev.}\ }\textbf {\bibinfo {volume}
  {D94}},\ \bibinfo {pages} {023527} (\bibinfo {year} {2016}{\natexlab{a}})},\
  \Eprint {http://arxiv.org/abs/1602.00006} {arXiv:1602.00006 [astro-ph.CO]}
  \BibitemShut {NoStop}%
\bibitem [{\citenamefont {Catena}\ and\ \citenamefont
  {Kouvaris}(2016{\natexlab{b}})}]{Catena:2016tlv}%
  \BibitemOpen
  \bibfield  {author} {\bibinfo {author} {\bibfnamefont {R.}~\bibnamefont
  {Catena}}\ and\ \bibinfo {author} {\bibfnamefont {C.}~\bibnamefont
  {Kouvaris}},\ }\href@noop {} {\  (\bibinfo {year} {2016}{\natexlab{b}})},\
  \Eprint {http://arxiv.org/abs/1608.07296} {arXiv:1608.07296 [astro-ph.CO]}
  \BibitemShut {NoStop}%
\bibitem [{\citenamefont {Cirelli}\ \emph {et~al.}(2013)\citenamefont
  {Cirelli}, \citenamefont {Del~Nobile},\ and\ \citenamefont
  {Panci}}]{DelNobile:2013sia}%
  \BibitemOpen
  \bibfield  {author} {\bibinfo {author} {\bibfnamefont {M.}~\bibnamefont
  {Cirelli}}, \bibinfo {author} {\bibfnamefont {E.}~\bibnamefont {Del~Nobile}},
  \ and\ \bibinfo {author} {\bibfnamefont {P.}~\bibnamefont {Panci}},\ }\href
  {\doibase 10.1088/1475-7516/2013/10/019} {\bibfield  {journal} {\bibinfo
  {journal} {JCAP}\ }\textbf {\bibinfo {volume} {1310}},\ \bibinfo {pages}
  {019} (\bibinfo {year} {2013})},\ \Eprint {http://arxiv.org/abs/1307.5955}
  {arXiv:1307.5955 [hep-ph]} \BibitemShut {NoStop}%
\bibitem [{\citenamefont {Catena}\ and\ \citenamefont
  {Gondolo}(2014)}]{Catena:2014uqa}%
  \BibitemOpen
  \bibfield  {author} {\bibinfo {author} {\bibfnamefont {R.}~\bibnamefont
  {Catena}}\ and\ \bibinfo {author} {\bibfnamefont {P.}~\bibnamefont
  {Gondolo}},\ }\href {\doibase 10.1088/1475-7516/2014/09/045} {\bibfield
  {journal} {\bibinfo  {journal} {JCAP}\ }\textbf {\bibinfo {volume} {1409}},\
  \bibinfo {pages} {045} (\bibinfo {year} {2014})},\ \Eprint
  {http://arxiv.org/abs/1405.2637} {arXiv:1405.2637 [hep-ph]} \BibitemShut
  {NoStop}%
\bibitem [{\citenamefont {Gresham}\ and\ \citenamefont
  {Zurek}(2014)}]{Gresham:2014vja}%
  \BibitemOpen
  \bibfield  {author} {\bibinfo {author} {\bibfnamefont {M.~I.}\ \bibnamefont
  {Gresham}}\ and\ \bibinfo {author} {\bibfnamefont {K.~M.}\ \bibnamefont
  {Zurek}},\ }\href@noop {} {\  (\bibinfo {year} {2014})},\ \Eprint
  {http://arxiv.org/abs/1401.3739} {arXiv:1401.3739 [hep-ph]} \BibitemShut
  {NoStop}%
\bibitem [{\citenamefont {Catena}\ and\ \citenamefont
  {Gondolo}(2015)}]{Catena:2015uua}%
  \BibitemOpen
  \bibfield  {author} {\bibinfo {author} {\bibfnamefont {R.}~\bibnamefont
  {Catena}}\ and\ \bibinfo {author} {\bibfnamefont {P.}~\bibnamefont
  {Gondolo}},\ }\href {\doibase 10.1088/1475-7516/2015/08/022} {\bibfield
  {journal} {\bibinfo  {journal} {JCAP}\ }\textbf {\bibinfo {volume} {1508}},\
  \bibinfo {pages} {022} (\bibinfo {year} {2015})},\ \Eprint
  {http://arxiv.org/abs/1504.06554} {arXiv:1504.06554 [hep-ph]} \BibitemShut
  {NoStop}%
\bibitem [{\citenamefont {Catena}(2015{\natexlab{b}})}]{Catena:2015iea}%
  \BibitemOpen
  \bibfield  {author} {\bibinfo {author} {\bibfnamefont {R.}~\bibnamefont
  {Catena}},\ }\href {\doibase 10.1088/1475-7516/2015/04/052} {\bibfield
  {journal} {\bibinfo  {journal} {JCAP}\ }\textbf {\bibinfo {volume} {1504}},\
  \bibinfo {pages} {052} (\bibinfo {year} {2015}{\natexlab{b}})},\ \Eprint
  {http://arxiv.org/abs/1503.04109} {arXiv:1503.04109 [hep-ph]} \BibitemShut
  {NoStop}%
\bibitem [{\citenamefont {Catena}(2017)}]{Catena:2016kro}%
  \BibitemOpen
  \bibfield  {author} {\bibinfo {author} {\bibfnamefont {R.}~\bibnamefont
  {Catena}},\ }\href {\doibase 10.1088/1475-7516/2017/01/059} {\bibfield
  {journal} {\bibinfo  {journal} {JCAP}\ }\textbf {\bibinfo {volume} {1701}},\
  \bibinfo {pages} {059} (\bibinfo {year} {2017})},\ \Eprint
  {http://arxiv.org/abs/1609.08967} {arXiv:1609.08967 [astro-ph.CO]}
  \BibitemShut {NoStop}%
\bibitem [{\citenamefont {Aprile}\ \emph
  {et~al.}(2017{\natexlab{b}})\citenamefont {Aprile} \emph
  {et~al.}}]{Aprile:2017aas}%
  \BibitemOpen
  \bibfield  {author} {\bibinfo {author} {\bibfnamefont {E.}~\bibnamefont
  {Aprile}} \emph {et~al.} (\bibinfo {collaboration} {XENON100}),\ }\href
  {\doibase 10.1103/PhysRevD.96.042004} {\bibfield  {journal} {\bibinfo
  {journal} {Phys. Rev.}\ }\textbf {\bibinfo {volume} {D96}},\ \bibinfo {pages}
  {042004} (\bibinfo {year} {2017}{\natexlab{b}})},\ \Eprint
  {http://arxiv.org/abs/1705.02614} {arXiv:1705.02614 [astro-ph.CO]}
  \BibitemShut {NoStop}%
\bibitem [{\citenamefont {Fan}\ \emph {et~al.}(2010)\citenamefont {Fan},
  \citenamefont {Reece},\ and\ \citenamefont {Wang}}]{Fan:2010gt}%
  \BibitemOpen
  \bibfield  {author} {\bibinfo {author} {\bibfnamefont {J.}~\bibnamefont
  {Fan}}, \bibinfo {author} {\bibfnamefont {M.}~\bibnamefont {Reece}}, \ and\
  \bibinfo {author} {\bibfnamefont {L.-T.}\ \bibnamefont {Wang}},\ }\href
  {\doibase 10.1088/1475-7516/2010/11/042} {\bibfield  {journal} {\bibinfo
  {journal} {JCAP}\ }\textbf {\bibinfo {volume} {1011}},\ \bibinfo {pages}
  {042} (\bibinfo {year} {2010})},\ \Eprint {http://arxiv.org/abs/1008.1591}
  {arXiv:1008.1591 [hep-ph]} \BibitemShut {NoStop}%
\bibitem [{\citenamefont {Agrawal}\ \emph {et~al.}(2010)\citenamefont
  {Agrawal}, \citenamefont {Chacko}, \citenamefont {Kilic},\ and\ \citenamefont
  {Mishra}}]{Agrawal:2010fh}%
  \BibitemOpen
  \bibfield  {author} {\bibinfo {author} {\bibfnamefont {P.}~\bibnamefont
  {Agrawal}}, \bibinfo {author} {\bibfnamefont {Z.}~\bibnamefont {Chacko}},
  \bibinfo {author} {\bibfnamefont {C.}~\bibnamefont {Kilic}}, \ and\ \bibinfo
  {author} {\bibfnamefont {R.~K.}\ \bibnamefont {Mishra}},\ }\href@noop {} {\
  (\bibinfo {year} {2010})},\ \Eprint {http://arxiv.org/abs/1003.1912}
  {arXiv:1003.1912 [hep-ph]} \BibitemShut {NoStop}%
\bibitem [{\citenamefont {Dienes}\ \emph {et~al.}(2014)\citenamefont {Dienes},
  \citenamefont {Kumar}, \citenamefont {Thomas},\ and\ \citenamefont
  {Yaylali}}]{Dienes:2013xya}%
  \BibitemOpen
  \bibfield  {author} {\bibinfo {author} {\bibfnamefont {K.~R.}\ \bibnamefont
  {Dienes}}, \bibinfo {author} {\bibfnamefont {J.}~\bibnamefont {Kumar}},
  \bibinfo {author} {\bibfnamefont {B.}~\bibnamefont {Thomas}}, \ and\ \bibinfo
  {author} {\bibfnamefont {D.}~\bibnamefont {Yaylali}},\ }\href {\doibase
  10.1103/PhysRevD.90.015012} {\bibfield  {journal} {\bibinfo  {journal} {Phys.
  Rev.}\ }\textbf {\bibinfo {volume} {D90}},\ \bibinfo {pages} {015012}
  (\bibinfo {year} {2014})},\ \Eprint {http://arxiv.org/abs/1312.7772}
  {arXiv:1312.7772 [hep-ph]} \BibitemShut {NoStop}%
\bibitem [{\citenamefont {Bishara}\ \emph {et~al.}(2017)\citenamefont
  {Bishara}, \citenamefont {Brod}, \citenamefont {Grinstein},\ and\
  \citenamefont {Zupan}}]{Bishara:2016hek}%
  \BibitemOpen
  \bibfield  {author} {\bibinfo {author} {\bibfnamefont {F.}~\bibnamefont
  {Bishara}}, \bibinfo {author} {\bibfnamefont {J.}~\bibnamefont {Brod}},
  \bibinfo {author} {\bibfnamefont {B.}~\bibnamefont {Grinstein}}, \ and\
  \bibinfo {author} {\bibfnamefont {J.}~\bibnamefont {Zupan}},\ }\href
  {\doibase 10.1088/1475-7516/2017/02/009} {\bibfield  {journal} {\bibinfo
  {journal} {JCAP}\ }\textbf {\bibinfo {volume} {1702}},\ \bibinfo {pages}
  {009} (\bibinfo {year} {2017})},\ \Eprint {http://arxiv.org/abs/1611.00368}
  {arXiv:1611.00368 [hep-ph]} \BibitemShut {NoStop}%
\bibitem [{\citenamefont {Hoferichter}\ \emph {et~al.}(2015)\citenamefont
  {Hoferichter}, \citenamefont {Klos},\ and\ \citenamefont
  {Schwenk}}]{Hoferichter:2015ipa}%
  \BibitemOpen
  \bibfield  {author} {\bibinfo {author} {\bibfnamefont {M.}~\bibnamefont
  {Hoferichter}}, \bibinfo {author} {\bibfnamefont {P.}~\bibnamefont {Klos}}, \
  and\ \bibinfo {author} {\bibfnamefont {A.}~\bibnamefont {Schwenk}},\ }\href
  {\doibase 10.1016/j.physletb.2015.05.041} {\bibfield  {journal} {\bibinfo
  {journal} {Phys. Lett.}\ }\textbf {\bibinfo {volume} {B746}},\ \bibinfo
  {pages} {410} (\bibinfo {year} {2015})},\ \Eprint
  {http://arxiv.org/abs/1503.04811} {arXiv:1503.04811 [hep-ph]} \BibitemShut
  {NoStop}%
\bibitem [{\citenamefont {Crivellin}\ \emph {et~al.}(2014)\citenamefont
  {Crivellin}, \citenamefont {D'Eramo},\ and\ \citenamefont
  {Procura}}]{Crivellin:2014qxa}%
  \BibitemOpen
  \bibfield  {author} {\bibinfo {author} {\bibfnamefont {A.}~\bibnamefont
  {Crivellin}}, \bibinfo {author} {\bibfnamefont {F.}~\bibnamefont {D'Eramo}},
  \ and\ \bibinfo {author} {\bibfnamefont {M.}~\bibnamefont {Procura}},\ }\href
  {\doibase 10.1103/PhysRevLett.112.191304} {\bibfield  {journal} {\bibinfo
  {journal} {Phys. Rev. Lett.}\ }\textbf {\bibinfo {volume} {112}},\ \bibinfo
  {pages} {191304} (\bibinfo {year} {2014})},\ \Eprint
  {http://arxiv.org/abs/1402.1173} {arXiv:1402.1173 [hep-ph]} \BibitemShut
  {NoStop}%
\bibitem [{\citenamefont {Aprile}\ \emph {et~al.}(2011)\citenamefont {Aprile}
  \emph {et~al.}}]{Aprile:2011hx}%
  \BibitemOpen
  \bibfield  {author} {\bibinfo {author} {\bibfnamefont {E.}~\bibnamefont
  {Aprile}} \emph {et~al.} (\bibinfo {collaboration} {XENON100
  Collaboration}),\ }\href {\doibase 10.1103/PhysRevD.84.052003} {\bibfield
  {journal} {\bibinfo  {journal} {Phys. Rev.}\ }\textbf {\bibinfo {volume}
  {D84}},\ \bibinfo {pages} {052003} (\bibinfo {year} {2011})},\ \Eprint
  {http://arxiv.org/abs/1103.0303} {arXiv:1103.0303 [hep-ex]} \BibitemShut
  {NoStop}%
\bibitem [{\citenamefont {Aprile}\ \emph {et~al.}(2016)\citenamefont {Aprile}
  \emph {et~al.}}]{Aprile:2015uzo}%
  \BibitemOpen
  \bibfield  {author} {\bibinfo {author} {\bibfnamefont {E.}~\bibnamefont
  {Aprile}} \emph {et~al.} (\bibinfo {collaboration} {XENON}),\ }\href
  {\doibase 10.1088/1475-7516/2016/04/027} {\bibfield  {journal} {\bibinfo
  {journal} {JCAP}\ }\textbf {\bibinfo {volume} {1604}},\ \bibinfo {pages}
  {027} (\bibinfo {year} {2016})},\ \Eprint {http://arxiv.org/abs/1512.07501}
  {arXiv:1512.07501 [physics.ins-det]} \BibitemShut {NoStop}%
\bibitem [{\citenamefont {Catena}\ and\ \citenamefont
  {Ullio}(2010)}]{Catena:2009mf}%
  \BibitemOpen
  \bibfield  {author} {\bibinfo {author} {\bibfnamefont {R.}~\bibnamefont
  {Catena}}\ and\ \bibinfo {author} {\bibfnamefont {P.}~\bibnamefont {Ullio}},\
  }\href {\doibase 10.1088/1475-7516/2010/08/004} {\bibfield  {journal}
  {\bibinfo  {journal} {JCAP}\ }\textbf {\bibinfo {volume} {1008}},\ \bibinfo
  {pages} {004} (\bibinfo {year} {2010})},\ \Eprint
  {http://arxiv.org/abs/0907.0018} {arXiv:0907.0018 [astro-ph.CO]} \BibitemShut
  {NoStop}%
\bibitem [{\citenamefont {Salucci}\ \emph {et~al.}(2010)\citenamefont
  {Salucci}, \citenamefont {Nesti}, \citenamefont {Gentile},\ and\
  \citenamefont {Martins}}]{Salucci:2010qr}%
  \BibitemOpen
  \bibfield  {author} {\bibinfo {author} {\bibfnamefont {P.}~\bibnamefont
  {Salucci}}, \bibinfo {author} {\bibfnamefont {F.}~\bibnamefont {Nesti}},
  \bibinfo {author} {\bibfnamefont {G.}~\bibnamefont {Gentile}}, \ and\
  \bibinfo {author} {\bibfnamefont {C.~F.}\ \bibnamefont {Martins}},\ }\href
  {\doibase 10.1051/0004-6361/201014385} {\bibfield  {journal} {\bibinfo
  {journal} {Astron. Astrophys.}\ }\textbf {\bibinfo {volume} {523}},\ \bibinfo
  {pages} {A83} (\bibinfo {year} {2010})},\ \Eprint
  {http://arxiv.org/abs/1003.3101} {arXiv:1003.3101 [astro-ph.GA]} \BibitemShut
  {NoStop}%
\bibitem [{\citenamefont {Catena}\ and\ \citenamefont
  {Ullio}(2012)}]{Catena:2011kv}%
  \BibitemOpen
  \bibfield  {author} {\bibinfo {author} {\bibfnamefont {R.}~\bibnamefont
  {Catena}}\ and\ \bibinfo {author} {\bibfnamefont {P.}~\bibnamefont {Ullio}},\
  }\href {\doibase 10.1088/1475-7516/2012/05/005} {\bibfield  {journal}
  {\bibinfo  {journal} {JCAP}\ }\textbf {\bibinfo {volume} {1205}},\ \bibinfo
  {pages} {005} (\bibinfo {year} {2012})},\ \Eprint
  {http://arxiv.org/abs/1111.3556} {arXiv:1111.3556 [astro-ph.CO]} \BibitemShut
  {NoStop}%
\bibitem [{\citenamefont {Pato}\ \emph {et~al.}(2015)\citenamefont {Pato},
  \citenamefont {Iocco},\ and\ \citenamefont {Bertone}}]{Pato:2015dua}%
  \BibitemOpen
  \bibfield  {author} {\bibinfo {author} {\bibfnamefont {M.}~\bibnamefont
  {Pato}}, \bibinfo {author} {\bibfnamefont {F.}~\bibnamefont {Iocco}}, \ and\
  \bibinfo {author} {\bibfnamefont {G.}~\bibnamefont {Bertone}},\ }\href
  {\doibase 10.1088/1475-7516/2015/12/001} {\bibfield  {journal} {\bibinfo
  {journal} {JCAP}\ }\textbf {\bibinfo {volume} {1512}},\ \bibinfo {pages}
  {001} (\bibinfo {year} {2015})},\ \Eprint {http://arxiv.org/abs/1504.06324}
  {arXiv:1504.06324 [astro-ph.GA]} \BibitemShut {NoStop}%
\bibitem [{\citenamefont {Benito}\ \emph {et~al.}(2017)\citenamefont {Benito},
  \citenamefont {Bernal}, \citenamefont {Bozorgnia}, \citenamefont {Calore},\
  and\ \citenamefont {Iocco}}]{Benito:2016kyp}%
  \BibitemOpen
  \bibfield  {author} {\bibinfo {author} {\bibfnamefont {M.}~\bibnamefont
  {Benito}}, \bibinfo {author} {\bibfnamefont {N.}~\bibnamefont {Bernal}},
  \bibinfo {author} {\bibfnamefont {N.}~\bibnamefont {Bozorgnia}}, \bibinfo
  {author} {\bibfnamefont {F.}~\bibnamefont {Calore}}, \ and\ \bibinfo {author}
  {\bibfnamefont {F.}~\bibnamefont {Iocco}},\ }\href {\doibase
  10.1088/1475-7516/2017/02/007} {\bibfield  {journal} {\bibinfo  {journal}
  {JCAP}\ }\textbf {\bibinfo {volume} {1702}},\ \bibinfo {pages} {007}
  (\bibinfo {year} {2017})},\ \Eprint {http://arxiv.org/abs/1612.02010}
  {arXiv:1612.02010 [hep-ph]} \BibitemShut {NoStop}%
\bibitem [{\citenamefont {Green}(2017)}]{Green:2017odb}%
  \BibitemOpen
  \bibfield  {author} {\bibinfo {author} {\bibfnamefont {A.~M.}\ \bibnamefont
  {Green}},\ }\href {\doibase 10.1088/1361-6471/aa7819} {\bibfield  {journal}
  {\bibinfo  {journal} {J. Phys.}\ }\textbf {\bibinfo {volume} {G44}},\
  \bibinfo {pages} {084001} (\bibinfo {year} {2017})},\ \Eprint
  {http://arxiv.org/abs/1703.10102} {arXiv:1703.10102 [astro-ph.CO]}
  \BibitemShut {NoStop}%
\bibitem [{\citenamefont {Mertig}\ \emph {et~al.}(1991)\citenamefont {Mertig},
  \citenamefont {Bohm},\ and\ \citenamefont {Denner}}]{Mertig:1990an}%
  \BibitemOpen
  \bibfield  {author} {\bibinfo {author} {\bibfnamefont {R.}~\bibnamefont
  {Mertig}}, \bibinfo {author} {\bibfnamefont {M.}~\bibnamefont {Bohm}}, \ and\
  \bibinfo {author} {\bibfnamefont {A.}~\bibnamefont {Denner}},\ }\href
  {\doibase 10.1016/0010-4655(91)90130-D} {\bibfield  {journal} {\bibinfo
  {journal} {Comput. Phys. Commun.}\ }\textbf {\bibinfo {volume} {64}},\
  \bibinfo {pages} {345} (\bibinfo {year} {1991})}\BibitemShut {NoStop}%
\bibitem [{\citenamefont {Shtabovenko}\ \emph {et~al.}(2016)\citenamefont
  {Shtabovenko}, \citenamefont {Mertig},\ and\ \citenamefont
  {Orellana}}]{Shtabovenko:2016sxi}%
  \BibitemOpen
  \bibfield  {author} {\bibinfo {author} {\bibfnamefont {V.}~\bibnamefont
  {Shtabovenko}}, \bibinfo {author} {\bibfnamefont {R.}~\bibnamefont {Mertig}},
  \ and\ \bibinfo {author} {\bibfnamefont {F.}~\bibnamefont {Orellana}},\
  }\href {\doibase 10.1016/j.cpc.2016.06.008} {\bibfield  {journal} {\bibinfo
  {journal} {Comput. Phys. Commun.}\ }\textbf {\bibinfo {volume} {207}},\
  \bibinfo {pages} {432} (\bibinfo {year} {2016})},\ \Eprint
  {http://arxiv.org/abs/1601.01167} {arXiv:1601.01167 [hep-ph]} \BibitemShut
  {NoStop}%
\bibitem [{\citenamefont {Kilian}\ \emph {et~al.}(2011)\citenamefont {Kilian},
  \citenamefont {Ohl},\ and\ \citenamefont {Reuter}}]{Kilian:2007gr}%
  \BibitemOpen
  \bibfield  {author} {\bibinfo {author} {\bibfnamefont {W.}~\bibnamefont
  {Kilian}}, \bibinfo {author} {\bibfnamefont {T.}~\bibnamefont {Ohl}}, \ and\
  \bibinfo {author} {\bibfnamefont {J.}~\bibnamefont {Reuter}},\ }\href
  {\doibase 10.1140/epjc/s10052-011-1742-y} {\bibfield  {journal} {\bibinfo
  {journal} {Eur. Phys. J.}\ }\textbf {\bibinfo {volume} {C71}},\ \bibinfo
  {pages} {1742} (\bibinfo {year} {2011})},\ \Eprint
  {http://arxiv.org/abs/0708.4233} {arXiv:0708.4233 [hep-ph]} \BibitemShut
  {NoStop}%
\end{thebibliography}
%

\end{document}